\documentclass[12pt]{article}
\pdfoutput=1
\usepackage{putex}
\usepackage{graphicx}
\usepackage{epstopdf}
\usepackage{enumerate}
\usepackage[nosort]{cite}
\usepackage{tensor}
\usepackage{slashed}
\usepackage{feynmf}
\usepackage{hyperref}
\usepackage{float}
\usepackage[bottom]{footmisc}
\usepackage[utf8]{inputenc}
\usepackage{amsmath}
\usepackage{xfrac}
\usepackage[usenames,dvipsnames,svgnames,table]{xcolor}
\usepackage[bf,hang,footnotesize]{caption}
\usepackage{bbold}

\usepackage{makecell}
\usepackage{graphicx,wrapfig,lipsum,float,subcaption,caption,sidecap}
\usepackage{hyperref,enumerate,array,multirow,tabularx,bm,soul,multicol}
\usepackage[toc,page]{appendix}
\usepackage{tikz}
\usetikzlibrary{intersections,shapes.geometric}
\usepackage{tkz-euclide}
\usepackage{pgfplots}
\usepgfplotslibrary{fillbetween}
\usepackage{tikz}
\usetikzlibrary{decorations.pathreplacing,decorations.markings}
\usetikzlibrary{calc}

\newcommand{\Hom}{\operatorname{Hom}}
\newcommand{\End}{\operatorname{End}}

\numberwithin{equation}{section}
\hypersetup{
	pdfstartview={FitH},    % fits the width of the page to the window
	pdftitle={Title},    % title
	pdfauthor={Kim, Kraus, Sun},     % authors
	colorlinks=true,       % false: boxed links; true: colored links
	linkcolor=blue,          % color of internal links
	citecolor=blue!59!black! ,        % color of links to bibliography
	filecolor=magenta,      % color of file links
	urlcolor=blue!75!black,           % color of external links
	linktoc=all
}

\makeatletter
\g@addto@macro\bfseries{\boldmath}
\makeatother

\numberwithin{equation}{section}

\newcommand{\arrow}{\rightarrow}
\newcommand{\lls}{[\![}
\newcommand{\rrs}{]\!]}

\newcommand{\RN}[1]{%
	\textup{\uppercase\expandafter{\romannumeral#1}}%
}
\newcommand{\dr}{\mathrm{d}}
\newcommand{\ca}{\mathcal{A}}
\newcommand{\lag}{\mathcal{L}}
\newcommand{\bd}{\mathbb{D}}
\newcommand{\bz}{\mathbb{Z}}
\newcommand{\bi}{\mathbb{1}}
\newcommand{\ba}{\mathbb{A}}

\numberwithin{equation}{section}

\usepackage[framemethod=TikZ]{mdframed} % fancy boxes
\usepackage{tikz-cd} 
\usetikzlibrary{decorations.markings}

\tikzset{line/.style={line width=0.25mm},
curve/.style={line,smooth,tension=1},
->-/.style={decoration={
  markings,
  mark=at position #1 with {\arrow[>=stealth]{>}}},postaction={decorate}},
-<-/.style={decoration={
  markings,
  mark=at position #1 with {\arrow[>=stealth]{<}}},postaction={decorate}},
}

\tikzset{
    partial ellipse/.style args={#1:#2:#3}{
        insert path={+ (#1:#3) arc (#1:#2:#3)}
    }
}

\definecolor{azure}{rgb}{0.0, 0.5, 1.0}
\definecolor{darkblue}{rgb}{0.15,0.35,0.7}
\definecolor{reddish}{rgb}{0.65, 0.2, 0.2}
\definecolor{brandeisblue}{rgb}{0.0, 0.44, 1.0}
\definecolor{ceruleanblue}{rgb}{0.16, 0.32, 0.75}
\definecolor{indigo(dye)}{rgb}{0.0, 0.25, 0.42}
\definecolor{dgrey}{rgb}{0.3,0.3,0.3}
\definecolor{grey}{rgb}{0.9,0.9,0.9}

\usepackage{comment}

\begin{document}

\institution{UCLA}{ \quad\quad\quad\quad\quad\quad\quad\ ~ \, $^{1}$Mani L. Bhaumik Institute for Theoretical Physics
		\cr Department of Physics \& Astronomy,\,University of California,\,Los Angeles,\,CA\,90095,\,USA}

    \institution{EPFL}{ \quad\quad\quad\quad\quad\quad\quad\ ~ \, $^{2}$Laboratory for Theoretical Fundamental Physics 
		\cr Institute of Physics, Ecole Polytechnique Federale
    de Lausanne (EPFL), %CH-1015 Lausanne, 
    Switzerland}

\title{Higher structure of non-invertible symmetries from Lagrangian descriptions
}

\authors{Seolhwa Kim$^1$, Orr Sela$^{2}$, Zhengdi Sun$^{1}$}
	
\abstract{The symmetry structure of a quantum field theory is determined not only by the topological defects that implement the symmetry and their fusion rules, but also by the topological networks they can form, which is referred to as the \textit{higher structure} of the symmetry. In this paper, we consider theories with non-invertible symmetries that have an explicit Lagrangian description, and use it to study their higher structure. Starting with the 2d free compact boson theory and its non-invertible duality defects, we will find Lagrangian descriptions of networks of defects and use them to recover all the $F$-symbols of the familiar Tambara-Yamagami fusion category $\operatorname{TY}(\mathbb{Z}_N,+1)$. We will then use the same approach in 4d Maxwell theory to compute $F$-symbols associated with its non-invertible duality and triality defects, which are 2d topological field theories. In addition, we will also compute some of the $F$-symbols using a different (group theoretical) approach that is not based on the Lagrangian description, and find that they take the expected form.}
	
	\date{}
	
	\maketitle
	\setcounter{tocdepth}{2}
	\begingroup
	\hypersetup{linkcolor=black}
	\tableofcontents
	\endgroup
	
%%%%%%%%%%%%%%%%%%%%%%%%%%%%%%%%%%%%%%%%%%%%%%%%%%%

\section{Introduction}

Global symmetries have proven to be invaluable in the study of quantum field theories, and in the modern perspective, they are viewed as implemented by topological operators \cite{Gaiotto:2014kfa}. In order to fully specify the symmetry structure of a given theory, one begins by identifying these topological operators and their fusion rules. However, this is not all, and some additional physical information is encoded in the topological junctions and networks that these defects can form, which is referred to as the \textit{higher structure} of the symmetry. This includes, in particular, the set of topological interfaces between pairs of defects, called 1-morphisms, which describe junctions (including fusion junctions) between defects and are also used to encode the associativity structure of the fusion product. This associativity 1-morphism, known as the associator or $F$-symbol, is an interface obtained by shrinking a certain topological network of defects that captures this associativity (see the next section for more details), and is an important piece of information in specifying the symmetry structure. In two dimensions, the $F$-symbols are numbers, but in a higher dimension $D$, they are $D-2$ dimensional topological field theories, and nontrivial interfaces (or 2-morphisms) might also be considered between them. This gives rise to the notion of higher associators, where the $F$-symbols are considered as 1-associators. Overall, we have $D-1$ higher associators, where the $(D-1)$-associators are numbers. All the associators are part of the mathematical structure describing the symmetry, which is called a fusion $(D-1)$-category, and contain physical information \cite{Copetti:2023mcq}. As an example, let us note that for ordinary symmetries with group-like fusion rules (that is, invertible symmetries), the $(D-1)$-associators are simply the ’t Hooft anomalies of the symmetries. 

In many occasions, the fusion of the topological defects does not follow a simple group-like law, and the corresponding symmetry is then referred to as non-invertible (see \cite{Schafer-Nameki:2023jdn,Shao:2023gho} for reviews). Our focus in this paper will be on studying the higher structure of non-invertible symmetries in theories in which an explicit Lagrangian description of the different defects is known. We will use this Lagrangian description to construct explicitly topological junctions between the defects, including fusion junctions in which two defects are fused into a third one, while accounting for the appropriate boundary conditions and boundary terms at the junctions. Then, we will use these junctions as building blocks to construct the topological configurations of defects required for the computation of the $F$-symbols, and obtain the desired results after carefully shrinking the configuration. We note that even though our focus here will be on the $F$-symbols, i.e. the 1-associators, the same method can similarly be employed for the computation of the higher associators.  

To illustrate our approach, we will analyze in detail two types of theories. The first will be that of the free compact scalar in two dimensions, whose non-invertible symmetries and associated fusion 1-category have been widely studied and are well known (see \cite{Thorngren:2021yso,Fuchs:2007tx,Kapustin:2009av,Bhardwaj:2017xup,Choi:2021kmx,Niro:2022ctq,Argurio:2024ewp} for a partial list of references). This will serve as a good starting point for demonstrating our new method and testing it against known results. Using the Lagrangian descriptions of the shift-symmetry defects and the non-invertible $T$-duality defects that are present in the theory, we will construct topological junctions and use them to study the networks of defects that capture the different $F$-symbols. We will compute all the $F$-symbols this way and find that they indeed match those of the familiar Tambara-Yamagami fusion category $\operatorname{TY}(\mathbb{Z}_N,+1)$. 

We will then turn to four dimensions and consider Maxwell theory, which was recently shown to host a variety of interesting non-invertible symmetries \cite{Choi:2021kmx,Niro:2022ctq,Kaidi:2021xfk,Roumpedakis:2022aik,Choi:2022zal,Hasan:2024aow,Paznokas:2025auw,Sela:2024okz,Shao:2025qvf}.  In this case, even though the corresponding topological defects and their fusions were studied, the higher-structure data was not computed explicitly (see \cite{Bhardwaj:2024xcx,DelZotto:2024ngj,Bah:2025oxi} for related work). Focusing on the triality and duality defects found in \cite{Choi:2022zal,Choi:2021kmx,Kaidi:2021xfk}, we will use their Lagrangian descriptions to find the actions describing the topological configurations of defects that capture the $F$-symbols of the corresponding fusion 3-categories. Then, we will shrink these networks to interfaces and find these $F$-symbols, which are 2d topological field theories and new results of this paper (see Table~\ref{tab: triality associators} and Eq.~\eqref{eq: duality associators}). In some cases, we will find that there is more than one possible 2d theory corresponding to a given $F$-symbol, depending on the choice of boundary conditions used in its computation. In addition to these calculations, as a complementary analysis and to check our results, we will also compute the $F$-symbols using a different (group theoretical) approach that is not based on the Lagrangian description, and find that the results obtained using the two methods indeed have the same form.\footnote{The triality and duality defects we consider are parameterized by a positive integer $N$, and in the case of the triality defects the Lagrangian approach will only be used for even $N$ while the group-theoretical one only for certain odd $N$ values. In this case, the two methods will match except for some cases where the final result depends on whether $N$ is even or odd.} 

The remainder of the paper is organized as follows. In section~\ref{sec: higher structure} we briefly review the higher structure of non-invertible symmetries, focusing on the definition of $F$-symbols in two and higher dimensions. In section~\ref{sec: compact boson} we consider the 2d free compact boson theory, and use the Lagrangian description of the symmetry defects to construct topological junctions and networks of them. We then use it to compute the $F$-symbols of the Tambara-Yamagami fusion category $\operatorname{TY}(\mathbb{Z}_N,+1)$ and recover the known expressions. In section~\ref{sec: 4dtri} we generalize this analysis to the triality defects of 4d Maxwell theory for even values of the integer that parameterizes them, computing the corresponding $F$-symbols while highlighting some differences compared to the previous section. Then, in section~\ref{Sec_Group}, we compute the $F$-symbols for certain odd values of the parameter using a different (group theoretical) approach and find that they take the expected form. A few appendices contain constructions of topological junctions and computations of $F$-symbols that have been deferred from the main body of the paper for brevity, and include the case of the duality defect of 4d Maxwell theory.

% ------------------------------------------
% ------------ new section -----------------
% ------------------------------------------

\section{Higher structure of symmetry}\label{sec: higher structure}
In this subsection, we briefly review the (higher) fusion category symmetries and their higher structures with a focus on the associators. We begin by reviewing the associators of fusion category symmetry in 2d to set the stage, and then move to explain the associators in higher fusion categories following \cite{Copetti:2023mcq}.

\subsection{Higher structure in fusion categories}
\label{Fusion}
Topological defect lines (TDLs) in a 2d QFT are captured by the mathematical structure known as a fusion category. Given two TDLs labeled by $\mathcal{L}_a$ and $\mathcal{L}_b$, we can fuse them by putting them close to each other and generate a new topological line, which then in general decomposes into a finite sum of other topological lines,
\begin{equation}
    \mathcal{L}_a \times \mathcal{L}_b = \bigoplus_{c} N_{ab}^c \mathcal{L}_c ~, \quad N_{ab}^c \in \mathbb{Z}_{\geq 0} ~.
\end{equation}
A TDL which cannot be written as a sum of other TDLs is known as a \textit{simple} TDL. 

When $N_{ab}^c \neq 0$, the locality of the QFT allows the configurations where $\mathcal{L}_a$ and $\mathcal{L}_b$ join locally and become $\mathcal{L}_c$ at a trivalent junction. Such topological junctions form a complex vector space whose dimension is $N_{ab}^c$. Similarly, the TDL $\mathcal{L}_c$ can locally split into $\mathcal{L}_a$ and $\mathcal{L}_b$, and the topological junctions also form a vector space with the same dimension. To proceed, we fix the basis for each vector space and denote them as
\begin{equation}\label{eq:f1c_basis}
    \begin{tikzpicture}[scale=0.8,baseline={([yshift=-.5ex]current bounding box.center)},vertex/.style={anchor=base,
    circle,fill=black!25,minimum size=18pt,inner sep=2pt},scale=0.50]
    \draw[thick, black] (-2,-2) -- (0,0);
    \draw[thick, black] (+2,-2) -- (0,0);
    \draw[thick, black] (0,0) -- (0,2);
    \draw[thick, black, -stealth] (-2,-2) -- (-1,-1);
    \draw[thick, black, -stealth] (+2,-2) -- (1,-1);
    \draw[thick, black, -stealth] (0,0) -- (0,1);
    \filldraw[thick, black] (0,0) circle (3pt);
    \node[black, right] at (0,0) {\footnotesize $\mu$};
    \node[black, below] at (-2,-2) {$\mathcal{L}_a$};
    \node[black, below] at (2,-2) {$\mathcal{L}_b$};
    \node[black, above] at (0,2) {$\mathcal{L}_c$};
    
\end{tikzpicture} ~, \quad \begin{tikzpicture}[scale=0.8,baseline={([yshift=-.5ex]current bounding box.center)},vertex/.style={anchor=base,
    circle,fill=black!25,minimum size=18pt,inner sep=2pt},scale=0.50]
    \draw[thick, black, ->-=0.5] (0,0) -- (-2,+2);
    \draw[thick, black, ->-=0.5] (0,0) -- (+2,+2);
    \draw[thick, black, ->-=0.5] (0,-2) -- (0,0);
    \filldraw[thick, black] (0,0) circle (3pt);
    \node[black, right] at (0,0) {\footnotesize $\mu$};
    \node[black, above] at (-2,+2) {$\mathcal{L}_a$};
    \node[black, above] at (+2,+2) {$\mathcal{L}_b$};
    \node[black, below] at (0,-2) {$\mathcal{L}_c$};
    
\end{tikzpicture} ~, \quad \mu = 1,2,\cdots, N^{c}_{ab} \,,
\end{equation}
satisfying the following conditions
\begin{equation}\label{eq:fus_spl_relation}
\begin{tikzpicture}[baseline={([yshift=-1ex]current bounding box.center)},vertex/.style={anchor=base,
    circle,fill=black!25,minimum size=18pt,inner sep=2pt},scale=0.5]
    \draw[black, line width = 0.4mm, ->-=0.5] (1.333,-2) -- (1.333,3);
    \draw[black, line width = 0.4mm, ->-=0.5] (2.667,-2) -- (2.667,3);
    \node[black, below] at (1.333,-2) {\footnotesize $a$};
    \node[black, below] at (2.667,-2) {\footnotesize $b$};
    \node[above, above] at (1.333,3) {\footnotesize $a$};
    \node[above, above] at (2.667,3) {\footnotesize $b$};
    \end{tikzpicture}
= \sum_{c,\mu} \sqrt{\frac{d_c}{d_a d_b}} \, \begin{tikzpicture}[baseline={([yshift=-1ex]current bounding box.center)},vertex/.style={anchor=base,
    circle,fill=black!25,minimum size=18pt,inner sep=2pt},scale=0.5]
        \draw[black, line width = 0.4mm, ->-=0.5] (2.667,2) -- (2.667,3);
        \draw[black, line width = 0.4mm, ->-=0.5] (1.333,2) -- (1.333,3);
        \draw[black, line width = 0.4mm] (1.333,2) arc (180:270:0.667);
        \draw[black, line width = 0.4mm] (2.667,2) arc (0:-90:0.667);
        \draw[black, thick, ->-=.5, line width = 0.4mm] (2,-0.333) -- (2,1.333);
        \draw[black, line width = 0.4mm] (1.333,-1) arc (180:90:0.667);
        \draw[black, line width = 0.4mm] (2.667,-1) arc (0:90:0.667);
        \draw[black, line width = 0.4mm, ->-=0.8] (1.333,-2) -- (1.333,-1);
        \draw[black, line width = 0.4mm, ->-=0.8] (2.667,-2) -- (2.667,-1);
        \filldraw[black] (2,1.333) circle (3pt);
        \filldraw[black] (2,-0.333) circle (3pt);
        \node[black, above] at (1.333,3) {\footnotesize $a$};
        \node[black, above] at (2.667,3) {\footnotesize $b$};
        \node[black, right] at (2,0) {\footnotesize $c$}; 
        \node[above, black] at (2,1.333) {\scriptsize $\mu$};
        \node[below, black] at (2,-0.333) {\scriptsize $\mu$};
        \node[black, below] at (1.333,-2) {\footnotesize $a$};
        \node[black, below] at (2.667,-2) {\footnotesize $b$}; 
    \end{tikzpicture} ~, \quad \quad \quad \begin{tikzpicture}[baseline={([yshift=-1ex]current bounding box.center)},vertex/.style={anchor=base,
    circle,fill=black!25,minimum size=18pt,inner sep=2pt},scale=0.5]
        \draw[black, thick, ->-=.8, line width = 0.4mm] (2,2.667) -- (2,4);
        \draw[black, line width = 0.4mm] (1.333,2) arc (180:90:0.667);
        \draw[black, line width = 0.4mm] (2.667,2) arc (0:90:0.667);
        \draw[black, line width = 0.4mm] (1.333,1) arc (180:270:0.667);
        \draw[black, line width = 0.4mm] (2.667,1) arc (0:-90:0.667);
    
        \draw[black, line width = 0.4mm, ->-=0.5] (1.333,1) -- (1.333,2);
        \draw[black, line width = 0.4mm, ->-=0.5] (2.667,1) -- (2.667,2);
        \draw[black, line width = 0.4mm, ->-=0.8] (2,-1) -- (2,0.333);
        \filldraw[black] (2,2.667) circle (3pt);
        \filldraw[black] (2,0.333) circle (3pt);
        
        \node[black, left] at (1.333,1.5) {\footnotesize $a$};
        \node[black, right] at (2.667,1.5) {\footnotesize $b$};
        \node[black, above] at (2,4) {\footnotesize $c'$}; 
        \node[black, below] at (2,-1) {\footnotesize $c$};
        \node[below, black] at (2,2.667) {\footnotesize $\mu$};
        \node[above, black] at (2,0.333) {\footnotesize $\nu$};
    \end{tikzpicture} = \delta_{c,c'} \delta_{\mu\nu} \sqrt{\frac{d_a d_b}{d_c}} \begin{tikzpicture}[baseline={([yshift=-1ex]current bounding box.center)},vertex/.style={anchor=base,
    circle,fill=black!25,minimum size=18pt,inner sep=2pt},scale=0.5]
    \draw[black, line width = 0.4mm, ->-=0.5] (2,-1) -- (2,4);
    \node[black, below] at (2,-1) {\footnotesize $c$};
    \node[above, above] at (2,4) {\footnotesize $c$};
    \end{tikzpicture} ~,
\end{equation} 
where $d_a$ is the quantum dimension of the TDL $a$ defined as
\begin{equation}
d_a = \quad \begin{tikzpicture}[baseline={([yshift=-1ex]current bounding box.center)},vertex/.style={anchor=base,
    circle,fill=black!25,minimum size=18pt,inner sep=2pt},scale=0.5]
    \draw[black, line width = 0.4mm, ->-=0] (0,1.5) arc (0:360:1.5);
    \node[black, right] at (0,1.5) {$a$};
\end{tikzpicture} ~.
\end{equation}
Consider a single TDL splitting into three TDLs, there are two ways in which this process can be done. They are related by the \textit{higher structure} of the fusion category known as the \textit{associativity map}. With the explicit basis \eqref{eq:f1c_basis}, the associativity map is characterized by a set of $\mathbb{C}$-numbers known as the $F$-symbols
\begin{equation}\label{eq:F-symbols}
    \begin{tikzpicture}[baseline={([yshift=-1ex]current bounding box.center)},vertex/.style={anchor=base,
    circle,fill=black!25,minimum size=18pt,inner sep=2pt},scale=0.7]
    \draw[line width = 0.4mm, black, ->-=0.7] (0,0.75) -- (-0.75,1.5);
    \draw[line width = 0.4mm, black, ->-=0.7] (0,0.75) -- (0.75,1.5);
    \draw[line width = 0.4mm, black, ->-=0.7] (0.75,0) -- (0,0.75);
    \draw[line width = 0.4mm, black, ->-=0.7] (0.75,0) -- (2.25,1.5);
	\draw[line width = 0.4mm, black, ->-=0.7] (1.5,-0.75) -- (0.75,0); 
 
	\node[above, black] at (-0.75,1.5) {\scriptsize $a$};
	\node[above, black] at (0.75,1.5) {\scriptsize $b$};
	\node[above, black] at (2.25,1.5) {\scriptsize $c$};
	\node[black] at (0.55,0.55) {\scriptsize $e$};
    \node[left, black] at (0,0.75) {\scriptsize $\mu$};
 	\node[left, black] at (0.75,0) {\scriptsize $\nu$};
	\node[below, black] at (1.5,-0.75) {\scriptsize $d$};
	\filldraw[black] (0.75,0) circle (2pt);
	\filldraw[black] (0,0.75) circle (2pt);
	\end{tikzpicture}
    = \sum_{f,\rho,\sigma} \left[F^{abc}_d\right]_{(e,\mu,\nu),(f,\rho,\sigma)}
    \begin{tikzpicture}[baseline={([yshift=-1ex]current bounding box.center)},vertex/.style={anchor=base,
    circle,fill=black!25,minimum size=18pt,inner sep=2pt},scale=0.7]
    \draw[line width = 0.4mm, black, ->-=0.7] (0.75,0) -- (-0.75,1.5);
    \draw[line width = 0.4mm, black, ->-=0.7] (1.5,0.75) -- (0.75,1.5);
    \draw[line width = 0.4mm, black, ->-=0.7] (0.75,0) -- (1.5,0.75);
    \draw[line width = 0.4mm, black, ->-=0.7] (1.5,0.75) -- (2.25,1.5);
	\draw[line width = 0.4mm, black, ->-=0.7] (1.5,-0.75) -- (0.75,0); 
 
	\node[above, black] at (-0.75,1.5) {\scriptsize $a$};
	\node[above, black] at (0.75,1.5) {\scriptsize $b$};
	\node[above, black] at (2.25,1.5) {\scriptsize $c$};
	\node[black] at (1,0.55) {\scriptsize $f$};
    \node[right, black] at (1.5,0.75) {\scriptsize $\rho$};
 	\node[right, black] at (0.75,0) {\scriptsize $\sigma$};
	\node[below, black] at (1.5,-0.75) {\scriptsize $d$};
	\filldraw[black] (0.75,0) circle (2pt);
	\filldraw[black] (1.5,0.75) circle (2pt);
	\end{tikzpicture} ~.
\end{equation}
As pointed out in \cite{Bhardwaj:2017xup,Copetti:2023mcq}, the $F$-symbols $\left[F^{abc}_d\right]_{(e,\mu,\nu),(f,\rho,\sigma)}$ can also be computed by the following relation
\begin{equation}\label{eq:TL_diag}
    \begin{tikzpicture}[baseline={([yshift=-.5ex]current bounding box.center)},vertex/.style={anchor=base,circle,fill=black!25,minimum size=18pt,inner sep=2pt},scale=0.75]
        \draw[line width = 0.4mm, ->-=0.5] (0,1.5)--(0,2.5);
        \node[above] at (0,2.5) {\footnotesize $\mathcal{L}_d$}; %a(bc)
        \draw[line width = 0.4mm, ->-=0.5] (-1.5,0)--(0,1.5); %a
        \node[left] at (-1.5,0) {\footnotesize $\mathcal{L}_a$};
        \draw[line width = 0.4mm, ->-=0.5] (0.75,0.75)--(0,1.5);
        \node[right] at (0.25,1.25) {\footnotesize $\mathcal{L}_f$}; %bc
        \draw[line width = 0.4mm, ->-=0.5] (1.5,0)--(0.75,0.75); 
        \draw[line width = 0.4mm, ->-=0.5] (-0.75,-0.75)--(0.75,0.75);
        \node[left] at (-0.1,0) {\footnotesize $\mathcal{L}_b$}; %b
        \draw[line width = 0.4mm, ->-=0.5] (0,-1.5)--(1.5,0);
        \node[right] at (1.5,0) {\footnotesize $\mathcal{L}_c$}; %c
        \draw[line width = 0.4mm, ->-=0.5] (0,-1.5)--(-0.75,-0.75);
        \node[left] at (-0.25,-1.25) {\footnotesize $\mathcal{L}_e$}; %ab
        \draw[line width = 0.4mm, ->-=0.5] (-0.75,-0.75)--(-1.5,0);
        \draw[line width = 0.4mm, ->-=0.5] (0,-2.5)--(0,-1.5);
        \node[below] at (0,-2.5) {\footnotesize $\mathcal{L}_d$}; %(ab)c
        \filldraw[black] (0,1.5) circle (2pt) node[below,black] () {\footnotesize $\sigma$};
        \filldraw[black] (0.75,0.75) circle (2pt) node[below,black] () {\footnotesize $\rho$};
        \filldraw[black] (-0.75,-0.75) circle (2pt) node[above,black] () {\footnotesize $\mu$};
        \filldraw[black] (0,-1.5) circle (2pt) node[above,black] () {\footnotesize $\nu$};
    \end{tikzpicture} = \sqrt{\frac{d_a d_b d_c}{d_d}} \left[F^{abc}_d\right]_{(e,\mu,\nu),(f,\rho,\sigma)} \begin{tikzpicture}[baseline={([yshift=-.5ex]current bounding box.center)},vertex/.style={anchor=base,circle,fill=black!25,minimum size=18pt,inner sep=2pt},scale=0.75]
        \draw[line width = 0.4mm, ->-=0.5] (0,-2.5)--(0,2.5);
        \node[below] at (0,-2.5) {\footnotesize $\mathcal{L}_d$}; %a(bc)
    \end{tikzpicture} 
\end{equation}
which can be derived first using \eqref{eq:F-symbols} and \eqref{eq:fus_spl_relation}. \eqref{eq:TL_diag} is more convenient for extracting $F$-symbols in the Lagrangian approach in 2d, as well as generalizing to higher dimensions.

To conclude, we consider an example of the Tambara-Yamagami fusion category $\operatorname{TY}(\mathbb{Z}_N,+1)$. This fusion category contains a $\mathbb{Z}_N$ invertible symmetry generated by the line $\eta$, as well as a non-invertible duality line $T$. The fusion rules between $\eta$ and $T$ are
\begin{equation}
    T \times \eta = \eta \times T = T ~, \quad T \times T = \sum_{k=0}^{N-1} \eta^k ~.
\end{equation}
The invertible lines $\eta^k$ have quantum dimension $d_{\eta^k} = 1$, while the duality defect has quantum dimension $d_T = \sqrt{N}$. Most of the $F$-symbols are equal to $1$ for this fusion category, with the non-trivial $F$-symbols given by
\begin{equation}
     F^T_{\eta^n,T,\eta^m} = F^{\eta^n}_{T,\eta^m,T}=\exp\left({2\pi i nm\over N}\right), \qquad \left[F^T_{T,T,T} \right]_{\eta^n,\eta^m} = {1\over\sqrt{N}}\exp\left(-{2\pi i nm\over N}\right) ~.
\end{equation}
%

% --------------------------- 
\subsection{Associator 1-morphisms in 4d QFT} \label{sec: higher category} 
Turning to 4d, we briefly review the higher structure of the fusion 3-category focusing on the associator 1-morphisms.  We refer to \cite{Copetti:2023mcq} for more details.

\paragraph{Objects} In a $4$d QFT, all symmetry operators are captured by a fusion $3$-category $\mathfrak{C}^{[3]}$. The objects in $\mathfrak{C}^{[3]}$ are $0$-form symmetry operators $\mathcal{D}$ supported on codim-$1$ surfaces. These objects admit an additive structure from the direct sum of the symmetry operators, as well as a tensor product coming from fusing two symmetry operators. We call an object \textit{simple} if it cannot be written as a direct sum of other objects.

\paragraph{Morphisms}
Given two codim-$1$ symmetry operators $\mathcal{D}_1$ and $\mathcal{D}_2$, a topological interfaces (or a junction) between them is a 1-morphism from $\mathcal{D}_1$ to $\mathcal{D}_2$. In particular, the morphisms between identity codim-$1$ operators are $1$-form symmetry operators supported on codim-$2$ surfaces. All such interfaces between $\mathcal{D}_1$ and $\mathcal{D}_2$ form a $2$-category $\Hom_{\mathfrak{C}^{[3]}}(\mathcal{D}_1,\mathcal{D}_2)$; and the fusion $2$-category $\End_{\mathfrak{C}^{[3]}}(\mathcal{D})\equiv \Hom_{\mathfrak{C}^{[3]}}(\mathcal{D},\mathcal{D})$ describes the symmetry operators living in the world volume of the operator $\mathcal{D}$.  Invoking the definition, the fusion $2$-category $\End_{\mathfrak{C}^{[3]}}(\mathcal{D})$ contains topological surface operators (as objects in $\End_{\mathfrak{C}^{[3]}}(\mathcal{D})$) and topological line operators (as morphisms between identity surface operators in $\End_{\mathfrak{C}^{[3]}}(\mathcal{D})$) in the 3-dim world-volume of $\mathcal{D}$.

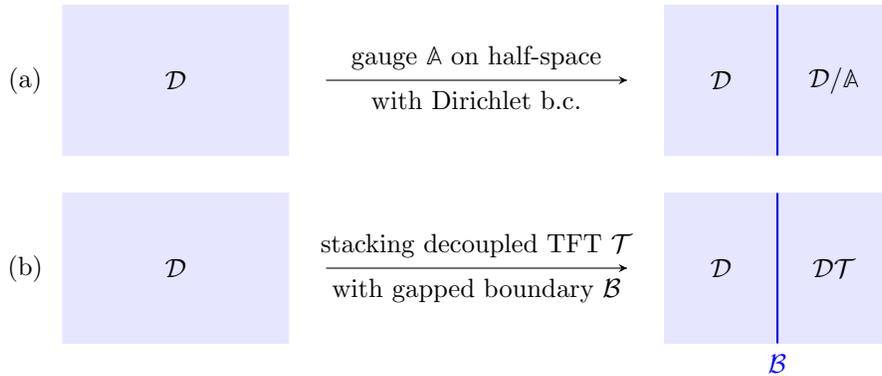
\begin{figure}[h]
    \centering
    \begin{tikzpicture}[baseline={([yshift=-.5ex]current bounding box.center)},vertex/.style={anchor=base,circle,fill=black!25,minimum size=18pt,inner sep=2pt},scale=1]
        %\filldraw[blue, opacity = 0.1] (-1.5,-1) -- (-1.5,+1) -- (-3.5,+1) -- (-3.5,-1) -- (-1.5,-1);
        \filldraw[blue, opacity = 0.1] (-2.5,-1) -- (-2.5,+1) -- (0.5,+1) -- (0.5,-1) -- (-2.5,-1);
        %\node[black] at (-2.5,0) {\footnotesize $\mathcal{D}$};
        %\node[black, below] at (-1.5,-1) {\footnotesize $\mathcal{I}$};
        \node[black] at (-1.0,0) {\footnotesize $\mathcal{D}$};
        \draw[black,->-=1] (1,0) -- (5,0);
        \node[black, above] at (3,0) {\footnotesize gauge $\mathbb{A}$ on half-space};
        \node[black, below] at (3,0) {\footnotesize with Dirichlet b.c.};
        %\filldraw[blue, opacity = 0.1] (3.5,-1) -- (3.5,+1) -- (6.5,+1) -- (6.5,-1) -- (3.5,-1);
        %\node[black] at (5,0) {\footnotesize $\mathcal{D}/\mathbb{A}$};
        %\node[black] at (7,0) {\footnotesize $:$};
        \filldraw[blue, opacity = 0.1] (5.5,-1) -- (8.5,-1) -- (8.5,+1) -- (5.5,+1) -- (5.5,-1);
        \draw[blue, thick] (7,-1) -- (7,+1);
        \node[black] at (6.25,0) {\footnotesize $\mathcal{D}$};
        \node[black] at (7.75,0) {\footnotesize $\mathcal{D}/\mathbb{A}$};
        \node[black] at (-3,0) {\footnotesize (a)};

        \node[black] at (-3,-2.5) {\footnotesize (b)};
        \filldraw[blue, opacity = 0.1] (-2.5,-3.5) -- (-2.5,-1.5) -- (0.5,-1.5) -- (0.5,-3.5) -- (-2.5,-3.5);
        \node[black] at (-1.0,-2.5) {\footnotesize $\mathcal{D}$};
        \node[black, above] at (3,-2.5) {\footnotesize stacking decoupled TFT $\mathcal{T}$};
        \node[black, below] at (3,-2.5) {\footnotesize with gapped boundary $\mathcal{B}$};
        \draw[black,->-=1] (1,-2.5) -- (5,-2.5);
        \filldraw[blue, opacity = 0.1] (5.5,-3.5) -- (8.5,-3.5) -- (8.5,-1.5) -- (5.5,-1.5) -- (5.5,-3.5);
        \draw[blue, thick] (7,-3.5) -- (7,-1.5);
        \node[black] at (6.25,-2.5) {\footnotesize $\mathcal{D}$};
        \node[black] at (7.75,-2.5) {\footnotesize $\mathcal{D}\mathcal{T}$};
        \node[blue, below] at (7,-3.5) {\footnotesize $\mathcal{B}$};
    \end{tikzpicture}
    \caption{(a) Gauging some symmetry $\mathbb{A}$ in $\End_{\mathfrak{C}^{[3]}}(\mathcal{D})$ on half-space with Dirichlet boundary condition creates a non-trivial topological interface between $\mathcal{D}$ and $\mathcal{D}/\mathbb{A}$. (b) Stacking a decoupled 3d TFT $\mathcal{T}$ with the gapped boundary $\mathcal{B}$ creates a non-trivial topological interface between $\mathcal{D}$ and $\mathcal{D}\mathcal{T}$. $\mathcal{D}$, $\mathcal{D}/\mathbb{A}$, and $\mathcal{D}\mathcal{T}$ are in the same equivalence class, i.e., $\lls \mathcal{D} \rrs = \lls \mathcal{D}/\mathbb{A} \rrs = \lls \mathcal{D} \mathcal{T} \rrs$.} 
    \label{fig:stc_gg}
\end{figure}

\paragraph{Witt equivalence class} Two codim-$1$ defects are said to be \textit{Witt equivalent} if there is a non-trivial morphism between them\footnote{Notice that unlike the 2d case where any simple line cannot have non-trivial junction to the identity operator $\mathbb{1}$ unless it is $\mathbb{1}$, in 4d, a non-identity simple $0$-form symmetry operator can have a non-trivial junction to $\mathbb{1}$.}. For instance, such junctions can arise by stacking decoupled TQFTs admitting gapped boundaries or gauging a non-anomalous symmetry in $\End_{\mathfrak{C}}(\mathcal{D})$ as shown in Figure \ref{fig:stc_gg}. We use $\lls \mathcal{D}\rrs$ to denote the Witt equivalence class of $\mathcal{D}$. Generically there is no preferred choice for a representative for each class $\lls \mathcal{D} \rrs$. But for the $0$-form symmetry operators considered in this paper, there is always a canonical choice for the representative, which contains minimal operator content on the world-volume. We denote them by $\lls\mathcal{D}\rrs_0$. A generic operator in $\lls \mathcal{D} \rrs$ can be turned into $\lls\mathcal{D}\rrs_0$ by gauging certain symmetries. 

\paragraph{Condensation defect} An important example is a class of defects known as the condensation defect $C$ \cite{Roumpedakis:2022aik,Bah:2025oxi,Kong:2013aya}.  They are defined as gauging $p$-form symmetry on an $n$-dimensional surface $\Sigma^{(n)}$ (where $n\geq p+1$) with a possible discrete torsion. For example, the following condensation defect gauges a 1-form $\mathbb{Z}_N$ symmetry generated by $U(\sigma^{(2)})$ on a 3-surface $\Sigma^{(3)}$.
\begin{equation}
    C_k(\Sigma^{(3)}) = \frac{1}{|H^0(\Sigma^{(3)},\mathbb{Z}_N)|}\sum_{\sigma^{(2)} \in H_2(\Sigma^{(3)},\mathbb{Z}_N)} e^{\frac{2\pi i}{N} k  Q(\Sigma^{(3)},\sigma^{(2)})} U(\sigma^{(2)}) ~,
\end{equation}
where $k \in H^3(B\mathbb{Z}_N,U(1)) \simeq \mathbb{Z}_N$ parameterizes the choice of discrete torsion in the higher gauging. The discrete torsion phase $Q(\Sigma^{(3)},\sigma^{(2)})$ is defined as
\begin{equation}
    Q(\Sigma^{(3)},\sigma^{(2)}) = \int a^{(1)} \cup \beta (a^{(1)}) ~,
\end{equation}
where $a^{(1)} = PD(\sigma^{(2)},\Sigma^{(3)}) \in H^1(\Sigma^{(3)},\mathbb{Z}_N)$ is the Poincare dual 1-cocycle of $\sigma^{(2)}$ on $\Sigma^{(3)}$ and $\beta:H^1(\Sigma^{(3)},\mathbb{Z}_N) \rightarrow H^2(\Sigma^{(3)},\mathbb{Z}_N)$ is the Bockstein homomorphism associated to the short exact sequence $\mathbb{Z}_N \rightarrow \mathbb{Z}_{N^2}\rightarrow \mathbb{Z}_N$.

Such a condensation defect has higher quantum symmetries generated by $\mathbb{Z}_N$ topological lines living in $\End_{\mathfrak{C}^{[3]}}(C_k)$. The higher quantum symmetries are gaugeable on the world-volume of the condensation defect, and gauging it turns the condensation defect to the identity operator.\footnote{Such a junction serves as a topological boundary with Dirichlet boundary conditions are imposed on the quantum symmetry gauge fields of the higher quantum symmetry.}  Hence, a condensation defect is always Witt equivalent to the identity operator and $\lls C_k\rrs_0=\bi$.\footnote{Consequently, fusion with a condensation defect will not change the equivalence class.}

\begin{figure}
    \centering
    \begin{tikzpicture}[baseline={([yshift=-.5ex]current bounding box.center)},vertex/.style={anchor=base,circle,fill=black!25,minimum size=18pt,inner sep=2pt},scale=0.5]
    \draw[black, line width = 0.4mm, ->-=0.5] (+2,-2) -- (+2,+2);
    \draw[black, line width = 0.4mm, ->-=0.5] (-2,-2) -- (-2,+2);
    \node[black, below] at (-2,-2) {\footnotesize $\mathcal{D}_a$};
    \node[black, below] at (+2,-2) {\footnotesize $\mathcal{D}_b$};
    \draw[black, ->-=1] (2.5,0) -- (8.5,0);
    \node[above] at (5.5,0) {\footnotesize fuse upper part};
    
    \draw[thick, black, line width = 0.4mm, ->-=0.5] (9,-2) -- (11,0);
    \draw[thick, black, line width = 0.4mm, ->-=0.5] (13,-2) -- (11,0);
    \draw[thick, black, line width = 0.4mm, ->-=0.5] (11,0) -- (11,2);
    \node[black, below] at (9,-2) {\footnotesize $\mathcal{D}_a$};
    \node[black, below] at (13,-2) {\footnotesize $\mathcal{D}_b$};
    \node[black, above] at (11,+2) {\footnotesize $\mathcal{D}_a \otimes \mathcal{D}_b$};
    \draw[black, ->-=1] (13.5,0) -- (19.5,0);
    \node[above] at (16.5,0) {\footnotesize gauging on $\mathcal{D}_a \otimes \mathcal{D}_b$};
    \node[below] at (16.5,0) {\footnotesize with Dirichlet b.c.};

    \draw[thick, black, line width = 0.4mm, ->-=0.5] (20,-2) -- (22,0);
    \draw[thick, black, line width = 0.4mm, ->-=0.5] (24,-2) -- (22,0);
    \draw[thick, black, line width = 0.4mm, ->-=0.5] (22,0) -- (22,2);
    \node[black, below] at (20,-2) {\footnotesize $\mathcal{D}_a$};
    \node[black, below] at (24,-2) {\footnotesize $\mathcal{D}_b$};
    \node[black, above] at (22,+2) {\footnotesize $\lls\mathcal{D}_a \otimes \mathcal{D}_b\rrs_0$};
    
    \end{tikzpicture}
    \caption{Given with two codim-$1$ defects $\mathcal{D}_a$ and $\mathcal{D}_b$ where $\mathcal{D}_a = \lls\mathcal{D}_a\rrs_0, \mathcal{D}_b = \lls \mathcal{D}_b\rrs_0$, one can always construct a topological local fusion junction into $\lls\mathcal{D}_a \otimes \mathcal{D}_b\rrs_0$ as follows. Starting with two parallel $\mathcal{D}_a$ and $\mathcal{D}_b$, fusing the upper part leads to a topological local fusion junction from $\mathcal{D}_a$ and $\mathcal{D}_b$ to $\mathcal{D}_a \otimes \mathcal{D}_b$. Then, gauging the corresponding symmetry on the world-volume of $\mathcal{D}_a \otimes \mathcal{D}_b$ with Dirichlet boundary condition on the fusion junction leads to a topological local fusion junction with the outgoing line being $\lls \mathcal{D}_a \otimes \mathcal{D}_b \rrs_0$.}
    \label{fig:4dlfc}
\end{figure}
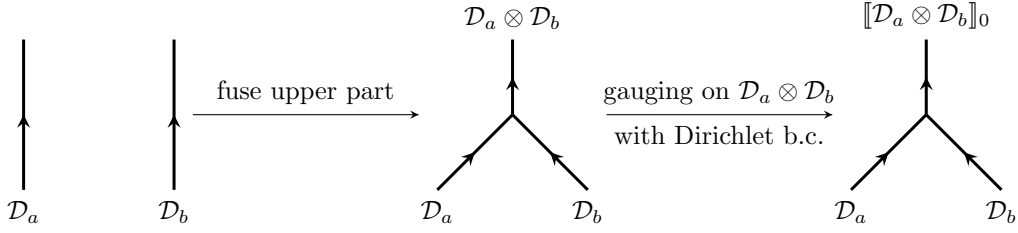

\paragraph{Local fusion junctions and associator 1-morphism}
The fusion of two 0-form symmetry operators $\mathcal{D}_a$ and $\mathcal{D}_b$ can be expanded as a sum of simple operators
\begin{equation}\label{eq:4dgf}
    \mathcal{D}_a \otimes \mathcal{D}_b = \sum_{\mathcal{D}_c: simple} \mathcal{C}_{ab}^c \mathcal{D}_c ~,
\end{equation}
where the coefficients $\mathcal{C}_{ab}^c$'s are valued in $3$-dimensional topological field theory.  In known cases (including the cases considered in this paper) in $d>2$, the fusion product of the equivalence class is group-like even if the $\mathcal{D}_a$'s themselves are non-invertible, i.e. there is always a single term appearing on the RHS of \eqref{eq:4dgf}.\footnote{This is more explicit after quotienting out the equivalence class in the fusion rule $\lls\mathcal{D}_a\rrs \otimes \lls\mathcal{D}_b\rrs = \sum N_{ab}^c \lls \mathcal{D}_c\rrs $ where the fusion coefficients $N_{ab}^c$ now take non-negative integer values. In known cases, there is always a single term appearing in the RHS. For the remainder of the discussion, we restrict ourselves to this case.}  In such cases, we can add a 1-morphism junction to the canonical representative of the Witt equivalence class after the local fusion junction.  By combining the fusion and morphism junctions, we can construct a new local fusion junction between the canonical representatives of the equivalence class as demonstrated in Figure~\ref{fig:4dlfc}.

We are now ready to define the generalization of the $F$-symbols in $\mathfrak{C}^{[3]}$. Let's consider the diagram
\begin{equation}
    \begin{tikzpicture}[baseline={([yshift=-.5ex]current bounding box.center)},vertex/.style={anchor=base,circle,fill=black!25,minimum size=18pt,inner sep=2pt},scale=1]
        \draw[line width = 0.4mm, ->-=0.5] (0,1.5)--(0,2.5);
        \node[above] at (0,2.5) {\footnotesize $\lls\mathcal{D}_a \otimes (\mathcal{D}_b \otimes \mathcal{D}_c)\rrs_0$}; %a(bc)
        \draw[line width = 0.4mm, ->-=0.5] (-1.5,0)--(0,1.5); %a
        \node[left] at (-1.5,0) {\footnotesize $\mathcal{D}_a$};
        \draw[line width = 0.4mm, ->-=0.5] (0.75,0.75)--(0,1.5);
        \node[right] at (0.25,1.25) {\footnotesize $\lls \mathcal{D}_b \otimes \mathcal{D}_c\rrs_0$}; %bc
        \draw[line width = 0.4mm, ->-=0.5] (1.5,0)--(0.75,0.75); 
        \draw[line width = 0.4mm, ->-=0.5] (-0.75,-0.75)--(0.75,0.75);
        \node[left] at (-0.1,0) {\footnotesize $\mathcal{D}_b$}; %b
        \draw[line width = 0.4mm, ->-=0.5] (0,-1.5)--(1.5,0);
        \node[right] at (1.5,0) {\footnotesize $\mathcal{D}_c$}; %c
        \draw[line width = 0.4mm, ->-=0.5] (0,-1.5)--(-0.75,-0.75);
        \node[left] at (-0.25,-1.25) {\footnotesize $\lls\mathcal{D}_a \otimes \mathcal{D}_b\rrs_0$}; %ab
        \draw[line width = 0.4mm, ->-=0.5] (-0.75,-0.75)--(-1.5,0);
        \draw[line width = 0.4mm, ->-=0.5] (0,-2.5)--(0,-1.5);
        \node[below] at (0,-2.5) {\footnotesize $\lls(\mathcal{D}_a \otimes \mathcal{D}_b) \otimes \mathcal{D}_c\rrs_0$}; %(ab)c
        \filldraw[red] (0,1.5) circle (2pt);
        \filldraw[red] (0.75,0.75) circle (2pt);
        \filldraw[blue] (-0.75,-0.75) circle (2pt);
        \filldraw[blue] (0,-1.5) circle (2pt);
    \end{tikzpicture} = \begin{tikzpicture}[baseline={([yshift=-.5ex]current bounding box.center)},vertex/.style={anchor=base,circle,fill=black!25,minimum size=18pt,inner sep=2pt},scale=1]
    \draw[line width = 0.4mm, ->-=0.5] (0,1.5)--(0,2.5);
    \draw[line width = 0.4mm, ->-=0.5] (0,-1.5)--(0,+1.5);
    \draw[line width = 0.4mm, ->-=0.5] (0,-2.5)--(0,-1.5);
    
    \node[above] at (0,2.5) {\footnotesize $\lls\mathcal{D}_a \otimes (\mathcal{D}_b \otimes \mathcal{D}_c)\rrs_0$};
    \node[below] at (0,-2.5) {\footnotesize $\lls(\mathcal{D}_a \otimes \mathcal{D}_b) \otimes \mathcal{D}_c\rrs_0$}; %(ab)c
    \node[right] at (0,0) {\footnotesize $\mathcal{D}_a \otimes \mathcal{D}_b \otimes \mathcal{D}_c$};

    \filldraw[red] (0,1.5) circle (2pt);
    \node[right, red] at (0,1.5) {\footnotesize $\mathcal{I}_+$};
    \filldraw[blue] (0,-1.5) circle (2pt);
    \node[right, blue] at (0,-1.5) {\footnotesize $\mathcal{I}_-$};
    \end{tikzpicture} = \begin{tikzpicture}[baseline={([yshift=-.5ex]current bounding box.center)},vertex/.style={anchor=base,circle,fill=black!25,minimum size=18pt,inner sep=2pt},scale=1]
    \draw[line width = 0.4mm, ->-=0.5] (0,0)--(0,2.5);
    \draw[line width = 0.4mm, ->-=0.5] (0,-2.5)--(0,0);
    
    \node[above] at (0,2.5) {\footnotesize $\lls\mathcal{D}_a \otimes (\mathcal{D}_b \otimes \mathcal{D}_c)\rrs_0$};
    \node[below] at (0,-2.5) {\footnotesize $\lls(\mathcal{D}_a \otimes \mathcal{D}_b) \otimes \mathcal{D}_c\rrs_0$}; %(ab)c

    \filldraw[black] (0,0) circle (2pt);
    \node[right, black] at (0,0) {\footnotesize $F_{\mathcal{D}_a,\mathcal{D}_b,\mathcal{D}_c}$};
    \end{tikzpicture} ~.
\end{equation}
In the first equal sign, we fuse all the defects in the middle and also fuse the fusion (splitting) junctions represented by red (blue) dots into a single interface $\mathcal{I}_+$ and $\mathcal{I}_-$. Then we fuse the interfaces $\mathcal{I}_+$ and $\mathcal{I}_-$ to create an interface from $\lls(\mathcal{D}_a \otimes \mathcal{D}_b) \otimes \mathcal{D}_c\rrs_0$ to $\lls\mathcal{D}_a \otimes (\mathcal{D}_b \otimes \mathcal{D}_c)\rrs_0$; this interface is interpreted as the generalization of the $F$-symbol 1-morphism in 4d \cite{Copetti:2023mcq}, which we denote by $F_{\mathcal{D}_a,\mathcal{D}_b,\mathcal{D}_c} \in \Hom_{\mathfrak{C}^{[3]}}(\lls(\mathcal{D}_a \otimes \mathcal{D}_b) \otimes \mathcal{D}_c\rrs_0, \lls\mathcal{D}_a \otimes (\mathcal{D}_b \otimes \mathcal{D}_c)\rrs_0)$.

% ------------------------------------------
% ------------ new section -----------------
% ------------------------------------------

\section{Compact boson in 2d}\label{sec: compact boson}
\subsection{Lagrangians for TY($\mathbb{Z}_N$,+1) symmetry defects}

We consider a single 2d compact scalar with periodicity $2\pi$, described by the action 
\begin{equation}
    S = {R^2\over 4\pi}\int_{M_2} \dr\phi \wedge \star \dr\phi\,, \quad \phi \sim \phi+2\pi\,. 
    \label{eq: aa}
\end{equation}
This theory is completely specified by the choice of radius $R$, and has the interesting property that it is dual to the theory with radius $1/R$. This duality, called $T$-duality, maps the operators between the two theories in a nontrivial way, and to see how it arises let us rewrite the action \eqref{eq: aa} in terms of an unconstrained 1-form field $X$ (instead of $\mathrm{d}\phi$) and a compact Lagrange multiplier $\widetilde{\phi}$, 
\begin{equation}
S=\int_{M_{2}}\left(\frac{R^{2}}{4\pi}X\wedge\star X+\frac{i}{2\pi}X\wedge d\widetilde{\phi}\right),\quad\widetilde{\phi}\sim\widetilde{\phi}+2\pi\,.
\label{eq: ab}
\end{equation}
The role of the Lagrange multiplier is to restrict $X$ to be closed and with holonomies valued in $2\pi\mathbb{Z}$ (which are the properties of $\mathrm{d}\phi$), and indeed integrating $\widetilde{\phi}$ out sets $X=\mathrm{d}\phi$ and results in the original action \eqref{eq: aa}. On the other hand, given the action \eqref{eq: ab}, we can integrate out $X$ using its equation of motion $iR^{2}\star X=\mathrm{d}\widetilde{\phi}$ and obtain an action for the Lagrange multiplier field $\widetilde{\phi}$, 
\begin{equation}
S=\frac{\widetilde{R}^{2}}{4\pi}\int_{M_{2}}\mathrm{d}\widetilde{\phi}\wedge\star \mathrm{d}\widetilde{\phi}
\label{eq: abc}
\end{equation}
where $\widetilde{R}=1/R$. We see that we have two dual descriptions, \eqref{eq: aa} and \eqref{eq: abc}, for the same theory, one with radius $R$ and the other with radius $1/R$. Moreover, there is a nontrivial map between the operators of these two descriptions, for example $\mathrm{d}\widetilde{\phi}=iR^{2}\star\mathrm{d}\phi$. 

Let us next examine some of the symmetries of the theory. First, at any radius $R$ there is a $U(1)_m \times U(1)_w\rtimes \mathbb{Z}_2^C$ symmetry, where $\mathbb{Z}_2^C$ is charge conjugation under which $\phi \to -\phi$ while $U(1)_m$ and $U(1)_w$ are the momentum (or shift) and winding symmetries, respectively, with charges  
\begin{equation}
Q_{m}=iR^{2}\oint\frac{\star\mathrm{d}\phi}{2\pi}\,,\quad Q_{w}=\oint\frac{\mathrm{d}\phi}{2\pi}\,.
\label{eq: Qs}
\end{equation}
Note that using the $T$-duality map of operators mentioned before, we can immediately see that these two charges map into each other under the duality. Indeed, the winding symmetry $U(1)_w$ of the scalar $\phi$ turns into the momentum symmetry $U(1)_m$ of the dual scalar $\widetilde{\phi}$ which acts by shifting it, while the momentum symmetry of $\phi$ turns into the winding symmetry of $\widetilde{\phi}$. 

Even though the momentum and winding symmetries have a mixed 't Hooft anomaly, each of them is non-anomalous by itself and can be gauged. Let us consider e.g. the effect of gauging a $\mathbb{Z}_N$ subgroup of the momentum symmetry. Upon such gauging, the periodicity of $\phi$ changes from $2\pi$ to $2\pi/N$, and to have again a compact scalar with periodicity $2\pi$ we can define $\phi=\phi'/N$ such that $\phi'$ has the desired periodicity. This, however, results in the radius of the $\phi'$ theory being $R'=R/N$ (as can be seen by substituting into \eqref{eq: aa}), and we hence see that the effect of gauging a $\mathbb{Z}_N$ subgroup of $U(1)_m$ amounts to changing $R\rightarrow R/N$. 

At this point, we observe that combining this gauging with $T$-duality maps the original radius $R$ to $N/R$, such that the radius $R^2=N$ is invariant under this combined operation. As a result, gauging $\mathbb{Z}_{N}\subset U(1)_{m}$ in half space (with Dirichlet boundary conditions for the corresponding discrete gauge field) followed by $T$-duality in the same half space results in a nontrivial topological defect as long as $R^2=N$ (for other values of $R$ we would obtain a topological interface between different theories). Denoting this defect by $T$ and the basic $\mathbb{Z}_{N}\subset U(1)_{m}$ defect by $\eta$, one can show that they generate together the  Tambara-Yamagami fusion category $TY(\mathbb{Z}_N,+1)$ \cite{Thorngren:2021yso}. The corresponding fusion algebra, which was mentioned in subsection \ref{Fusion} and reproduced here for convenience, is given by 
\begin{equation}
    \eta^n\times \eta^m = \eta^{n+m}\,, \quad \eta^N=\mathbb{1}\,,\quad
    T\times \eta =\eta\times T = T\,, \quad T\times T = \sum_{k=0}^{N-1}\eta^k
\end{equation}
and the nontrivial $F$-symbols by 
\begin{equation}
    F^T_{\eta^n,T,\eta^m} = F^{\eta^n}_{T,\eta^m,T}=\exp\left({2\pi i nm\over N}\right)\,, \qquad \left[F^T_{T,T,T} \right]_{\eta^n,\eta^m} = {1\over\sqrt{N}}\exp\left(-{2\pi i nm\over N}\right)\,. 
\label{eq: ada}
\end{equation}

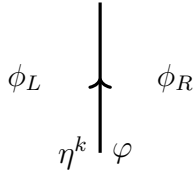
\begin{figure}[H]
	\centering
\begin{tikzpicture}
		\begin{scope}[very thick,decoration={
				markings,
				mark=at position 0.5 with {\arrow{>}}}
			] 
			\draw[postaction={decorate}] (0,-1)--(0,1) node[at start,left] (l) {$\eta^k$} node[at start ,right] () {$\varphi$};

            \node at (-1,0) () {$\phi_L$};
            \node at (1,0) () {$\phi_R$};

            \node at (2,0) (f) {};
		\end{scope}
        \end{tikzpicture}
	\caption{Description of an $\eta^k$ line defect using two fields, $\phi_L$ and $\phi_R$, from the two sides of the defect and a field $\varphi$ which is supported on the defect.}
	\label{fig: 2d line}
\end{figure}

As discussed in \cite{Choi:2021kmx,Choi:2022zal,Niro:2022ctq,Arias-Tamargo:2025xdd}, one can also represent the $T$ and $\eta^k$ defects using the actions\footnote{Let us comment that since the quantum dimension of $T$ is not $1$, but instead $\sqrt{N}$ (see subsection \ref{Fusion}), the full insertion in the partition function that describes it is $\sqrt{N}\exp\left(\frac{iN}{2\pi}\int_{x=0}\phi_{L}\mathrm{d}\phi_{R}\right)$.\label{T_qd}} 
\begin{equation}
T:\quad S=\frac{N}{4\pi}\int_{x<0}\mathrm{d}\phi_{L}\wedge\star\mathrm{d}\phi_{L}+\frac{iN}{2\pi}\int_{x=0}\phi_{L}\mathrm{d}\phi_{R}+\frac{N}{4\pi}\int_{x>0}\mathrm{d}\phi_{R}\wedge\star\mathrm{d}\phi_{R}
\label{T_L}
\end{equation}
and 
\begin{equation}
\eta^{k}:\quad S=\frac{N}{4\pi}\int_{x<0}\mathrm{d}\phi_{L}\wedge\star\mathrm{d}\phi_{L}+\frac{i}{2\pi}\int_{x=0}\mathrm{d}\varphi\left(\phi_{L}-\phi_{R}+\frac{2\pi k}{N}\right)+\frac{N}{4\pi}\int_{x>0}\mathrm{d}\phi_{R}\wedge\star\mathrm{d}\phi_{R}\,,
\label{eta_L}
\end{equation}
where $\phi_L$ and $\phi_R$ denote the compact boson on the left and right sides of the defects, which we placed at $x=0$. Moreover, the field $\varphi$ in Eq. \eqref{eta_L} is another compact scalar which is supported only on the $\eta^{k}$ defect worldvolume, as shown in Figure~\ref{fig: 2d line}.
In general, in order to find what defect actions of the type \eqref{T_L} or \eqref{eta_L} describe, one should examine the part of the variation of the action which is localized on the defect and deduce from it the way the defect acts on operators. For the $T$ defect, we have 
\begin{equation}
\delta S|_{x=0}=\frac{N}{2\pi}\int_{x=0}\left[\delta\phi_{L}\left(\star\mathrm{d}\phi_{L}+i\mathrm{d}\phi_{R}\right)-\delta\phi_{R}\left(\star\mathrm{d}\phi_{R}+i\mathrm{d}\phi_{L}\right)\right]
\end{equation}
and therefore $\mathrm{d}\phi_{R}|_{x=0}=i\star\mathrm{d}\phi_{L}|_{x=0}$ and $\mathrm{d}\phi_{L}|_{x=0}=i\star\mathrm{d}\phi_{R}|_{x=0}$. Using the relation between the compact scalar $\phi$ and its $T$-dual $\widetilde{\phi}$, which for the radius $R^2=N$ we consider is given by $\mathrm{d}\widetilde{\phi}=iN\star\mathrm{d}\phi$ (see the discussion below Eq. \eqref{eq: abc}), we have $\mathrm{d}\phi_{L}|_{x=0}=(1/N)\,\mathrm{d}\widetilde{\phi}_{R}|_{x=0}$ which indeed corresponds to gauging $\mathbb{Z}_{N}\subset U(1)_{m}$ in the right half-space followed by $T$-duality, as expected. For the $\eta^k$ defect, we notice that integrating out $\varphi$ in \eqref{eta_L} sets $(\phi_{R}-\phi_{L})|_{x=0}=2\pi k/N+2\pi\mathbb{Z}$, as expected. 

In the rest of this section, we will use the Lagrangian descriptions \eqref{T_L} and \eqref{eta_L} of the $T$ and $\eta^k$ defects to construct topological junctions and networks of defects, paying special attention to the boundary conditions and boundary terms at the junctions. Then, we will use these ingredients to compute the $F$-symbols, recovering the expressions in \eqref{eq: ada}.

\subsection{Local fusion junctions} \label{sec: local fusion junctions}
\subsubsection{Topological construction of local fusion junctions} \label{sec: top construction of local fusion junction}
For topological defects $a,b$ and $c$, the local fusion junction Hom($a\times b, c$) can be obtained by taking two parallel defects $a$ and $b$ and fusing the upper parts, as depicted in the figure below.
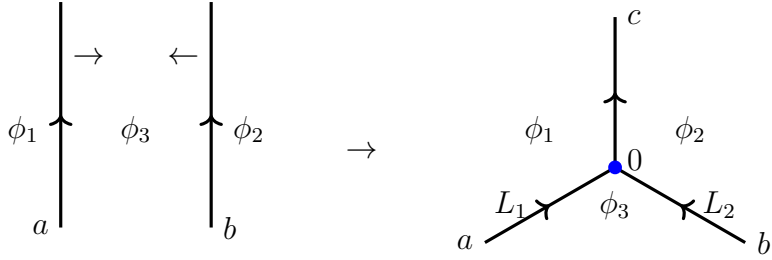
\begin{figure}[H]
	\centering
	\begin{tikzpicture}
		\begin{scope}[very thick,decoration={
				markings,
				mark=at position 0.5 with {\arrow{>}}}
			] 
			\draw[postaction={decorate}] (-1,-1)--(-1,2) node[at start,left] () {$a$} node[near end,right] () {$\to$};
			\draw[postaction={decorate}] (1,-1)--(1,2) node[at start,right] () {$b$} node[near end,left] () {$\leftarrow$};
			\node at (-1.5,0.3) () {$\phi_1$};
			\node at (1.5,0.3) () {$\phi_2$};
			\node at (0,0.3) () {$\phi_3$};
			\node at (3,0) (f) {$\to$};
		\end{scope}
		\begin{scope}[shift={($(f.east)+(3,0.8)$)}, very thick,decoration={
				markings,
				mark=at position 0.5 with {\arrow{>}}}
			] 
			\draw[postaction={decorate}] (-1.73,-2)--(0,-1) node[at start,left] (l) {$a$} node[midway,left] () {$L_1~$};
			\draw[postaction={decorate}] (1.73,-2)--(0,-1) node[at start,right] () {$b$} node[midway,right] () {$~L_2$};
			\draw[postaction={decorate}] (0,-1)--(0,1) node[at end,right] () {$c$};
			\node at (-1,-0.5) () {$\phi_1$};
			\node at (1,-0.5) () {$\phi_2$};
			\node at (0,-1.5) () {$\phi_3$};
			\node[right] at (0,-0.9) () {$0$};
			\filldraw[blue] (0,-1) circle (2pt) node[right] () {};
		\end{scope}
	\end{tikzpicture}
	\caption{Topological construction of the local fusion junction for Hom($a\times b,c$).  $L_{1,2}$ in the second diagram are the semi-infinite lines from the bottom that end on the local fusion junction at $t=0$ (blue dot). }
	\label{fig: local fusion junction}
\end{figure}
The part of the action which is supported on the defects in this configuration (like the middle terms in Eqs. \eqref{T_L} and \eqref{eta_L}) is given by 
\begin{equation}
    S = \int_{L_1}\mathcal{L}_{a}[\phi_1,\phi_3] + \int_{L_2}\mathcal{L}_{b}[\phi_3,\phi_2] +  \int_{t=0}^{t=\infty} \left(\mathcal{L}_{a}[\phi_1,\phi_3] +\mathcal{L}_{b}[\phi_3,\phi_2] \right)
\end{equation}
where $L_{1,2}$ denote the two bottom lines that terminate at the local fusion junction, and we use the coordinate $t>0$ for the upper part of the fused defects.  The first two terms describe the defects $a$ and $b$ below the fusion junction, and the last term describes the fusion of the two. One can then manipulate this last term in the upper part into the action of the topological defect(s) expected from the fusion algebra (in this case, the action of the defect $c$), which in general may result in
boundary terms at the junction and a decoupled trivial $\mathbb{Z}_1$ TQFT at the fused segment, i.e.
\begin{equation}
    \int_{t=0}^{t=\infty} \left(\mathcal{L}_{a}[\phi_1,\phi_3] +\mathcal{L}_{b}[\phi_3,\phi_2] \right) = \int_{t=0}^{t=\infty} \mathcal{L}_{c}[\phi_1,\phi_2,\phi_3] + {i\over 2\pi}\int^{t=\infty}_{t=0} \widetilde{\phi} \, \dr\widetilde{\varphi} + S_{bdry} 
\label{Fusion_2d}
\end{equation}
where $\widetilde{\phi}$ and $\widetilde{\varphi}$ in the second term are fields supported only along the defect in the upper part $t>0$, such as $\phi_3$ in Figure~\ref{fig: local fusion junction} or the defect field $\varphi$ in \eqref{eta_L} used in the expression for the $\eta^k$ defect. Hence, the second term on the RHS describes a trivial decoupled TQFT. We stress that both the boundary terms and the decoupled trivial TQFT are not
always present. 

Since we are interested in fusion junctions as in Figure~\ref{fig: local fusion junction}, in which each defect (i.e. $a$, $b$ and $c$ in the figure) is not accompanied by a decoupled TQFT, when such a TQFT
does arise we will need to add a topological boundary for it along the worldvolume of the $c$ defect (i.e. at some $t_{0}\in\left(0,\infty\right)$) and then push it to the
junction at $t=0$, resulting overall in a topological junction between defects with no decoupled TQFT. The topological boundary condition at the boundary we add along the $t>0$ segment is that of Dirichlet for both $\widetilde{\phi}$ and $\widetilde{\varphi}$ in \eqref{Fusion_2d}, as we explain next in more detail.

\paragraph{Boundary conditions of trivial TQFT}
Since we consider defects and fusion junctions that lie within a bulk where the periodicity of the compact scalars is maintained, we impose $\phi \sim \phi+2\pi$ at all the topological boundaries we add when constructing such junctions. For the trivial TQFT in \eqref{Fusion_2d}, this implies that the well-defined operators are the vertex operators $e^{in\widetilde\phi}$ and $e^{im\widetilde\varphi}$, both in its bulk and at the topological boundary we add for it. Now, using the equal-time commutation relation 
\begin{equation}
 [\widetilde\phi(t),\widetilde\varphi(t)] = 2\pi i 
\end{equation}
we see that these vertex operators generate the (gauge) transformations $\widetilde\varphi\to \widetilde\varphi+2\pi m$ and $\widetilde\phi\to \widetilde\phi+2\pi n$ and commute with each other. As a result, they are identified with the identity operator and the topological boundary condition at the boundary we add along the $t>0$ segment at some $t=t_0$ is given by\footnote{Note also that the path integral over each of the fields $\widetilde{\phi}$ or $\widetilde{\varphi}$ restricts the other one to vanish modulo $2\pi$, such that the boundary condition \eqref{eq: aaa} is the only one which is compatible with a non-vanishing partition function.} 
\begin{equation}
    \left. \widetilde\phi, \widetilde\varphi \right|_{t=t_0} = 0 \quad \text{mod $2\pi$}\,. 
    \label{eq: aaa}
\end{equation}
These boundary conditions are consistent with the variational principle, $\left. \delta S \right|_{t=t_0}=0$, and the periodicity (or gauge symmetry) of the compact scalars, $\Delta S = 0$ mod $2\pi$. 

Now that the topological boundary has been added at $t=t_0$, we push it to $t=0$ and obtain the desired junction. In this junction we have, on top of the other junction data, the boundary condition \eqref{eq: aaa}.

\begin{figure}[H]
	\centering
	\begin{tikzpicture}
		\begin{scope}[very thick,decoration={
				markings,
				mark=at position 0.5 with {\arrow{>}}}
			] 
			\draw[postaction={decorate}] (-1,-1)--(-1,2) node[at start,left] () {$T$} node[near end,right] () {$\to$};
			\draw[postaction={decorate}] (1,-1)--(1,2) node[at start,right] () {$\eta^k$} node[near end,left] () {$\leftarrow$};
			\node at (-1.5,0.3) () {$\phi_1$};
			\node at (1.5,0.3) () {$\phi_2$};
			\node at (0,0.3) () {$\phi_3$};
			\node at (3,0) (f) {$\to$};
		\end{scope}
		\begin{scope}[shift={($(f.east)+(3,0.8)$)}, very thick,decoration={
				markings,
				mark=at position 0.5 with {\arrow{>}}}
			] 
			\draw[postaction={decorate}] (-1.73,-2)--(0,-1) node[at start,left] (l) {$T$} node[midway,left] () {$L_1~$};
			\draw[postaction={decorate}] (1.73,-2)--(0,-1) node[at start,right] () {$\eta^k$} node[midway,right] () {$~L_2$};
			\draw[postaction={decorate}] (0,-1)--(0,1) node[at end,right] () {$T$};
			\node at (-1,-0.5) () {$\phi_1$};
			\node at (1,-0.5) () {$\phi_2$};
			\node at (0,-1.5) () {$\phi_3$};
			\node[right] at (0,-0.9) () {$0$};
			\filldraw[blue] (0,-1) circle (2pt) node[right] () {};
		\end{scope}
	\end{tikzpicture}
	\caption{Topological construction of the local fusion junction for Hom($T\times \eta^k,T$).}
	\label{fig: T-eta fusion}
\end{figure}
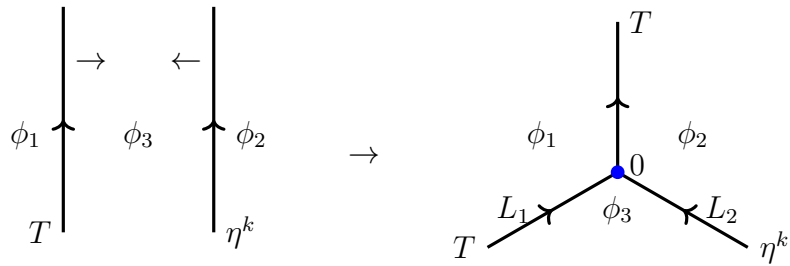

\paragraph{Example: $T\times \eta^k = T$} \label{sec: local fusion junction}
As an example, let us consider fusing the two defects $T$ and $\eta^k$ along the segment $t>0$ as shown in Figure~\ref{fig: T-eta fusion}.
Using \eqref{T_L} and \eqref{eta_L}, the defect part of the action associated with this configuration is
\begin{equation}
\begin{aligned}
S =& {iN\over 2\pi} \int_{L_1}\phi_1 \dr\phi_3 + {i\over 2\pi}\int_{L_2}\dr\varphi\left(\phi_3-\phi_2 + {2\pi k \over N}\right) \\
&+ {iN\over 2\pi}\int_{t>0}  \phi_1\dr\phi_2 + {i\over 2\pi}\int_{t>0} \widetilde{\phi} \dr\widetilde{\varphi} + S_{bdry} ~,
\end{aligned}
\end{equation}
where the trivial TQFT fields are
\begin{equation}
\widetilde{\phi} = \phi_3 - \phi_2 + {2\pi k \over N}\,, \quad \widetilde{\varphi} = \varphi - N \phi_1 ~, \label{eq: ae}
\end{equation}
and the boundary term is
\begin{equation}
    S_{bdry} = -\left. {iN\over 2\pi} \phi_1 \widetilde{\phi} \right|_0 \,.
\end{equation}
The action on the first line describes the $T$ and $\eta^k$ defects along the lines $L_1$ and $L_2$ below the fusion junction $t=0$, the first term on the second line is the $T$ defect along the fused segment $t>0$, as expected from the fusion algebra $T\times \eta^k = T$, and the last term is the boundary term at the fusion junction $t=0$.

Since, for $t>0$, $\phi_3$ and $\varphi$ are supported only along the fused segment of the defect, the trivial TQFT is decoupled from the bulk; hence, as discussed above, one should impose the Dirichlet boundary conditions
\begin{equation}
    \left. \widetilde{\phi}_3 \right|_{t=0} =0 \quad \text{mod $2\pi$} \,, \quad \left. \widetilde{\varphi} \right|_{t=0} = 0 \quad \text{mod $2\pi$}
\end{equation}
(obtained by adding a topological boundary for the TQFT on the $t>0$ segment and pushing it to the $t=0$ junction). Note that we deliberately added the constant ${2\pi k \over N}$ to the expression of $\widetilde{\phi}$ so that the boundary condition is consistent with the relation between the bulk fields set along the $\eta^k$ line, $\phi_2=\phi_3+{2\pi k \over N}$ mod $2\pi$.  If we did not include this constant in $\widetilde{\phi}$, the boundary term would differ by ${ik\over N}\widetilde{\varphi}(t=0)$, which is not trivial in general. In summary, the boundary conditions and boundary terms at the fusion junction $t=0$ are given by 
\begin{equation}
\begin{aligned}
    \text{Hom}(T\times \eta^k, T): \quad &\left. \phi_3-\phi_2+{2\pi k \over N}\right|_{t=0} = 0\,, \quad \left. \varphi - N\phi_1\right|_{t=0} = 0 \quad \text{mod $2\pi$}\,,\\
    & \quad S_{bdry} = -{iN\over 2\pi}\phi_1\left(\phi_3-\phi_2+{2\pi k \over N}\right)\,.
\end{aligned} \label{eq: aj}
    \end{equation}

Similarly, the junction data at the corresponding splitting junction $\text{Hom}(T,T\times \eta^k)$ is the same except for the boundary term, which has the opposite sign.

\subsubsection{Extra junction}
When the fusion outcome involves the direct sum of more than one simple defect, a specific fusion channel can be obtained by adding an extra junction between the direct sum and the particular simple defect of interest.  At the junction, we impose the well-defined variational principle and gauge invariance to get the boundary conditions and required boundary terms.  In the following, we illustrate this process using the example of the $T\times T$ fusion.

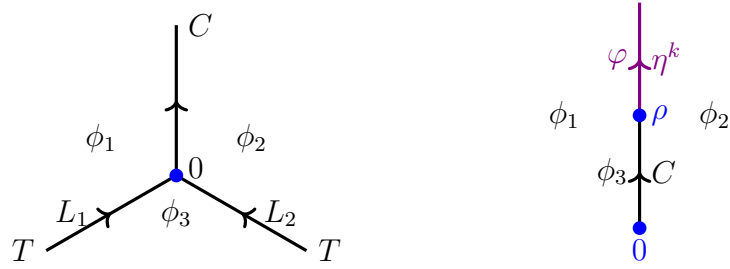
\begin{figure}[H]
	\centering
	\begin{tikzpicture}
		\begin{scope}[very thick,decoration={
				markings,
				mark=at position 0.5 with {\arrow{>}}}
			] 
			\draw[postaction={decorate}] (-1.73,-2)--(0,-1) node[at start,left] (l) {$T$} node[midway,left] () {$L_1~$};
			\draw[postaction={decorate}] (1.73,-2)--(0,-1) node[at start,right] () {$T$} node[midway,right] () {$~L_2$};
			\draw[postaction={decorate}] (0,-1)--(0,1) node[at end,right] () {$C$};
			\node at (-1,-0.5) () {$\phi_1$};
			\node at (1,-0.5) () {$\phi_2$};
			\node at (0,-1.5) () {$\phi_3$};
			\node[right] at (0,-0.9) () {$0$};
			\filldraw[blue] (0,-1) circle (2pt) node[right] () {};
            \node at (3,-2.5) (f) {};
		\end{scope}
		\begin{scope}[shift={($(f.east)+(3,0.8)$)}, very thick,decoration={
				markings,
				mark=at position 0.5 with {\arrow{>}}}
			] 
	\draw[postaction={decorate}] (0,0)--(0,1.5) node[midway,right] () {$C$} node[midway,left] () {$\phi_3$};
    \draw[Purple,postaction={decorate}] (0,1.5)--(0,3) node[midway,right] () {$\eta^k$} node[midway,left] () {${\varphi}$};
		
		\node at (-1,1.5) () {$\phi_1$};
		\node at (1,1.5) () {$\phi_2$};

        \filldraw[blue] (0,0) circle (2pt) node[below,blue] () {$0$};
		\filldraw[blue] (0,1.5) circle (2pt) node[right,blue] () {$\rho$};	
		\end{scope}
	\end{tikzpicture}
	\caption{The topological fusion junction for Hom($T\times T,C$) at $t=0$ (left) and the junction Hom($C,\eta^k$) at $t=\rho$ (right).}
	\label{fig: T-T fusion}
\end{figure}

\paragraph{Example: $T\times T= C \to \eta^k$}
Using the method discussed in the previous subsection, a local fusion junction can be constructed for $T\times T = \sum_{n=0}^{N-1} \eta^n\equiv C$, where $C$ denotes the condensation defect\footnote{Notice that this condensation defect is simply the projector for the $\mathbb{Z}_{N}\subset U(1)_{m}$ symmetry, and is not simple as a defect.} for the $\mathbb{Z}_{N}\subset U(1)_{m}$ shift symmetry, as shown in Figure~\ref{fig: T-T fusion}. 
The defect action in the fused segment is then as follows, 
\begin{equation}
    \begin{aligned}
        S&={iN\over 2\pi}\int_{t>0} \left( \phi_1 \dr\phi_3 + \phi_3 \dr \phi_2 \right)\\
        &= {iN\over 2\pi }\int_{t>0} \dr\phi_3 (\phi_1-\phi_2) - \left. {iN\over 2\pi}\phi_3\phi_2\right|_{t=0} ~,
    \end{aligned}\label{eq: af}
\end{equation}
where the first term on the second line corresponds to the condensation defect and the second term is a boundary term. To see that the first term indeed describes the condensation defect, let us note that the sum of the $\mathbb{Z}_{N}\subset U(1)_{m}$ defects used in the definition of $C$ is given by  
\begin{equation}
\sum_{n=0}^{N-1}\exp\left[\frac{i}{2\pi}\int_{L}\mathrm{d}\varphi\left(\phi_{1}-\phi_{2}+\frac{2\pi n}{N}\right)\right]=\exp\left[\frac{i}{2\pi}\int_{L}\mathrm{d}\varphi\left(\phi_{1}-\phi_{2}\right)\right]\sum_{n=0}^{N-1}\exp\left(\frac{in}{N}\int_{L}\mathrm{d}\varphi\right) ~,
\label{C_as_sum}
\end{equation}
and the last sum restricts the winding number of $\varphi$ to be an integer multiple of $2\pi N$. As a result, we can write $\varphi = N\phi$ for some $2\pi$-periodic compact scalar $\phi$ and have the following equality under the path integral, 
\begin{equation}
\sum_{n=0}^{N-1}\exp\left[\frac{i}{2\pi}\int_{L}\mathrm{d}\varphi\left(\phi_{1}-\phi_{2}+\frac{2\pi n}{N}\right)\right]=N\exp\left[\frac{iN}{2\pi}\int_{L}\mathrm{d}\phi\left(\phi_{1}-\phi_{2}\right)\right],
\label{C_equality}
\end{equation}
which results in precisely the first term on the second line of Eq. \eqref{eq: af} with $\phi=\phi_3$.\footnote{The factor on $N$ on the right hand side of \eqref{C_equality} is expected and corresponds to the two factors of $\sqrt{N}$ that are present in the two $T$ defects which are part of the fusion junction of Figure~\ref{fig: T-T fusion} (see footnote \ref{T_qd}).}

\paragraph{Selection of a fusion channel}
A specific fusion channel such as $T\times T \to \eta^k$ can be obtained by adding an additional junction $\rho$ between the condensation defect and $\eta^k$ as in the right diagram in Figure~\ref{fig: T-T fusion}.  To describe this junction, we start by examining the action  
\begin{equation}
    S = {iN\over 2\pi}\int_{t=0}^{t=\rho} \dr\phi_3(\phi_1-\phi_2) + {i\over 2\pi}\int_{t>\rho} \dr\varphi \left(\phi_1-\phi_2+{2\pi k \over N}\right)\,. \label{eq: afb}
\end{equation}
Under the gauge transformations $\phi_i \to \phi_i + 2\pi n_i$, 
\begin{equation}
   \Delta S = \left. i(n_1-n_2)(N\phi_3-\varphi)\right|_{t=\rho} ~,
\end{equation}
and in order to have gauge invariance, we set
\begin{equation}
    \left. (N\phi_3-\varphi)\right|_{t=\rho} \in 2\pi \mathbb{Z}\,. 
\label{eq: afa}
\end{equation}
Additionally, we impose
\begin{equation}
    \left. \phi_1-\phi_2 +{2\pi k \over N}\right|_{t=\rho} \in 2\pi\mathbb{Z} ~,
\label{eq: aga}
\end{equation}
since $t=\rho$ is a part of the $\mathbb{Z}_N$ defect $\eta^k$ on which this relation holds.

%These boundary conditions are consistent with the variational principle.  To see this, note that the total variation at the junction $t=\rho$ is given by 
Turning to the variational principle, the total variation at the junction $t=\rho$ is given by 
\begin{equation}
   \left. \delta S\right|_{t=\rho} = \left. {i\over 2\pi}\delta\left(N\phi_3-\varphi\right) \left(\phi_1-\phi_2 +{2\pi k \over N}\right)\right|_{t=\rho} -{ik}\delta\phi_3(\rho) = 0\,.
\end{equation}
Now, the first term vanishes due to the boundary condition \eqref{eq: afa}. However, in order to remove the second term we need to add an extra boundary term to \eqref{eq: afb} at the junction $\rho$, 
\begin{equation}
    S_{bdry} = ik\phi_3(\rho)\,.
\end{equation}
In summary, the boundary conditions and the boundary term at the junction Hom($C,\eta^k$) are 
\begin{equation}
\begin{aligned}
    \text{Hom}(C,\eta^{k}):\quad&\left.(N\phi_{3}-\varphi)\right|_{t=\rho}=0\,\;\text{mod \ensuremath{2\pi}}\,,\quad\left.\left(\phi_{1}-\phi_{2}+\frac{2\pi k}{N}\right)\right|_{t=\rho}=0\,\;\text{mod \ensuremath{2\pi}}\,,\\
    &S_{bdry}=ik\phi_{3}(\rho)\,.
\label{eq: ag}
\end{aligned}
\end{equation}

Now, to obtain the local fusion junction $T\times T \to \eta^k$ we can shrink the condensation defect on the interval $0<t<\rho$ by taking $\rho \to 0$.  Combining the boundary terms from $T\times T \to C$~\eqref{eq: af} and $C\to \eta^k$~\eqref{eq: ag}, the boundary data at the resulting junction is given by 
\begin{equation}
\begin{aligned}
    \text{Hom}(T\times T, \eta^k):\quad&\left.(N\phi_{3}-\varphi)\right|_{t=0}=0\,\;\text{mod \ensuremath{2\pi}}\,,\quad\left.\left(\phi_{1}-\phi_{2}+\frac{2\pi k}{N}\right)\right|_{t=0}=0\,\;\text{mod \ensuremath{2\pi}}\,,\\
    &S_{bdry}=-\left. {iN\over 2\pi}\phi_3\left(\phi_2 -{2\pi k \over N}\right) \right|_{t=0}\,.
\end{aligned}
\label{eq: ak}
\end{equation}

The analyses above can be generalized to any extra junction between two different topological defects: the gauge invariance and $\delta S=0$ at the junction fix the boundary conditions and boundary terms.

In Appendix~\ref{app: local fusion junctions}, we provided an alternative derivation for the same results~\eqref{eq: ak} using the previously obtained $T\times \eta^{-k}\to T$ junction data.

\paragraph{Other local fusion junctions} In Appendix~\ref{app: local fusion junctions} we provide the junction data for all the other local fusion junctions involving the $T$ and $\eta$ defects.

%===========================

\subsection{$F$-symbols using Lagrangian}
\subsubsection{The method} \label{sec: F-symbol method}
\begin{figure}[H]
	\centering
	\begin{tikzpicture}
    \begin{scope}[scale=0.8,every node/.style={scale=0.8}]
		\begin{scope}[very thick,decoration={
				markings,
				mark=at position 0.5 with {\arrow{>}}}
			]
            \draw[postaction={decorate}] (0,1.5)--(0,3) node[midway,left] () {$a\times(b\times c)$};%a(bc)
            \draw[postaction={decorate}] (-1.5,0)--(0,1.5) node[midway,left] () {$a~$};%a
            \draw[postaction={decorate}] (0.75,0.75)--(0,1.5) node[at end,right] () {$~b\times c$};%bc
            \draw[postaction={decorate}] (1.5,0)--(0.75,0.75);
            \draw[postaction={decorate}] (-0.75,-0.75)--(0.75,0.75) node[midway,left] () {$b~$};%b
            \draw[postaction={decorate}] (0,-1.5)--(1.5,0) node[midway,right] () {$~c$};%c
            \draw[postaction={decorate}] (0,-1.5)--(-0.75,-0.75) node[at start,left] () {$a\times b~$};%ab
            \draw[postaction={decorate}] (-0.75,-0.75)--(-1.5,0);
            \draw[postaction={decorate}] (0,-3)--(0,-1.5) node[midway,right] () {$(a\times b)\times c$};%(ab)c
            			
	\node[blue] at (-1,-2) () {$\phi_L$};
        \node[blue] at (1,2) () {$\phi_R$};
        \node[blue] at (0.5,-0.25) () {$\phi_I'$};
        \node[blue] at (0,0.75) () {$\phi_I$};
			
            \filldraw[blue] (0,1.5) circle (2pt) node[left,blue] () {$\alpha$};
            \filldraw[blue] (0.75,0.75) circle (2pt) node[right,blue] () {$\beta$};
            \filldraw[blue] (-0.75,-0.75) circle (2pt) node[left,blue] () {$\gamma$};
            \filldraw[blue] (0,-1.5) circle (2pt) node[right,blue] () {$\delta$};

			\node at (2,0) (f) {$\to$};
		\end{scope}
        \begin{scope}[shift={($(f.east) + (1.7,0)$)},very thick,decoration={
				markings,
				mark=at position 0.5 with {\arrow{>}}}
			] 
			\draw[postaction={decorate}] (0,1.5)--(0,3) node[midway,left] () {$a\times(b\times c)$};
            \draw[postaction={decorate}] (-1.5,0)--(0,1.5) node[midway,left] () {$a~$};
            \draw[postaction={decorate}] (1.5,0)--(0,1.5) node[midway, right] () {$~c$}; 
            \draw[postaction={decorate}] (0,-1.5)--(1.5,0) node[midway,right] () {$~c$};
            \draw[postaction={decorate}] (0,-1.5)--(0,1.5);
            \draw[postaction={decorate}] (0,-1.5)--(-1.5,0);
            \draw[postaction={decorate}] (0,-3)--(0,-1.5) node[midway,right] () {$(a\times b)\times c$};

         \node at (0.25,0.75) () {$b$};
	\node[blue] at (-1,-2) () {$\phi_L$};
        \node[blue] at (1,2) () {$\phi_R$};
        \node[blue] at (0.5,0) () {$\phi_I'$};
        \node[blue] at (-0.5,0) () {$\phi_I$};
			
            \filldraw[blue] (0,1.5) circle (2pt) node[left,blue] () {$\alpha=\beta$};
            \filldraw[blue] (0,-1.5) circle (2pt) node[right,blue] () {$\gamma=\delta$};
			\node at (2,0) (ff) {$\to$};
		\end{scope}
        \begin{scope}[shift={($(ff.east) + (1,0)$)},very thick,decoration={
				markings,
				mark=at position 0.5 with {\arrow{>}}}
			] 
			
			\draw[postaction={decorate}] (0,-3)--(0,-1.5) node[midway,right] () {$(a\times b)\times c$};
            \draw[postaction={decorate}] (0,-1.5)--(0,1.5) node[midway,right] at (0.5,0.2) () {$a\times b\times c$};
            \draw[postaction={decorate}] (0,1.5)--(0,3) node[midway,right] () {$a\times (b\times c)$};
			\filldraw[blue] (0,1.5) circle (2pt) node[right,blue] () {$\alpha=\beta$};
            \filldraw[blue] (0,-1.5) circle (2pt) node[right,blue] () {$\gamma=\delta$};
			
			\node at (-0.5,-0.75) () {$\phi_L$};
			\node at (0.5,-0.75) () {$\phi_R$};
			\node at (2.5,0) (fff) {$\to$};
		\end{scope}
		\begin{scope}[shift={($(fff.east) + (1,0)$)},very thick,decoration={
				markings,
				mark=at position 0.5 with {\arrow{>}}}
			] 
			
			\draw[postaction={decorate}] (0,-3)--(0,0) node[midway,right] () {$(a\times b)\times c$};
            \draw[postaction={decorate}] (0,-0)--(0,3) node[midway,right] () {$a\times (b\times c)$};
			\filldraw[blue] (0,0) circle (2pt) node[above right,blue] () {$\swarrow$};

            \node[blue] at (0,0.7) () {$\alpha=\beta=\gamma=\delta$}; 
			\node at (-0.5,-0.75) () {$\phi_L$};
			\node at (0.5,-0.75) () {$\phi_R$};
			\node at (2,0) (ffff) {$=F^d_{a,b,c}\times$};
		\end{scope}
        \begin{scope}[shift={($(ffff.east) + (0,0)$)},very thick,decoration={
				markings,
				mark=at position 0.5 with {\arrow{>}}}
			] 
		
			\draw[postaction={decorate}] (0,-3)--(0,3);
		\end{scope}
        \end{scope}
	\end{tikzpicture}
	\caption{The associator $F^d_{a,b,c}$ can be obtained by collapsing the diamond in the middle.  First arrow: Merging the two nodes $\alpha$ and $\beta$ and also the two nodes $\gamma$ and $\delta$. Second arrow: Parallel fusion of the $a,b,c$ defects in the middle. Third arrow: Reducing the interval between the two nodes $\alpha=\beta$ and $\gamma=\delta$.}
	\label{fig: associator}
\end{figure}
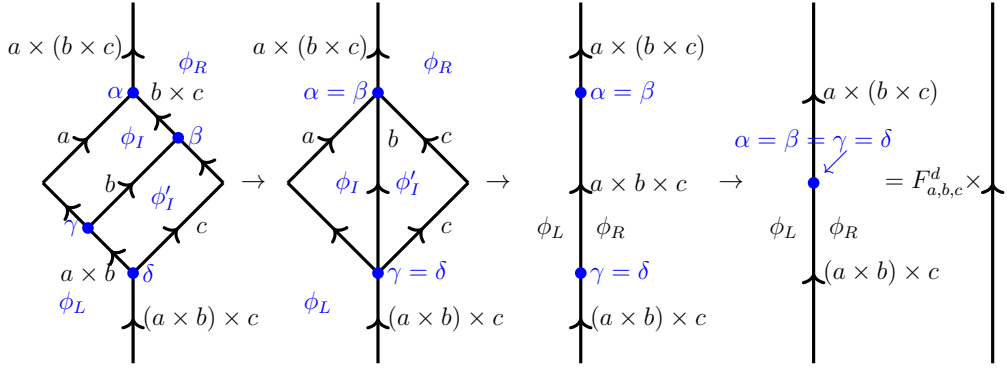

A specific component of an associator, which is a $\mathbb{C}$-number in 2d, can be computed by collapsing the diamond-shaped region in a corresponding defect network, as described in Figure~\ref{fig: associator}~\cite{Bhardwaj:2017xup}.  

We collapse the associator region by taking the following steps:
\begin{enumerate}[Step 1:]
    \item \textbf{Junction analysis}:  Using the method explained in Section~\ref{sec: local fusion junctions}, compute the boundary conditions and boundary terms at each local fusion/splitting junction $\alpha,\beta,\gamma,\delta$ as well as at possible extra junctions.
    
    \item \textbf{Merge} the two upper junctions $\alpha$ and $\beta$ by combining the boundary conditions and boundary terms.  Repeat for the lower junctions $\gamma$ and $\delta$.

    \item \textbf{Global fusion}:  Fuse the three defects $a,b,c$ by adding their actions along the interval between the $\alpha=\beta$ and $\gamma=\delta$ junctions.  The result will in general be the sum of the expected defect, some decoupled TQFT (if present) and a boundary term,
    \begin{equation}
        S= S_{a\times b\times c}+ S_{decoupled~TQFT} + S_{bdry;global}\,. 
    \label{eq: apa}
    \end{equation}
%    If there is a decoupled TQFT, it can be either trivial or a $\mathbb{Z}_N$ gauge theory (${iN\over 2\pi}\int^{\beta}_{\gamma} \tilde\phi \, \dr\tilde\varphi$). 

    \item \textbf{Interval reduction}:  Compute the path integral of the decoupled TQFT as an inner product between the states corresponding to the boundary conditions and some boundary terms. 
    
    \item \textbf{Associator}:  Evaluate all the boundary terms using the boundary conditions in Step~2.  The associator is
    \begin{equation}
        F^d_{a,b,c} = (\text{Inner product})\times e^{-S_{bdry}}\,\,. \label{eq: ap}
    \end{equation}

\end{enumerate}

\subsubsection{Path integral of $\mathbb{Z}_N$ TQFT}\label{sec: PI of ZN}
The result of the global fusion~\eqref{eq: apa} may include a 1d decoupled trivial or $\mathbb{Z}_N$  TQFT.  After reducing the interval $[\gamma,\beta]$, they are replaced with their path integral subject to the given boundary conditions plus some boundary terms.  For the trivial TQFT, the required boundary terms are fixed values.  However, for the decoupled $\mathbb{Z}_N$ TQFT  
\begin{equation}
    S = {iN\over 2\pi}\int^\beta_\gamma \phi \, \dr\varphi 
\label{eq: am}
\end{equation}
the boundary terms might involve unfixed boundary values of the defect fields.  However, the unfixed part of the boundary terms will drop out once it is combined with all other boundary terms -- those from each junction and those from the global fusion part -- and the total boundary term will only involve the fixed boundary values.  Therefore, the boundary terms do not affect the path integral over the decoupled defect fields.  In this subsection, we illustrate the reduction of the interval when a decoupled $\mathbb{Z}_N$ TQFT shows up after the global fusion.

To perform the path integral, one can add and subtract an appropriate boundary term to write the action as the sum of the part that has a vanishing variation on the boundary with respect to the boundary conditions and the boundary terms.  The former part is path integrated to the inner product between the states corresponding to the boundary conditions on the upper and lower junctions. 

\paragraph{Example: Distinct boundary conditions}
As an example, consider the boundary conditions\footnote{The Neumann conditions can be written in terms of an unfixed integer $0\le k < N$ as $\phi(\gamma) = {2\pi k \over N}$.}
\begin{equation}
\begin{aligned}
    \phi(\beta) = {2\pi m \over N}\,, \quad \varphi(\gamma) = {2\pi n \over N} \quad \text{mod $2\pi$} \quad \text{(Dirichlet)} ~, \\
    N\phi(\gamma) =0\,, \quad N\varphi(\beta)=0 \quad \text{mod $2\pi$} \quad \text{(Neumann)\,.}
\end{aligned}
\end{equation}
Rewriting the action~\eqref{eq: am} as
\begin{equation}
\begin{aligned}
    S &= \left( {iN\over 2\pi}\int^{\beta}_{\gamma} \phi\dr\varphi -{iN\over 2\pi}\phi(\beta)\varphi(\beta) \right) + {iN\over 2\pi}\phi(\beta)\varphi(\beta) ~,
    \label{eq: al}
\end{aligned}
\end{equation}
the terms in the bracket have trivial variation on the boundary with respect to the given boundary conditions.  The path integral of this part results in the inner product between the states corresponding to Dirichlet boundary conditions at each end. Specifically,\footnote{The LHS should be understood as the analytic continuation to the Lorentzian spacetime.}\footnote{The noamalisation constant can be obtained by demanding the orthonormality: $\left\langle \phi={2\pi i m\over N}\right| \left. \phi={2\pi i n \over N}\right\rangle = \sum_{k=0}^{N-1}\left\langle \phi={2\pi i m\over N}\right| \left. \varphi={2\pi i k \over N}\right\rangle \left\langle \varphi={2\pi i k \over N}\right|\left.\phi = {2\pi i n \over N}\right\rangle  = \delta_{m,n}$.}
\begin{equation}
    \int_{DBC} [\mathcal{D}\varphi][\mathcal{D}\phi] \, e^{-S_{PI}} = \langle \phi(\beta)\, |\,\varphi(\gamma)\rangle = {1\over \sqrt{N}}\exp\left({iN\over 2\pi}\phi(\beta)\varphi(\gamma) \right) ~,
\label{Int_alt}
\end{equation}
where $S_{PI}$ is given by the terms in brackets in Eq.~\eqref{eq: al}.  

The boundary term (last term in ~\eqref{eq: al}) is carried forward.  After combined with other boundary terms from the junction analysis or global fusion, the unfixed part in this term will either be removed or be a part of a fixed boundary value.

To summarize, the contribution to the $F$-symbol from the decoupled $\mathbb{Z}_N$ theory is the combination of the inner product of the boundary states~\eqref{Int_alt} and the boundary terms, as stated in Eq.~\eqref{eq: ap}:
\begin{equation}
{1\over \sqrt{N}}\exp\left[ {iN\over 2\pi}\phi(\beta)\left( \varphi(\gamma)-\varphi(\beta) \right) \right] \label{eq: an} = {1 \over \sqrt{N}}e^{{2\pi i mn\over N}}\exp\left( -{iN\over 2\pi}\phi(\beta)\varphi(\beta) \right)\,.
\end{equation}

\paragraph{General cases}
It is straightforward to generalize the above argument to other boundary conditions.  We list the resulting inner products and boundary terms in Table~\eqref{eq: ao}, where the first two columns are the boundary conditions for the fields at the specified boundary, and $D,N$ denote Dirichlet and Neumann boundary conditions, respectively.
\begin{equation}
    \begin{aligned}
    \begin{array}{cc|ccc}
    \hline
     \phi(\gamma,\beta)    & \varphi(\gamma,\beta) & S_{bdry} & \langle DBC_\beta | DBC_\gamma\rangle & \text{Contribution to $F$} \\
     \hline
     ND & DN & {iN\over 2\pi}\phi(\beta)\varphi(\beta) & {1\over \sqrt{N}} \exp\left[{iN\over 2\pi}\phi(\beta)\varphi(\gamma) \right] & {1\over \sqrt{N}} \exp\left[{iN\over 2\pi}\phi(\beta)(\varphi(\gamma) - \varphi(\beta)) \right] \\
     DN & ND & -{iN\over 2\pi}\phi(\gamma)\varphi(\gamma) & {1\over \sqrt{N}} \exp\left[-{iN\over 2\pi}\varphi(\beta) \phi(\gamma) \right] & {1\over \sqrt{N}}\exp\left[{iN\over 2\pi}\phi(\gamma) (\varphi(\gamma) - \varphi(\beta) )\right] \\
     DD & NN & -\left. {iN\over2\pi}\phi\varphi\right|^\beta_\gamma & \delta_{\phi(\beta),\phi(\gamma)} & \delta_{\phi(\beta),\phi(\gamma)} \exp\left[ \left. {iN\over2\pi}\phi\varphi\right|^\beta_\gamma \right] \\
     NN & DD & 0 & \delta_{\varphi(\beta),\varphi(\gamma)} & \delta_{\varphi(\beta),\varphi(\gamma)}\\
     \hline 
    \end{array}
    \end{aligned} \label{eq: ao}
\end{equation}

\subsubsection{Example: $\left[F^T_{T,T,T} \right]_{\eta^m,\eta^n}$}
\begin{figure}[H]
	\centering
	\begin{tikzpicture}
		\begin{scope}[very thick,decoration={
				markings,
				mark=at position 0.5 with {\arrow{>}}}
			]
            \draw[postaction={decorate}] (0,1.5)--(0,3) node[midway,left] () {$T$}; %a(bc)
            \draw[postaction={decorate}] (-1.5,0)--(0,1.5) node[midway,left] () {$T~$}; %a
            \draw[postaction={decorate},purple] (0.75,0.75)--(0,1.5) node[at end,right] () {$~\eta^n,\varphi_1$}; %bc
            \draw[postaction={decorate}] (1.5,0)--(0.75,0.75); 
            \draw[postaction={decorate}] (-0.75,-0.75)--(0.75,0.75) node[midway,left] () {$T~$}; %b
            \draw[postaction={decorate}] (0,-1.5)--(1.5,0) node[midway,right] () {$~T$}; %c
            \draw[postaction={decorate},purple] (0,-1.5)--(-0.75,-0.75) node[at start,left] () {$\eta^m,\varphi_2~$}; %ab
            \draw[postaction={decorate}] (-0.75,-0.75)--(-1.5,0);
            \draw[postaction={decorate}] (0,-3)--(0,-1.5) node[midway,right] () {$T$}; %(ab)c
            			
	\node[blue] at (-1,-2) () {$\phi_L$};
        \node[blue] at (1,2) () {$\phi_R$};
        \node[blue] at (0.5,-0.25) () {$\phi_I'$};
        \node[blue] at (0,0.75) () {$\phi_I$};
			
            \filldraw[blue] (0,1.5) circle (2pt) node[left,blue] () {$\alpha$};
            \filldraw[blue] (0.75,0.75) circle (2pt) node[right,blue] () {$\beta$};
            \filldraw[blue] (-0.75,-0.75) circle (2pt) node[left,blue] () {$\gamma$};
            \filldraw[blue] (0,-1.5) circle (2pt) node[right,blue] () {$\delta$};

        \end{scope}
	\end{tikzpicture}
	\caption{The configuration used for computing $[F^T_{T,T,T}]_{\eta^m,\eta^n}$.}
	\label{fig: FT-TTT}
\end{figure}
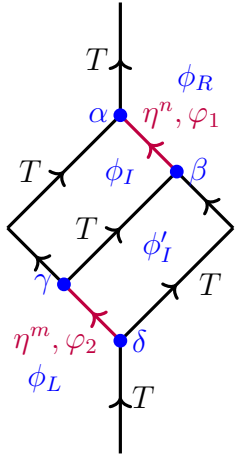

To illustrate the procedure, let us compute $[F^T_{T,T,T}]_{\eta^m,\eta^n}$ using the method explained in Sections~\ref{sec: F-symbol method} and \ref{sec: PI of ZN}, namely by collapsing the diamond region in Figure~\ref{fig: FT-TTT}.  
As a first step, we collect the boundary conditions and boundary terms at each junction $\alpha,\beta,\gamma,\delta$, taken from already computed results.  Specifically, the $\alpha$ junction is a fusion junction $T\times \eta^n \to T$ whose data is given in Eq.~\eqref{eq: aj}, the $\beta$ and $\gamma$ junctions are fusion and splitting junctions for $T\times T \to \eta^k$ with $k=n$ and $k=m$ respectively, with the data given in Eq.~\eqref{eq: ak}\footnote{For the splitting junction, the boundary term changes the sign.}, and $\delta$ is the splitting junction $T \to \eta^m\times T$ analyzed in Eq.~\eqref{eq: eta-T=T}.  Explicitly, 
\begin{equation}
    \begin{aligned}
        \alpha: &\quad \varphi_1 - N\phi_L = 0\,, \quad \phi_I - \phi_R = -{2\pi n\over N}\,, \quad S(\alpha) = -{iN \over 2\pi} \phi_L \left( \phi_I - \phi_R + {2\pi n \over N} \right), \\
        \beta: &\quad \varphi_1-N\phi_I'=0\,, \quad S(\beta) = -{iN\over 2\pi}\phi_I'\left( \phi_R -{2\pi n\over N}\right),\\
        \gamma: &\quad \varphi_2 - N\phi_I=0\,, \quad S(\gamma) = {iN\over 2\pi}\phi_I\left( \phi_I'-{2\pi m\over N} \right),\\
        \delta: &\quad \phi_L-\phi_I'=-{2\pi m \over N} \,, \quad \varphi_2 - N\phi_R = 0\,, \quad S(\delta) = im\phi_R(\delta) ~, \label{eq: aq}
    \end{aligned}
\end{equation}
where all the boundary conditions are written modulo $2\pi$, and $S(\alpha)$ denotes the boundary term at the junction $\alpha$ (and similarly for other junctions).  Next, we shrink the interval $[\beta,\alpha]$ to obtain a single upper junction $\beta=\alpha$.  As a consequence, we can impose the boundary conditions at the $\alpha$ and $\beta$ junctions simultaneously.  Similarly, we shrink the interval $[\delta,\gamma]$ to get the single lower junction $\gamma=\delta$.

Next, we fuse the three defects $T\times T\times T$ in the middle by adding the corresponding defect actions, 
\begin{equation}
    \begin{aligned}
        S& = {iN\over 2\pi}\int^\beta_\gamma \left( \phi_L\dr\phi_I + \phi_I\dr\phi_I' + \phi_I'\dr\phi_R \right)\\
       &= {iN\over 2\pi}\int^\beta_\gamma\phi_L \dr\phi_R+ {iN\over 2\pi}\int^\beta_\gamma (\phi_L-\phi_I')\dr\left(\phi_I-\phi_R\right) + {iN\over 2\pi}\left. \phi_I'\phi_I\right|^\beta_\gamma
    \end{aligned} \label{eq: ar}
\end{equation}
where the first term is the action for the resulting $T$ defect, the second term is a decoupled $\mathbb{Z}_N$ TQFT and the last term is a boundary term $S_{bdry;global}$.  The decoupled TQFT fields have the following boundary conditions,
\begin{equation}
\begin{aligned}
    N(\phi_L - \phi_I')(\beta) = 0\,, \quad &(\phi_I-\phi_R)(\beta) = -{2\pi n \over N} \quad \text{mod $2\pi$}\,, \\
    (\phi_L-\phi_I')(\gamma) = -{2\pi m \over N}\,, \quad &N(\phi_I-\phi_R)(\gamma) = 0 \quad \text{mod $2\pi$\,,}
\end{aligned}
\end{equation}
namely $\phi_L-\phi_I'$ has a Dirichlet boundary condition at $\gamma$ and a Neumann one at $\beta$, while $\phi_I-\phi_R$ has the opposite boundary conditions.  Using the second row in Table~\eqref{eq: ao}, the contribution from this $\mathbb{Z}_N$ theory is hence  
\begin{equation}
{1\over \sqrt{N}} e^{-S_{\mathbb{Z}_N}}\,, \quad S_{\mathbb{Z}_N} = \left. -im (\phi_I-\phi_R)\right|^\beta_\gamma \,.
\label{eq: at}
\end{equation}

As a last step, we combine the boundary terms at the upper and lower junctions, including the phase contribution $S_{\mathbb{Z}_N}$ from the $\mathbb{Z}_N$ TQFT and the boundary term $S_{bdry;global}$ in \eqref{eq: ar}. At the upper junction,
\begin{equation}
    \begin{aligned}
        S(\alpha) + S(\beta)+  \left(S_{bdry;global} + S_{\mathbb{Z}_N} \right)_{\beta} &= {iN\over 2\pi}\left( -\phi_L + \phi_I' - {2\pi m \over N} \right)\left(\phi_I-\phi_R+{2\pi n \over N}\right) + {2\pi i mn\over N}
    \end{aligned} \label{eq: as}
\end{equation}
and using the boundary conditions at the upper junction~\eqref{eq: aq}, ${N\over 2\pi}\left( -\phi_L + \phi_I' - {2\pi m \over N} \right) \in \mathbb{Z}$ and $\left(\phi_I-\phi_R+{2\pi n \over N}\right) \in 2\pi \mathbb{Z}$, we find that the first term in Eq.~\eqref{eq: as} is trivial modulo $2\pi i$.  Thus, the total phase contribution from the upper junction is given by 
\begin{equation}
    \begin{aligned}
        S(\alpha) + S(\beta)+  \left(S_{bdry;global} + S_{\mathbb{Z}_N} \right)_{\beta} &=  {2\pi i mn\over N}\,.
    \end{aligned} \label{eq: asa}
\end{equation}

Turning to the lower junction, the sum of all the boundary terms vanishes, i.e.
\begin{equation}
    S(\gamma) + S(\delta) + \left(S_{bdry;global} + S_{\mathbb{Z}_N} \right)_{\gamma} = 0\,. \label{eq: asb}
\end{equation}
Then, putting everything together we substitute the inner product and the boundary terms (given in Eqs.~\eqref{eq: at}, \eqref{eq: asa}-\eqref{eq: asb}) into Eq.~\eqref{eq: ap} and obtain the $F$-symbol
\begin{equation}
    [F^T_{T,T,T}]_{\eta^m,\eta^n} = {1\over\sqrt{N}} \exp\left( -{2\pi i mn \over N} \right)\,,
\end{equation}
which agrees with the known value~\eqref{eq: ada}.

\paragraph{Other $F$-symbols} All the other $F$-symbols of the compact boson theory are computed in Appendix~\ref{app: F of compact bosons} and agree with Eq.~\eqref{eq: ada}.

% -------------- New section --------------------

\section{Maxwell theory in 4d}\label{sec: 4dtri}
\subsection{Non-invertible triality symmetry and defect Lagrangians}

We next turn to four dimensions and consider Maxwell theory, which has the Lagrangian 
\begin{equation}
    \mathcal{L}=\frac{1}{2e^{2}}F\wedge\star F+\frac{i\theta}{8\pi^{2}}F\wedge F\,.
\end{equation}
Normalizing the 1-from gauge field $A$ so that the corresponding field strength $F=dA$ has fluxes $2\pi\mathbb{Z}$, the theory is specified by the choice of complexified coupling 
\begin{equation}
\tau=\frac{\theta}{2\pi}+\frac{2\pi i}{e^{2}}
\end{equation}
and has an $SL(2,\mathbb{Z})$ duality group (on spin manifolds) which is generated by 
\begin{equation}
\mathbb{S}:\;\tau\rightarrow-1/\tau\quad,\quad\mathbb{T}:\;\tau\rightarrow\tau+1
\end{equation}
with the relations 
\begin{equation}
\mathbb{S}^{2}=\mathbb{C}\quad,\quad\left(\mathbb{S}\mathbb{T}\right)^{3}=1\,,
\end{equation}
where $\mathbb{C}$ is the charge-conjugation symmetry under which $A\rightarrow-A$.  In addition to charge conjugation, the theory also has two 1-form symmetries $U(1)_{E}^{(1)}\times U(1)_{M}^{(1)}$ with charges 
\begin{equation}
Q_{E}=\frac{1}{2\pi}\oint\left(-\frac{2\pi i}{e^{2}}\star F+\frac{\theta}{2\pi}F\right)\quad,\quad Q_{M}=\frac{1}{2\pi}\oint F\,,
\label{1form_charges}
\end{equation}
the charged objects being Wilson and ’t Hooft lines, respectively. Upon gauging a $\mathbb{Z}_{N}^{(1)}\subset U(1)_{E}^{(1)}$ subgroup of the electric 1-form symmetry we cut out all Wilson lines (at least as genuine line operators) whose charges are not integer multiples of $N$, resulting in the charge quantization $Q_{E}\in N\mathbb{Z}$ and $Q_{M}\in\mathbb{Z}/N$. To have the standard quantization $Q_{E,M}\in\mathbb{Z}$ as in the theory before the gauging (i.e. having the same normalization for the gauge field), we rewrite $A=A'/N$ and $\tau=\tau'N^{2}$ which then results in a theory which is similar to the one before the gauging but with the new coupling $\tau'=\tau/N^{2}$. We therefore see that gauging $\mathbb{Z}_{N}^{(1)}\subset U(1)_{E}^{(1)}$ is equivalent to changing $\tau\rightarrow\tau/N^{2}$. Similarly, gauging a $\mathbb{Z}_{N}^{(1)}\subset U(1)_{M}^{(1)}$ subgroup of the magnetic 1-form symmetry would amount to changing $\tau\rightarrow\tau N^{2}$. 

In addition to simply gauging a (non-anomalous) $\mathbb{Z}_{N}^{(1)}$ subgroup of the 1-form symmetry, an operation usually denoted by $S$, we can also consider combining it with stacking an SPT phase for the corresponding background 2-form gauge field, an operation usually denoted by $T$ (see \cite{Gaiotto:2014kfa,Bhardwaj:2020ymp,Choi:2022zal}). Denoting the background field by $B\in H^{2}\left(M_{4};\mathbb{Z}_{N}\right)$ where $M_4$ is the space in which the theory lives, this SPT phase is given by $\exp(\frac{2\pi i}{2N}\int_{M_{4}}q\left(B\right))$ for even $N$ (where $q:H^{2}\left(M_{4};\mathbb{Z}_{N}\right)\rightarrow H^{4}\left(M_{4};\mathbb{Z}_{2N}\right)$ is the Pontryagin square operation) and by $\exp(\frac{2\pi i}{N}\int_{M_{4}}\frac{N+1}{2}B^{2})$ for odd $N$. Then, considering e.g. the gauging corresponding to $ST$, i.e. first stacking the SPT phase and then gauging the $\mathbb{Z}_{N}^{(1)}$ symmetry, we have $A=A'/N$ and $\theta'=\theta+2\pi/N$ for even $N$ which amounts to changing $\tau\rightarrow\tau/N^{2}+1/N$ in the original theory, and $A=A'/N$ and $\theta'=\theta+2\pi\left(N+1\right)/N$ for odd $N$ which amounts to the change $\tau\rightarrow\tau/N^{2}+\left(N+1\right)/N$. 

The two ingredients discussed above, i.e. gauging a subgroup of the 1-form symmetry (possibly with stacking an SPT phase) and performing a duality operation, were used recently to construct many new non-invertible defects in the theory \cite{Choi:2021kmx,Kaidi:2021xfk,Roumpedakis:2022aik,Choi:2022zal,Niro:2022ctq,Hasan:2024aow,Paznokas:2025auw,Shao:2025qvf}. For simplicity, we will focus in this section on the triality defects at $\tau=e^{2\pi i \over 3}N$ found in \cite{Choi:2022zal} and on condensation defects \cite{Roumpedakis:2022aik,Choi:2022zal} associated with electric 1-form gauging, which can be constructed as follows (in appendix \ref{app: duality} we will also discuss duality defects).

Starting with the triality defect, we consider the coupling $\tau = e^{2\pi i \over 3}N$ (for any $N\in\mathbb{N}$) and perform an $ST$ gauging of a $\mathbb{Z}_{N}^{(1)}$ subgroup of the electric 1-form symmetry in half space (with Dirichlet boundary conditions for the corresponding discrete gauge field), resulting in a topological interface between the theory with coupling $\tau = e^{2\pi i \over 3}N$ and the theory with coupling $\tau/N^{2}+1/N=e^{\pi i/3}/N$ for even $N$ and coupling $\tau/N^{2}+\left(N+1\right)/N=e^{\pi i/3}/N+1$ for odd $N$. However, since the coupling $e^{2\pi i \over 3}N$ is obtained by performing an $\mathbb{S}$ transformation to $e^{\pi i/3}/N$ and an $\mathbb{S}\mathbb{T}^{-1}$ transformation to $e^{\pi i/3}/N+1$, the two sides are in fact equivalent and we get a defect in a single theory, obtained by performing $ST$ $\mathbb{Z}_{N}^{(1)}$ gauging and $\mathbb{S}$ duality (for even $N$) or $\mathbb{S}\mathbb{T}^{-1}$ duality (for odd $N$) in half space. As shown in \cite{Choi:2022zal}, one can describe this defect, which is denoted by $D_3$, by the following action, 
\begin{equation}
\begin{aligned}
    D_3:\quad  S =& \int_{x<0}\left(\frac{\sqrt{3}N}{8\pi}F_{L}\wedge\star F_{L} -{iN\over 8\pi} F_L\wedge F_L \right)  +\int_{x>0}\left( \frac{\sqrt{3}N}{8\pi} F_{R}\wedge\star F_{R} -{iN\over 8\pi}\int F_R\wedge F_R \right)\\
    &+{iN\over 2\pi}\int_{x=0} \left( A_L\wedge \dr A_R + {1\over 2}A_L\wedge \dr A_L\right)
\end{aligned}
\label{D3_L}
\end{equation}
where $A_L$ and $A_R$ denote the Maxwell gauge field from the two sides of the defect, which is located at $x=0$. The equations of motion on the defect that follow from \eqref{D3_L}, then, determine the matching conditions between the left and right gauge fields, which in turn results in an action of the defect on operators (whether local or not) which is indeed similar to that of the gauging and duality used in its construction. The corresponding orientation reversal defect, denoted by $\overline{D}_{3}$, is given by 
\begin{equation}
    \overline{D}_{3}:\quad S=-\frac{iN}{2\pi}\int_{x=0}\left(A_{L}\wedge dA_{R}+\frac{1}{2}A_{R}\wedge dA_{R}\right)
\end{equation}
where we omit the bulk action on both sides of the defect here and below for brevity.  

Turning to the condensation defects we are interested in, at any $\tau$ we can gauge a $\mathbb{Z}_{N}^{(1)}$ subgroup of the electric 1-form symmetry with a discrete torsion labeled by $H^{3}(B\mathbb{Z}_{N};U(1))\cong\mathbb{Z}_{N}$ over a codimension-1 manifold and obtain a nontrivial defect. As shown in \cite{Roumpedakis:2022aik,Choi:2022zal}, the part of the action describing this defect is given by\footnote{Our defect action for $C_k$ (as well as for the duality defect discussed in appendix~\ref{app: duality}) differs from that in \cite{Choi:2022zal} by a total derivative term, which we choose this way in order to avoid extra boundary terms when considering junctions between defects. 
} 
\begin{equation}
\label{Ck}
C_{k}:\quad S=\int_{M_3}\left[\frac{iN}{2\pi}da\wedge(A_{L}-A_{R})+\frac{ik}{2\pi}(A_{L}-A_{R})\wedge d(A_{L}-A_{R})\right]\,,
\end{equation}
where $A_L$ and $A_R$ are as before the Maxwell gauge field from the two sides of the defect, $a$ is a $U(1)$ gauge field living on the worldvolume $M_3$ and $k\in\mathbb{Z}_{N}$ is the discrete-torsion label. Integrating the defect field $a$ out sets $A_L - A_R$ to be a flat $\mathbb{Z}_N$ connection,\footnote{To see this, let us decompose $a$ as $a = \overline{a}+a'$ where $a'$ has trivial fluxes and $\overline{a}$ is a choice of representative for each flux sector labeled by $H^2(M_3,\mathbb{Z})$. Integrating out $a'$ sets $A_L-A_R$ to be flat, and the sum over the flux sector $\overline{a}$ corresponds to summing over the 1-cycle $\gamma \in H_1(M_3,\mathbb{Z})$ where $\gamma$ is the Poincar\'e dual of the first Chern class $c_1 = \frac{d\overline{a}}{2\pi} \in H^2(M_3,\mathbb{Z})$. Decomposing $\gamma = n_i\gamma_i$ where $\{\gamma_i\}$ is the bases of 1-cycles, the remaining path integral becomes

\begin{equation}
    \prod_i\sum_{\{n_i\}} e^{iN n_i\oint_{\gamma_i}(A_L-A_R)} 
    %= \prod_{i}\sum_{k_i=1,\cdots, N}\delta({N\over 2\pi} \oint_{\gamma_i}(A_L-A_R) - k_i) ~,
\end{equation}

and the sum over $n_i$ implies that the holonomy $e^{i\oint (A_L - A_R)}$ on any $\gamma_i$ takes values in $\left\{e^{\frac{2\pi i}{N}k}\right\}_{k=0}^{N-1}$.} with $e^{i\oint_\gamma (A_L - A_R)}$ generating $\mathbb{Z}_N$ topological line operators which are identified as the higher quantum symmetry on $C_k$.

Let us also describe what the identity defect looks like in the Lagrangian description in terms of left and right fields. It is given by the action 
\begin{equation}
    \mathbb{1}: \quad S = {i\over 2\pi}\int_{M_3} \dr b \wedge (A_L-A_R) 
    \label{eq: Id}
\end{equation}
where $b$ is again a $U(1)$ gauge field living on $M_3$ which upon integrating it out sets $A_L-A_R$ to be a pure $U(1)$ gauge transformation on $M_3$.

The descriptions of the defects above can be easily used to find their fusions. For example, one finds \cite{Choi:2022zal} 
\begin{equation}
D_{3}\times D_{3}=U(1)_{N}\overline{D}_{3}\;\,,\;\,\overline{D}_{3}\times\overline{D}_{3}=U(1)_{-N}D_{3}\;\,,\;\,\overline{D}_{3}\times D_{3}=C_{0}\;\,,\;\,D_{3}\times\overline{D}_{3}=C_{\frac{N}{2}}
\end{equation}
for even $N$ and 
\begin{equation}
D_{3}\times D_{3}=SU(N)_{-1}\overline{D}_{3}\;\,,\;\,\overline{D}_{3}\times\overline{D}_{3}=SU(N)_{1}D_{3}\;\,,\;\,\overline{D}_{3}\times D_{3}=D_{3}\times\overline{D}_{3}=C_{0}
\end{equation}
for odd $N$.

In the next two subsections, we will show how the Lagrangian descriptions \eqref{D3_L}-\eqref{eq: Id} can be used to characterize fusion and morphism junctions between these defects, as well as finding associator theories. We will focus on the triality associator $\left[F^{D_3}_{{D_3},D_3,\overline{D}_3}\right]_{\mathbb{1},U(1)_N\overline{D}_3}$ for even values of $N$ and the junctions that appear in its computation, deferring the rest of the associators to appendix \ref{app: triality}. Then, in section \ref{Sec_Group}, we will compute the associators using a different (group-theoretical) approach for certain odd values of $N$ and see the agreement with the results of this section, except for certain cases where the final result depends on whether $N$ is even or odd. 

In Appendix~\ref{app: duality} we will also discuss the non-invertible duality defects of \cite{Choi:2021kmx} and compute all the associators using both the Lagrangian (which is valid for all $N$) and group-theoretical methods and see their agreement.

% ----- new subsection ----- 

\subsection{Local fusion and morphism junctions}
In this subsection, we will describe the fusion and morphism junctions of the non-invertible symmetry defects we discussed above using their Lagrangian descriptions, emphasizing the way the boundary conditions and the boundary terms at these junctions arise. Notice that one can also describe the same junctions alternatively by adding additional boundary degrees of freedom as demonstrated in Appendix \ref{app:altj} (see also \cite{Kapustin:2014gua,Roumpedakis:2022aik,Bah:2025oxi}). We choose our current approach to simplify the number of fields we need to keep track of.

\subsubsection{Topological construction of the local fusion junction} \label{sec: top construction of local fusion junction 4d}
As for the 2d case, we obtain the local fusion junction Hom($\mathcal{D}_a\times \mathcal{D}_b, \mathcal{D}_c$) of the topological defects $\mathcal{D}_a,\mathcal{D}_b$ and $\mathcal{D}_c$ by fusion of the upper part of the two parallel defects $\mathcal{D}_a$ and $\mathcal{D}_b$, as depicted in the figure below.
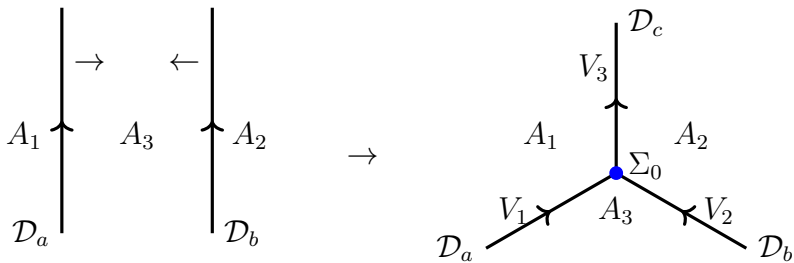
\begin{figure}[H]
	\centering
	\begin{tikzpicture}
		\begin{scope}[very thick,decoration={
				markings,
				mark=at position 0.5 with {\arrow{>}}}
			] 
			\draw[postaction={decorate}] (-1,-1)--(-1,2) node[at start,left] () {$\mathcal{D}_a$} node[near end,right] () {$\to$};
			\draw[postaction={decorate}] (1,-1)--(1,2) node[at start,right] () {$\mathcal{D}_b$} node[near end,left] () {$\leftarrow$};
			\node at (-1.5,0.3) () {$A_1$};
			\node at (1.5,0.3) () {$A_2$};
			\node at (0,0.3) () {$A_3$};
			\node at (3,0) (f) {$\to$};
		\end{scope}
		\begin{scope}[shift={($(f.east)+(3,0.8)$)}, very thick,decoration={
				markings,
				mark=at position 0.5 with {\arrow{>}}}
			] 
			\draw[postaction={decorate}] (-1.73,-2)--(0,-1) node[at start,left] (l) {$\mathcal{D}_a$} node[midway,left] () {$V_1~$};
			\draw[postaction={decorate}] (1.73,-2)--(0,-1) node[at start,right] () {$\mathcal{D}_b$} node[midway,right] () {$~V_2$};
			\draw[postaction={decorate}] (0,-1)--(0,1) node[at end,right] () {$\mathcal{D}_c$};
			\node at (-1,-0.5) () {$A_1$};
			\node at (1,-0.5) () {$A_2$};
			\node at (0,-1.5) () {$A_3$};
            \node at (-0.3,0.4) () {$V_3$};
			\node[right] at (0,-0.9) () {$\Sigma_0$};
			\filldraw[blue] (0,-1) circle (2pt) node[right] () {};
		\end{scope}
	\end{tikzpicture}
	\caption{Topological construction of the local fusion junction for Hom($\mathcal{D}_a\times \mathcal{D}_b,\mathcal{D}_c$).  $V_{1,2,3}$ are the 3-manifolds that end on the 2-dimensional local fusion junction at $\Sigma_0$ (blue dot).}
	\label{fig: local fusion junction 2}
\end{figure}
The action associated with this configuration is
\begin{equation}
    S = \int_{V_1}\mathcal{L}_{\mathcal{D}_a}[A_1,A_3] + \int_{V_2}\mathcal{L}_{\mathcal{D}_b}[A_3,A_2] +  \int_{V_3} \left(\mathcal{L}_{\mathcal{D}_a}[A_1,A_3] +\mathcal{L}_{\mathcal{D}_b}[A_3,A_2] \right)
\end{equation}
where $V_{1,2} (V_3)$ are 3-manifolds from the bottom (top) that end on the 2-dimensional local fusion junction $\Sigma_0$. The first two terms are the actions of the defects $\mathcal{D}_a$ and $\mathcal{D}_b$ below the fusion junction, and the last term describes the fusion of the two. One can then manipulate this last term in the upper part into the action of the topological defect(s) expected from the fusion algebra (in this case, the action of the defect $\mathcal{D}_c$), which in general may result in boundary terms at the junction and a decoupled trivial $\mathbb{Z}_{1}$ TQFT at the fused segment, i.e.
\begin{equation}
    \int_{V_3} \left(\mathcal{L}_{\mathcal{D}_a}[A_1,A_3] +\mathcal{L}_{\mathcal{D}_b}[A_3,A_2] \right) = \int_{V_3} \mathcal{L}_{\mathcal{D}_c}[A_1,A_2,A_3] + {i\over 2\pi}\int_{V_3} \widetilde{A} \wedge \dr\widetilde{B} + S_{bdry} \label{eq: aab}
\end{equation}
where $\widetilde{A}$ and $\widetilde{B}$ are the fields supported only along the defect in the upper half $(V_3)$, such as $A_3$ in Figure~\ref{fig: local fusion junction 2} or the defect fields of the condensation and identity defects (i.e. $a$ and $b$ in \eqref{Ck} and \eqref{eq: Id}).  Hence, the second term on the right-hand side describes a trivial decoupled TQFT. We stress that both the boundary terms and the decoupled trivial TQFT are not always present. As we will see below, however, both do appear e.g. in the construction of the local fusion junctions $\mathbb{1}\times D_3=D_3$, $D_3\times \mathbb{1} = D_3$, and their orientation reversals. 

Since we are interested in fusion junctions as in Figure~\ref{fig: local fusion junction 2}, in which each defect (i.e. $\mathcal{D}_a$, $\mathcal{D}_b$ and $\mathcal{D}_c$ in the figure) is not accompanied with a decoupled TQFT, when such a TQFT does arise we will need to add a topological boundary for it along $V_3$ and then push it to the junction $\Sigma_0$, resulting overall in a topological junction between defects with no decoupled TQFT. The topological boundary condition at the boundary we add along $V_3$ is that of Dirichlet for both $\widetilde{A}$ and $\widetilde{B}$ in \eqref{eq: aab}, as we explain next in more detail.

\paragraph{Boundary conditions of decoupled trivial TQFT}
In the 3d decoupled TQFT in \eqref{eq: aab}, the gauge-invariant operators are the Wilson lines $e^{i\oint_\gamma \widetilde{A}}$ and $e^{i\oint_{\gamma'} \widetilde{B}}$, which commute with each other as can be seen from the commutation relation between $\widetilde{A}$ and $\widetilde{B}$. As a result, on the topological boundary which we add along $V_3$, denoted by $\Sigma'_0$, both of these operators can be fixed. In fact, they have to be fixed to the identity operator (corresponding to trivial holonomies for $\widetilde{A}$ and $\widetilde{B}$) since the path integral over each of the fields restricts the other one to be a pure $U(1)$ gauge (and hence nontrivial boundary holonomies would result in a vanishing partition function). The boundary conditions for $\widetilde{A}$ and $\widetilde{B}$ at $\Sigma'_0$ are therefore given by 
\begin{equation}
    \left. \widetilde{A}, \widetilde{B}\right|_{\Sigma'_0} =  \dr\widetilde{\phi}_{A,B} 
\label{eq: decoupled TQFT bc}
\end{equation}
for some compact scalars $\widetilde{\phi}_{A,B}$, and are consistent with the variational principle ($\left. \delta S \right|_{\Sigma'_0}=0$) and gauge invariance ($\Delta S=0$). 

Now that the topological boundary has been added at $\Sigma'_0$, we push it to $\Sigma_0$ and obtain the desired junction. In this junction we have, on top of the other junction data, the boundary condition \eqref{eq: decoupled TQFT bc}.

\paragraph{Example: $D_3\times \mathbb{1} = D_3$} To illustrate the procedure, let's consider the local fusion junction for $D_3\times \mathbb{1} = D_3$ shown in the figure below.
\begin{figure}[H]
	\centering
	\begin{tikzpicture}
		\begin{scope}[very thick,decoration={
				markings,
				mark=at position 0.5 with {\arrow{>}}}
			] 
			\draw[postaction={decorate}] (-1.73,-2)--(0,-1) node[at start,left] (l) {${D}_3$} node[at start,right] () {$~V_1$};
			\draw[postaction={decorate},dotted] (1.73,-2)--(0,-1) node[at start,right] () {$\mathbb{1},b$} node[at start,left] () {$V_2~$};
			\draw[postaction={decorate}] (0,-1)--(0,1) node[at end,left] () {${D}_3$} node[at end,right] () {$V_3$};
			\node at (-1,-0.5) () {$A_1$};
			\node at (1,-0.5) () {$A_2$};
			\node at (0,-1.7) () {$A_3$};

			\filldraw[blue] (0,-1) circle (2pt) node[left,blue] () {$\Sigma_0$};
		\end{scope}
	\end{tikzpicture}
	\caption{Hom(${D}_3\times \mathbb{1},{D}_3$) junction. $b$ is the defect field on $V_2$ used in the Lagrangian description of the identity defect.}
	\label{fig: D3-id-D3 junction}
\end{figure}
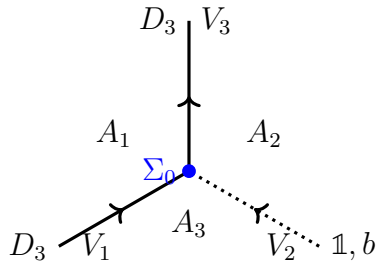
%The action associated with this configuration is
The part of the action along the $V_3$ segment obtained by fusing the defects $D_3$ and $\mathbb{1}$ in the upper part (as in Figure~\ref{fig: local fusion junction 2}) is given by\footnote{For the brevity of the notation, we leave the wedge product between a 1-form and an external derivative of a 1-form implicit hereafter.}
\begin{equation}
\begin{aligned}
    S&= {i\over 2\pi}\int_{V_3} \left[N A_1 \dr A_3 + {N\over 2}A_1  \dr A_1 + \dr b  (A_3-A_2) \right]\\
    &= {iN\over 2\pi}\int_{V_3}\left( A_1 \dr A_2 + {1\over 2}A_1  \dr A_1 \right)  + {i\over 2\pi}\int_{V_3} (b+NA_1)  \dr (A_3-A_2) - {i\over 2\pi}\int_{\Sigma_0} b\wedge (A_3-A_2)\,,
\end{aligned}
\end{equation}
where the first term on the last line describes the triality defect $D_3$ as expected from the fusion algebra, the second term is a trivial $\mathbb{Z}_{1}$ TQFT and the last term is a boundary term.  Since the fields $b+NA_1$ and $A_3-A_2$ of the trivial TQFT are defect fields living on $V_3$ (as may become more apparent by shifting $b$ by $-NA_1$ and $A_3$ by $A_2$) and do not appear in the first term describing $D_3$, this TQFT is decoupled from the bulk and from $D_3$.  Thus, as discussed above, %we impose the boundary conditions~\eqref{eq: decoupled TQFT bc}. 
we introduce a topological boundary for it along $V_3$ by imposing Dirichlet boundary conditions for both of its fields and then push this boundary to the junction $\Sigma_0$, effectively setting both $b+NA_1$ and $A_3-A_2$ to be pure $U(1)$ gauges at the junction. 

In summary, the boundary conditions and boundary term at this fusion junction are
\begin{equation}
\begin{aligned}
    \text{Hom}(D_3\times \mathbb{1}, D_3): \quad &\left. b+NA_1\right|_{\Sigma_0}, \left. A_3-A_2\right|_{\Sigma_0} = \text{pure $U(1)$ gauge}\,,\\
    & S_{bdry} =  - {i\over 2\pi}\int_{\Sigma_0} b\wedge (A_3-A_2)\,. \label{eq: bo}
\end{aligned}
\end{equation}

\paragraph{Other local fusion junctions}  All other local fusion junctions involving the triality defect are worked out in Appendix~\ref{app: triality}.

\subsubsection{Morphism junctions of condensation defects}\label{sec: codensation morphisms}
As explained in Section~\ref{sec: higher category}, condensation defects and defects containing a decoupled TQFT that admits topological boundary conditions are Witt equivalent to the identity defect. Demanding well-defined variational principle and gauge invariance at the corresponding morphism junctions fixes the boundary conditions and the extra boundary action at these junctions.  In this subsection, we illustrate this using the examples of the morphism junctions between the identity defect and the condensation defects $C_0$ and $C_{N\over 2}$, and then turn in the next subsection to consider the morphism junction between the defect containing the TQFT $U(1)_N\times U(1)_{-N}$ and the identity defect.

\begin{figure}[H]
	\centering
	\begin{tikzpicture}
		\begin{scope}[very thick,decoration={
				markings,
				mark=at position 0.5 with {\arrow{>}}}
			] 
			\draw[postaction={decorate}] (0,-1)--(0,0) node[near start,left] (l) {$C_0$ or $C_{N\over 2}$} node[near start ,right] () {$a$};
			\draw[postaction={decorate},dotted,Purple] (0,0)--(0,1) node[near end,left] () {$\mathbb{1}$} node[near end,right] () {$b$};

            \filldraw[blue] (0,0) circle (2pt) node[right,blue] () {$\Sigma_0$};
            \node at (-1,0) () {$A_1$};
            \node at (1,0) () {$A_2$};
            \node[] at (0.3,-1.4) () {$\small{V_{lower}}$};
            \node[Purple] at (0,1.4) () {$\small{V_{upper}}$};
            \node at (2,0) (f) {};
		\end{scope}
	\end{tikzpicture}
	\caption{Hom($C_0,\mathbb{1}$) junction: The lower part $V_{lower}$ supports $C_0$ (or $C_{N\over 2}$ later) and the upper part $V_{upper}$ supports $\mathbb{1}$.}
	\label{fig: C0-id junction}
\end{figure}
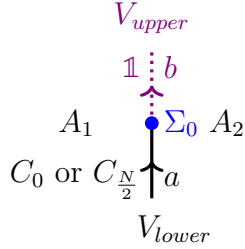

\paragraph{$\text{Hom}(C_0,\mathbb{1})$}
Consider a 2-dimensional morphism junction $\Sigma_0$ between the condensation defect $C_0$ and the identity defect supported at $V_{lower}$ and $V_{upper}$, respectively, as depicted in Figure~\ref{fig: C0-id junction}. We denote by $a$ and $b$ the defect fields on $C_0$ and $\mathbb{1}$, respectively. 
The action for this configuration is given by 
\begin{equation}
    \begin{aligned}
        S&= {iN\over 2\pi}\int_{V_{lower}} \dr a \, (A_1-A_2) + {i\over 2\pi}\int_{V_{upper}}\dr b \, (A_1-A_2)\,.
    \end{aligned} \label{eq: bb}
\end{equation}
To find the junction data at $\Sigma_0$, we demand gauge invariance and a well-defined variational principle at the junction. First, under the gauge transformations $A_{1,2} \to A_{1,2} + \dr\lambda_{1,2}$ we have 
\begin{equation}
    \begin{aligned}
        \Delta_\lambda S &= {i\over 2\pi}\int_{\Sigma_0} \left( Na -b \right)\wedge \dr(\lambda_1-\lambda_2)\,.
    \end{aligned} \label{eq: da}
\end{equation}
For small gauge transformations, i.e. $\lambda_{1,2}$ are single-valued such that $\oint \dr \lambda_{1,2}=0$, one can integrate by parts and obtain $\left. \dr(Na-b)\right|_{\Sigma_0}=0$ for gauge invariance. Then, for the large gauge transformations, i.e $\lambda_{1,2}$ have nontrivial winding modes such that $\oint \dr \lambda_{1,2}\ne 0$, one can use the Poincar\'e dual to demand $\Delta_\lambda S = i\int_{PD(\dr(\lambda_1-\lambda_2),\Sigma_0)}(Na-b) \overset{!}{\in}2\pi i \bz$ for gauge invariance. In other words, $Na-b$ is a pure $U(1)$ gauge at $\Sigma_0$. Let us then write $\left.Na-b\right|_{\Sigma_0}=\dr\phi$ for some compact scalar $\phi$, and get for the variation at $\Sigma_0$ the following
\begin{equation}
\left. \delta S \right|_{\Sigma_0} = {i\over 2\pi}\int_{\Sigma_0} \dr \delta\phi \wedge (A_1-A_2)\,.
\end{equation}
Now, since $\left. A_1-A_2\right|_{V_2}$ is a pure $U(1)$ gauge on the worldvolume $V_2$ of the identity defect which includes the junction $\Sigma_0$, $\left. \delta S\right|_{\Sigma_0}=0$ without needing any extra boundary terms. 

In summary, we have the following junction data:
\begin{equation}
    \text{Hom}(C_0,\mathbb{1}): \quad \left. b-Na, \quad A_1-A_2 \right|_{\Sigma_0} = \text{pure $U(1)$ gauge}, \quad S_{bdry} = 0\,. \label{eq: bj}
\end{equation}

Before moving on, let us describe another approach for arguing \eqref{eq: bj}. Taking advantage of the topological nature of the configuration in Figure~\ref{fig: C0-id junction}, we turn the upper part $V_{upper}$ along which the identity defect lies clockwise until it is fused with the condensation defect, resulting in a single defect along $V_{lower}$ that ends topologically at $\Sigma_0$. The bulk field $A_2$ (as well as the defect fields $a$ and $b$) then becomes a defect field living only on $V_{lower}$, and the action of this new defect is given by 
\begin{equation}
\frac{i}{2\pi}\int_{V_{lower}}\left(Nda-db\right)\left(A_{1}-A_{2}\right)
\label{S_folded}
\end{equation}
where the bulk field is now only $A_1$. To put this action in a more familiar form, we can denote $c=Na-b$ and $\widetilde{A}=A_{1}-A_{2}$ and notice that they are both standard $U(1)$ gauge fields (with fluxes $2\pi\mathbb{Z}$) which live only on $V_{lower}$ (alternatively, we can shift $b$ by $Na$ and $A_2$ by $A_1$). That is, we obtain a decoupled trivial $\mathbb{Z}_{1}$ BF theory on $V_{lower}$ with Lagrangian $(i/2\pi)\widetilde{A}dc$. Now, since this defect ends at $\Sigma_0$ topologically, the boundary conditions for the corresponding worldvolume TQFT should be topological, and as discussed in subsection \ref{sec: top construction of local fusion junction 4d} they are given by Dirichlet boundary conditions for both of the fields $c$ and $\widetilde{A}$. This, in turn, yields Eq. \eqref{eq: bj}.

\paragraph{Example: Hom($C_{N\over 2},\mathbb{1}$)} The discussion above straightforwardly applies to the case with a condensation defect with a nontrivial discrete torsion.  For example, for $C_{N\over 2}$ (for even $N$)
\begin{equation}
    S= {iN\over 2\pi}\int_{V_{lower}} \left[ (A_1-A_2)\dr a + {1\over 2}(A_1-A_2) \dr(A_1-A_2) \right] + {i\over 2\pi}\int_{V_{upper}} (A_1-A_2) \dr b
\end{equation}
and the analysis above can be used to find the following boundary conditions without any boundary term:\footnote{The boundary condition for $\mathcal{C}_k$ is given by the expression below after replacing $\frac{N}{2}$ with $k$.}
\begin{equation}
    \text{Hom}\left(C_{N\over 2},\mathbb{1}\right): \quad \left. Na + {N\over 2}(A_1-A_2) - b\,, \quad A_1-A_2\right|_{\Sigma_0} = \text{pure $U(1)$ gauge}\,, \quad S_{bdry}=0\,. \label{eq: bp}
\end{equation}

\subsubsection{Morphism junctions of $U(1)_N\times U(1)_{-N}$} \label{sec: u(1)N theories}
When a TQFT has a topological boundary condition, there exists a morphism between a defect on which this TQFT lives and the identity, obtained by imposing the topological boundary condition at the morphism junction.\footnote{Later, we will also refer to the gapped boundary of $U(1)_N\times U(1)_{-N}$ as a junction $U(1)_N\times U(1)_{-N}\to \bi$.  The $\bi$ here denotes the trivial theory and should not be confused with the identity defect~\eqref{eq: Id} in the Maxwell theory.}  Although a single $U(1)_N$ Chern-Simons theory does not have a topological boundary condition, a $U(1)_N \times U(1)_{-N}$ Chern-Simons theory does.  One obvious choice is identifying the $U(1)_N$ and $U(1)_{-N}$ gauge fields, including their gauge transformations and variations, at the junction.\footnote{This can be thought of as the folding trick---folding the 3-manifold supporting one $U(1)_N$ theory along a 2-dimensional junction and relabeling the $U(1)_N$ field with the opposite parity as a $U(1)_{-N}$ field.}  However, such a boundary condition is not suitable for the $U(1)_N\times U(1)_{-N}$ theory when the two $U(1)$ fields are made of different bulk gauge fields that are independently gauge invariant.  Is there a topological boundary condition that respects both $U(1)$ gauge symmetries?  What is the correct action, including the boundary terms, that is compatible with such a boundary condition?

To answer these questions, we note that the $U(1)_N \times U(1)_{-N}$ theory
\begin{equation}
    S = {iN\over 4\pi}\int_{V}\left(a\dr a- b\dr b \right)
\end{equation}
has two topological boundary conditions for any integer $N$ corresponding to condensing the two Lagrangian algebras\footnote{When $N=2M$ for an odd prime $M$, the given sets are the only Lagrangian algebras.} 
\begin{equation}
    \mathcal{A}_\pm = \left\{l(1,\pm 1)\, |\, l=0,\cdots, N-1\right\} \,.
\end{equation}
The gauge-invariant way of condensing $\ca_\pm$ is to impose
\begin{equation}
    \left. a\pm b\right|_{\partial V} = \text{pure $U(1)$ gauge}
\end{equation}
which are consistent with the bulk on-shell relations which set $a,b$ to be flat $U(1)$ connections with $\mathbb{Z}_N$ holonomies.

To impose these boundary conditions, one has to couple the two $U(1)$ fields by adding a boundary term\footnote{One can add the boundary term with the minus sign instead, but that will turn out to be equivalent using the resulting boundary conditions.} to the action as follows,\footnote{On the other hand, the non-gauge invariant condition $\left.a\pm b\right|_{\partial V}=0$ can be imposed on the bulk action without a boundary term, $S={iN\over 4\pi}\int_{V}\left( a\dr a- b\dr b\right)$.  This can be seen by the gauge transformation and the variation under these boundary conditions, $\Delta S = {iN\over 4\pi}\int_{\partial V} \dr\lambda\wedge (a\pm b)$ and the variation $\left. \delta S\right|_{\partial V}={iN\over 4\pi}\int_{\partial V}(a\pm b)\wedge\delta a$, which are trivial.}
\begin{equation}
\begin{aligned}
S &=  {iN\over 4\pi}\int_{V}\left( a\dr a - b\dr b\right) + {iN\over 4\pi}\int_{\partial V}a\wedge b\\
&= {iN\over 4\pi}\int_{V} (a+ b)\wedge \dr (a- b)\,.
\end{aligned} \label{eq: db}
\end{equation}
Then, using the invariance under gauge transformations and an infinitesimal variation on the boundary,
\begin{equation}
    \begin{aligned}
        \Delta_\lambda S &= {iN\over 4\pi}\int_{\partial V}\dr(-\lambda_a+ \lambda_b) \wedge (a + b)\,, \\
        \delta S|_{\partial V} &= {iN\over 4\pi}\int_{\partial V} (a - b) \wedge \delta(a + b)
    \end{aligned}
\end{equation}
and following the analysis around Eq.~\eqref{eq: da}, the topological boundary conditions that this action can accommodate should satisfy
\begin{equation}
    \left. {N\over 2}(a + b) \right|_{\partial V} = k\dr\varphi\,, \qquad  \left. a - b \right|_{\partial V}= \text{flat $U(1)$ connections} \label{eq: be}
\end{equation}
for some $k|{N\over 2}$.\footnote{This is because under the gauge transformation $a\to a+\dr\lambda$, $a={2k\over N}\dr\phi \to {2k\over N}\dr(\phi +{N\over 2k}\lambda )$ and for $\phi +{N\over 2k}\lambda $ to be a $2\pi$-periodic boson, $k|{N\over 2}$.}  We immediately see that this action is not compatible with odd integer $N$.\footnote{We leave finding an action compatible with the gauge invariant topological boundary conditions for odd $N$ to a future work.}  For even $N$, these boundary conditions always include a condensation of the Lagrangian algebras $\mathcal{A}_\pm$ for $k={N\over 2},1$, given by
\begin{equation}
\begin{aligned}
  \text{Hom($U(1)_N\times U(1)_{-N},\bi$)}: \quad \left. a\pm b\,\right|_{\partial V}=\text{pure $U(1)$ gauge}, \quad S_{bdry} = {iN\over 4\pi}\int_{\partial V}a\wedge b\,.
\end{aligned}
\end{equation}
Using the equations of motion $Na,Nb=$ pure $U(1)$ gauge, this is equivalent to\footnote{More specifically, $
    \left. {N\over 2}(a\pm b)\right|_{\partial V} = \left. {N\over 2}(a\mp b) \pm Nb\, \right|_{\partial V} = \text{pure $U(1)$ gauge} $.} 
\begin{equation}
\begin{aligned}
  \text{Hom($U(1)_N\times U(1)_{-N},\bi$)}: \quad &\left.a\pm b, \quad {N\over 2}(a\mp b) \right|_{\partial V}=\text{pure $U(1)$ gauge}, \\ & S_{bdry} = {iN\over 4\pi}\int_{\partial V}a\wedge b\,.
\end{aligned} \label{eq: u(1)s}
\end{equation}

%On the other hand, 
Alternatively, one can introduce some scalar fields at the junction that cancel the gauge transformations of the $U(1)$ fields $a$ and $b$ and obtain the same boundary conditions by integrating these scalars out, as explained in Section 6.2.2 of \cite{Roumpedakis:2022aik}.  However, instead of keeping track of these scalars, we will impose Eq.~\eqref{eq: u(1)s} at the Hom$(U(1)_N\times U(1)_{-N},\bi)$ junctions for simplicity.

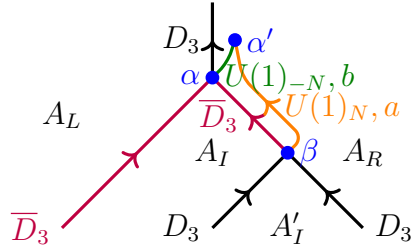
\begin{figure}[H]
	\centering
	\begin{tikzpicture}
		\begin{scope}[very thick,decoration={
				markings,
				mark=at position 0.5 with {\arrow{>}}}
			]
                \draw[postaction={decorate}] (0,0)--(0,1) node[midway,left] () {$D_3$};
			\draw[postaction={decorate},purple] (-2,-2)--(0,0) node[at start,left] (l) {$\overline{D}_3$};
                
                \draw[postaction={decorate},purple] (1,-1)--(0,0) node[midway, left] () {$\overline{D}_3$};

                \draw[postaction={decorate},rounded corners,orange] (1,-1)--(1.2,-0.8)--(0.4,0) node[midway,right] () {$U(1)_N,a$}--(0.3,0.5);
                \draw[rounded corners,Green] (0,0)--(0.2,0.2)node[at start,right] () {$U(1)_{-N},b$}--(0.3,0.5);
                % \draw[Purple,dotted] (0.3,0.5)--(0.3,1) node[at end,right] () {$\bi$};

                \draw[postaction={decorate}] (0,-2)--(1,-1) node[at start,left] () {$D_3$};
                \draw[postaction={decorate}] (2,-2)--(1,-1) node[at start,right] () {$D_3$};

			\node at (-2,-0.5) () {$A_L$};
			\node at (2,-1) () {$A_R$};
			\node at (0,-1) () {$A_I$};
            \node at (1,-2) () {$A_I'$};
	   \filldraw[blue] (1,-1) circle (2pt) node[right,blue] () {$\beta$};
       \filldraw[blue] (0,0) circle (2pt) node[left,blue] () {$\alpha$};
       \filldraw[blue] (0.3,0.5) circle (2pt) node[right,blue] () {$\alpha'$};
		\end{scope}
	\end{tikzpicture}
	\caption{The $\beta$: Hom(${D}_3\times {D}_3,\overline{D}_3 U(1)_N$), $\alpha$: Hom($\overline{D}_3\times \overline{D}_3,D_3U(1)_{-N}$) and $\alpha'$: Hom($U(1)_N\times U(1)_{-N},\bi$) junctions.}
	\label{fig: two U(1)N junctions}
\end{figure}

\paragraph{Example: $\overline{D}_3\times(D_3\times D_3) = {D}_3~U(1)_N\times U(1)_{-N} \to {D}_3$} 
Consider two consecutive fusions, $\overline{D}_3\times(D_3\times D_3)$ as shown in Figure~\ref{fig: two U(1)N junctions}.  

First, we construct the local fusion junction $D_3\times D_3$ by fusing the two parallel defects along the upper part above the 2d local fusion junction $\beta$.  This fused upper part is described by the action
\begin{equation}
    \begin{aligned}
\text{Hom}(D_3\times D_3,\overline{D}_3U(1)_N):\quad         S &=  \frac{iN}{2\pi} \int^\infty_\beta \left(A_I \dr A_I' + \frac{1}{2}A_I \dr A_I + A_I'\dr A_R + \frac{1}{2}A_I'\dr A_I' \right) \\
        &=-\frac{iN}{2\pi}\int^\infty_\beta \left(A_I \dr A_R + \frac{1}{2}A_R \dr A_R \right) \quad {\color{blue}(\leftarrow \, \overline{D}_3)}\\
		&\quad +  \frac{iN}{4\pi}\int^\infty_\beta (A_I + A_I'+ A_R)\dr (A_I + A_I'+ A_R) \quad {\color{blue}(\leftarrow \, U(1)_N)}\\
		&\quad + \frac{iN}{4\pi}\int_{\Sigma_\beta}  \left[ A_I \wedge A_R + (A_I-A_R)\wedge A_I' \right] \quad {\color{blue}(\leftarrow\, S_{bdry})}\\
    \end{aligned} \label{eq: bf}
\end{equation}
where we denote by the integral from $\beta$ to $\infty$ the 3d manifold above the 2d junction $\beta$, and by $\Sigma_\beta$ the 2d manifold corresponding to the junction $\beta$.  The fusion outcome has the expected $\overline{D}_3\times U(1)_N$ along with a boundary term at $\Sigma_\beta$.  Since $A_I'$ is supported only on the defect above $\beta$, the $U(1)_N$ is decoupled from the bulk (this can be seen by shifting $A_I'\to A_I'-A_I-A_R$).

Next, we take another copy of $\overline{D}_3$ and fuse it with the $\overline{D}_3$ we obtained in the previous fusion in the segment above the $\alpha$ junction.  The fusion outcome above $\alpha$ is described by the action
\begin{equation}
    \begin{aligned}
\text{Hom}(\overline{D}_3\times \overline{D}_3,U(1)_{-N}D_3):\quad         S =& -\frac{iN}{2\pi} \int^\infty_\alpha \left(A_L \dr A_I + \frac{1}{2}A_I \dr A_I + A_I \dr A_R + \frac{1}{2}A_R \dr A_R \right) \\
        &= \frac{iN}{2\pi}\int^\infty_\alpha \left(A_L \dr A_R + \frac{1}{2}A_L \dr A_L\right) \qquad {\color{blue}(\leftarrow D_3)} \\
		&\quad - \frac{iN}{4\pi}\int^\infty_\alpha (A_L+A_I+A_R)\dr (A_L+A_I+A_R)   \quad {\color{blue}(\leftarrow U(1)_{-N})}\\
		&\quad -\frac{iN}{4\pi}\int_{\Sigma_\alpha} \left[ A_L\wedge A_R + (A_L-A_R)\wedge A_I \right]\,. \quad {\color{blue}(\leftarrow S_{bdry})}
    \end{aligned} \label{eq: bff}
\end{equation}

Since they are decoupled from the bulk, the $U(1)_N\times U(1)_{-N}$ factors above $\alpha$ commute with all other defects, and so it is Witt equivalent to the identity defect on a 3-manifold.  Hence, there is a morphism junction between them where the two $U(1)$ theories end at $\alpha'$ in Figure~\ref{fig: two U(1)N junctions} with a topological boundary condition.  As explained above, this requires coupling the two theories at the morphism junction by adding a boundary term and choosing a boundary condition out of two (or more) options:
\begin{equation}
    \begin{aligned}
        \text{Hom}\left[U(1)_N\times U(1)_{-N},\mathbb{1}\right]: \quad & S_{bdry}(\alpha') = {iN\over 4\pi}\int_{\Sigma_{\alpha'}} a\wedge b,\\
        &\text{B.c.}\quad \begin{cases}
            (1) & \left. {N \over 2}(a+b),\quad a-b\right|_{\Sigma_{\alpha'}}=\text{pure $U(1)$ gauge}\\
        (2) & \left. a+b,\quad {N\over 2}(a-b)\right|_{\Sigma_{\alpha'}}=\text{pure $U(1)$ gauge}\\
        \end{cases}         
    \end{aligned} \label{eq: bi}
\end{equation}
where $a$ and $b$ denote the $U(1)_N$ and $U(1)_{-N}$ gauge fields which, in our case, are
\begin{equation}
    a=A_I+A_I'+A_R\,, \quad b=A_L+A_I+A_R\,.
\end{equation}
The coefficients ${N\over 2}$ in the boundary conditions can be replaced with $N$ if we further specify $Na, Nb = \text{pure $U(1)$ gauge}$.

\subsection{Associator 1-morphisms} 
\subsubsection{The method} \label{sec: 4d method}
In four dimensions, the associator 1-morphism between three codimension-1 defects $(a,b,c)$ is described by a 2-dimensional TQFT living at the interface between the two different fusions $(a\times b)\times c$ and $a\times (b\times c)$.  The first three steps in the method used in Section~\ref{sec: F-symbol method} to compute these associators are still valid in higher-dimensional cases.  In particular, the global fusion outcome (Step 3) may contain some decoupled 3d TQFT, either a $\bz_N$ gauge theory or a trivial TQFT.  However, since the junctions (points in Figure~\ref{fig: associator}) are now 2-manifolds, the last two steps need a slight modification. 
\begin{enumerate}
    \item[Step 4] \textbf{Interval reduction}: Using the boundary conditions, obtain the 2d theory from the decoupled TQFT in Step~3 after reducing the 3d interval $[\gamma,\beta]$ to the 2d manifold $\Sigma_{\beta\gamma}$.

    \item[Step 5] \textbf{Associator}:  Evaluate all the boundary actions at the combined junction using the boundary conditions in Step~2 and combine with the 2d theory after the reduction to obtain the associator 1-morphism as
    \begin{equation}
        F^d_{a,b,c} = (\text{2d theory after reduction}) \otimes (\text{2d theory at the combined junctions})\,. \label{eq: bn}
    \end{equation}
\end{enumerate}

In Step~4, to get the corresponding 2d theory after reducing the interval, we use a prescription based on the spectrum of the local and line operators on the boundary, as explained in Ref.~\cite{Bhardwaj:2023idu}. The fate of a topological Wilson line $e^{i\oint n_i A_i}$ in 3d after the interval reduction is determined by the boundary conditions of $n_i A_i$ as listed in Table~\ref{tab:spectrum of 2d theory}.
\begin{table}[H]
    \centering
    \begin{tabular}{c|c}
    \hline
    Boundary conditions of $n_i A_i$ & Operator from $e^{i\oint n_i A_i}$ after reduction \\
    \hline
     DBC at both ends    &  Untwisted local operator\\
     NBC at both ends    &  Line operator\\
     DBC at one and NBC at the other end & Twisted local operator\\
     \hline
    \end{tabular}
    \caption{Spectrum of the 2d theory based on the boundary conditions. We say $n_i A_i$ has Dirichlet boundary condition (DBC) if $n_i A_i = \text{pure $U(1)$ gauge}$ on the boundary, and $n_i A_i$ has Neumann boundary condition otherwise. Notice that two line operators after reduction arising from $e^{i\oint n_i A_i}$ and $e^{i\oint n_i' A_i}$ are identified if $(n_i - n_i')A_i$ has DBC on at least one boundary.}
    \label{tab:spectrum of 2d theory}
\end{table}

For the reduction of an Abelian 3d TQFT, if the 2d theory has untwisted local operators charged under line operators, it is a 2d $G$-gauge theory ($G$-SSB phases) where $G$ is the group formed by untwisted local operators. In addition, the theory also contains additional 0-form symmetries (besides $G$) which act trivially on local operators but may act non-trivially on twisted local operators via the Lasso action. Thus, strictly speaking, the reduced theory is a 2d $G$-gauge theory enriched by the additional trivially acting symmetry. In the special case where there are no untwisted local operators and $G$ is trivial, this is nothing but an SPT phase of those additional symmetries. In the following, we will restrict ourselves to describe only the untwisted local operators and the line operators, and only keep track of the $G$-gauge theory part. We will then refer the reduced theory as a trivial theory when there are no untwisted local operators.

\subsubsection{Interval reduction of $(\mathcal{Z}_N)_N$ (even $N$)}\label{sec:ZNNred}
In some cases, the global fusion (Step 3) of three triality defects contains a $\mathbb{Z}_N$ gauge theory with twist $N$ (denoted as $(\mathcal{Z}_N)_N$),
\begin{equation}\label{eq: ZNN}
\begin{aligned}
    S &= \frac{i}{2\pi} \int_V N a' db + \frac{N}{2}a'da' \\
    &=\frac{iN}{4\pi} \int_V (a'+b)d(a'+b) - bdb + \frac{iN}{4\pi} \int_{\partial V} b\wedge a' 
\end{aligned}
\end{equation}
which is equivalent to the $U(1)_N \times U(1)_{-N}$ theory together with the required boundary term, upon redefining $a = a'+b$. Thus, the boundary conditions of $U(1)_N \times U(1)_{-N}$ will apply in this case.

Let us briefly describe the interval reduction of this theory when placed on $V=[\gamma,\beta]\times M_2$ for the two boundary conditions $\mathcal{A}_\pm$ discussed in Section \ref{sec: u(1)N theories}, with the result summarized in Table~\ref{tab: ZNN reduction}. Apparently, choosing the same boundary conditions on both ends leads to 2d $\mathbb{Z}_N$ gauge theory after reduction. On the other hand, choosing $\mathcal{A}_+$ on one end and $\mathcal{A}_-$ on the other leads to 2d $\mathbb{Z}_2$ gauge theory after reduction. The untwisted local operators have the $\mathbb{Z}_2$ fusion rule and the generator comes from the line $e^{i\oint \frac{N}{2}(a+b)}$. The line operators also have the $\mathbb{Z}_2$ fusion rule and are generated by the reduction of $e^{i\oint a}$, and act non-trivially on the untwisted local operators. 

\begin{table}[H]
    \centering
    \begin{tabular}{c|c|c|c|c}
    \hline
      B.C. at $\gamma$  & B.C. at $\beta$ & local operators & line operators & 2d theory\\
    \hline
    $\mathcal{A}_+$ & $\mathcal{A}_+$  & $e^{ik\oint (a+b)}$ ($k\in \mathbb{Z}_N$) & $e^{ik\oint a}$ ($k\in \mathbb{Z}_N$) & \multirow{2}{*}{$\mathbb{Z}_N$ gauge theory} \\
    \cline{1-4}
    $\mathcal{A}_-$ & $\mathcal{A}_-$  & $e^{ik\oint (a-b)}$ ($k\in \mathbb{Z}_N$) & $e^{ik\oint a}$ ($k\in \mathbb{Z}_N$) & \\ 
    \hline
    $\mathcal{A}_+$ & $\mathcal{A}_-$ & \multirow{2}{*}{$e^{i\frac{Nk}{2}\oint (a+b)}$ ($k\in \mathbb{Z}_2$)} & \multirow{2}{*}{$e^{ik\oint a}$ ($k\in \mathbb{Z}_2$)} & \multirow{2}{*}{$\mathbb{Z}_2$ gauge theory}\\
    \cline{1-2}
    $\mathcal{A}_-$ & $\mathcal{A}_+$ & & \\
    \hline
    \end{tabular}
    \caption{The 2d theories obtained after the interval reduction along $[\gamma,\beta]$ for the theory \eqref{eq: ZNN} with different combinations of the boundary conditions $\mathcal{A}_\pm$ at the left boundary $\alpha$ and the right boundary $\beta$.}
    \label{tab: ZNN reduction}
\end{table}

\subsubsection{Example: Triality defect associator $\left[F^{D_3}_{{D_3},D_3,\overline{D}_3}\right]_{\mathbb{1},U(1)_{N}\overline{D}_3}$}  
To illustrate the method discussed above, let us compute a particular triality defect associator, $[F^{D_3}_{{D_3},D_3,\overline{D}_3}]_{\mathbb{1},U(1)_N\overline{D}_3}$, by collapsing the diamond region in Figure~\ref{fig: FD3-D3D3D3b} to a 2d interface between the two fusions 
\begin{equation}
    \begin{aligned}
    \text{Upper part}:\quad & {D}_3 \times (D_3\times \overline D_3) = {D}_3\times C_{N\over 2} \to  D_3 \times \bi  = D_3\,, \\
    \text{Lower part}: \quad &({D}_3\times D_3)\times \overline D_3 =U(1)_N~\overline D_3\times \overline D_3 = U(1)_N\times U(1)_{-N}~D_3 \to D_3\,,
    \end{aligned}
\end{equation}
where we have added the morphism junctions $\beta': C_{N\over 2}\to \mathbb{1}$ and $\delta': U(1)_N\times U(1)_{-N}\to \mathbb{1}$. For simplicity, we restrict here to $N=2\times\text{odd prime}$.\footnote{This is in order to have only two topological boundary conditions for the $U(1)_N\times U(1)_{-N}$ theory compatible with the Lagrangian.}
\begin{figure}[H]
	\centering
	\begin{tikzpicture}
		\begin{scope}[shift={($(ff.east)+(2,0)$)},very thick,decoration={
				markings,
				mark=at position 0.5 with {\arrow{>}}}
			]
        % a(bc)
            \draw[postaction={decorate}] (0,1.5)--(0,3) node[midway,left] () {$D_3$}; % a(bc)

        % bc
            \draw[postaction={decorate},dotted,Purple] (0.6,0.9)--(0,1.5) node[at end,right] () {$~\bi$}; % bc 
            \draw[] (0.75,0.75)--(0.6,0.9) node[at end,right] () {$~C_{N\over 2}$} ; % bc

        % a
            \draw[postaction={decorate}] (-0.75,-0.75)--(-1.5,0)--(0,1.5) node[midway,left] () {${D}_3~$}; % a

        % b           
            \draw[postaction={decorate}] (-0.75,-0.75)--(0.75,0.75) node[midway,left] () {$ D_3~$}; % b

        % c
        \draw[postaction={decorate},purple] (0,-1.5)--(1.5,0) node[midway,right] () {$~\overline{D}_3$}--(0.75,0.75); % c

        % ab            
            \draw[postaction={decorate},purple] (0,-1.5)--(-0.75,-0.75) node[near end, right] () {$\overline{D}_3$}; %ab
            \draw[postaction={decorate},Green,rounded corners] (-0.2,-2.25)--(-0.25,-1.65)--(-0.9,-0.9) node[at start,left] () {$U(1)_N,a_1$}--(-0.75,-0.75);

        % (ab)c    
            \draw[postaction={decorate},rounded corners,orange] (-0.2,-2.25)--(-0.2,-1.65) node[near start,left] () {$U(1)_{-N},a_2$}--(0,-1.5); % (ab)c
            % \draw[postaction={decorate},dotted,Purple] (-0.2,-3)--(-0.2,-2.25) node[at start,left] () {$\bi$};
            
            \draw[postaction={decorate}] (0,-3)--(0,-1.5) node[near start,right] () {$D_3$}; % (ab)c
            			
	\node[blue] at (-1.5,-1) () {$A_L$};
        \node[blue] at (1,2) () {$A_R$};
        \node[blue] at (0.5,-0.25) () {$A_I'$};
        \node[blue] at (-0.15,0.6) () {$A_I$};

            \filldraw[blue] (0,1.5) circle (2pt) node[left,blue] () {$\alpha$};
            \filldraw[Purple] (0.6,0.9) circle (2pt) node[left] () {$\beta'$};
            \filldraw[blue] (0.75,0.75) circle (2pt) node[below,blue] () {$\beta$};
            \filldraw[blue] (-0.75,-0.75) circle (2pt) node[above,blue] () {$\gamma$};
            \filldraw[blue] (0,-1.5) circle (2pt) node[right,blue] () {$\delta$};
            \filldraw[Purple] (-0.2,-2.25) circle (2pt) node[below] () {$\delta'$};
    \node at (3,0) (fff) {};
    \end{scope}
	\end{tikzpicture}
	\caption{The diagram for $\left[F^{D_3}_{{D_3},D_3,\overline{D}_3}\right]_{\mathbb{1},U(1)_N\overline{D}_3}$. $a_1$,$a_2$ are the gauge fields of the $U(1)_N\times U(1)_{-N}$ theory.}
	\label{fig: FD3-D3D3D3b}
\end{figure}
As we did in the 2d case, we first collect the boundary conditions and boundary terms at each junction $\alpha,\beta',\beta,\gamma,\delta,\delta'$ from our junction analyses.  Specifically, Eq.~\eqref{eq: bi} for $\delta': U(1)_N\times U(1)_{-N} \to \mathbb{1}$, Eq.~\eqref{eq: bff} for $\delta: D_3~U(1)_{-N} \to \overline{D}_3\times \overline{D}_3$, Eq.~\eqref{eq: bf} for $\gamma: U(1)_N~\overline{D}_3\to D_3\times D_3 $, Eq.~\eqref{eq: we} for $\beta: D_3\times \overline{D}_3\to C_{N\over 2}$, Eq.~\eqref{eq: bp} for $\beta': {C_{N\over 2}}\to  \mathbb{1}$ and Eq.~\eqref{eq: bo} for $\alpha: D_3\times \mathbb{1}\to D_3$. Then, we shrink the interval between the upper junctions $\alpha,\beta'$ and $\beta$ and combine the boundary conditions and boundary terms there. We repeat analogously for the lower junctions $\gamma,\delta$ and $\delta'$. Putting them all together,
\begin{equation}
    \begin{aligned}
        \alpha,\beta,\beta': \quad & N(A_L+A_I'+A_R), \quad A_I-A_R = \text{pure $U(1)$ gauge}\,,\\
         &S(\alpha) = {iN\over 2\pi}\int_{\Sigma_\alpha}A_L\wedge (A_I-A_R), \quad S(\beta') = {iN\over 2\pi}\int_{\Sigma_\alpha}\left(A_R\wedge A_I' - {1\over 2}A_I \wedge A_R\right), \\
        \gamma,\delta,\delta': \quad & a_1=A_L+A_I'+A_I, \quad a_2=A_L+A_I'+A_R,\\
        & \begin{cases}
            (1) & \underbrace{A_I-A_R}_{=a_1-a_2}, \quad N(A_L+A_I'+A_R) = \text{pure $U(1)$ gauge}\\
        \text{or }(2) & \underbrace{2(A_L+A_I'+A_R)+A_I-A_R}_{=a_1+a_2} , \quad {N\over 2}(A_I-A_R) = \text{pure $U(1)$ gauge} 
        \end{cases}\\
        &S(\gamma) = -{iN\over 4\pi}\int_{\Sigma_\gamma}\left[ A_L \wedge A_I' + (A_L-A_I')\wedge A_I \right], \\
        & S(\delta) = {iN\over 4\pi}\int_{\Sigma_\delta} \left( A_L \wedge A_R + A_L\wedge A_I' +A_I'\wedge A_R \right),\\
        & S(\delta') = {iN\over 4\pi}\int_{\Sigma_{\delta'}} \left(A_L\wedge A_R + A_I'\wedge A_R + A_I\wedge A_L + A_I\wedge A_I'+A_I\wedge A_R \right)
    \end{aligned} \label{eq: bx}
\end{equation}
where $S(\alpha)$ denotes the boundary term at the junction $\alpha$ (and similarly for the other junctions), $a_1$ and $a_2$ are the gauge fields of the $U(1)_N\times U(1)_{-N}$ theory and (1) and (2) are the two possible boundary conditions at the junction $\delta':~U(1)_N\times U(1)_{-N} \to \mathbb{1}$.

Next, we fuse the three defects ${D}_3\times D_3\times \overline{D}_3$ in the middle (along the interval $[\gamma,\beta]$) by summing their actions,
\begin{equation}
    \begin{aligned}
		S &= {iN\over 2\pi}\int^\beta_\gamma \left(A_L \dr A_I + {1\over 2}A_L\dr A_L +A_I \dr A_I' +{1\over 2}A_I \dr A_I - A_I' \dr A_R -{1\over 2}A_R \dr A_R\right) \\
        &= {iN\over 2\pi}\int^\beta_\gamma \left(A_L \dr A_R + {1\over 2}A_L\dr A_L\right) \quad {\color{blue}(\leftarrow D_3)}\\
        &\quad + {iN\over 2\pi}\int^\beta_\gamma \left[ (A_L+A_I'+A_R)\dr (A_I-A_R) + {1\over 2} (A_I-A_R)\dr  (A_I-A_R) \right] \quad {\color{blue}(\leftarrow (\mathcal{Z}_N)_N)}\\
        & \quad + {iN\over 2\pi}\int_{\Sigma_\beta - \Sigma_\gamma} \left( A_I'\wedge A_I +{1\over 2}A_R\wedge A_I \right) \quad {\color{blue}(\leftarrow S_{bdry;global})}
	\end{aligned} \label{eq: bl1}
\end{equation}
where the second line is the action for $D_3$ as expected from the fusion algebra, the third line is a decoupled $\mathbb{Z}_N$ gauge theory with twist $N$ (denoted by $(\mathcal{Z}_N)_N$) and the last line is the boundary term.

At this stage, we are left with the configuration
\begin{equation}
\begin{tikzpicture}[baseline={([yshift=-.5ex]current bounding box.center)},vertex/.style={anchor=base,circle,fill=black!25,minimum size=18pt,inner sep=2pt},scale=1]
\begin{scope}[shift={($(ff.east)+(2,0)$)},very thick,decoration={
				markings,
				mark=at position 0.5 with {\arrow{>}}}
			]
    \draw[postaction={decorate}] (0,-2) -- (0,-1);
    \draw[postaction={decorate}] (0,-1) -- (0,+1);
    \draw[postaction={decorate}] (0,+1) -- (0,+2);
    \filldraw[black] (0,-1) circle (2pt);
    \filldraw[black] (0,+1) circle (2pt);
    \node[below] at (0,-2) {\footnotesize $S_{D_3}$};
    \node[above] at (0,+2) {\footnotesize $S_{D_3}$};
    \node[right] at (0,0) {\footnotesize $S_{D_3} + S_{(\mathcal{Z}_N)_N}$};
    \node[right] at (0,-1) {\footnotesize $\gamma$};
    \node[right] at (0,+1) {\footnotesize $\beta$};
    \end{scope}
\end{tikzpicture} ~.
\end{equation}
The total boundary term at the (combined) upper junction $\beta$ is
\begin{equation}
\begin{aligned}
    S_{bdy,\beta} &= S(\alpha)+S(\beta')+\left. S_{global}\right|_{\alpha\beta} \\
    &= {iN\over 2\pi}\int_{\Sigma_\alpha}(A_L+A_I'+A_R) \wedge (A_I-A_R)\in2\pi i\mathbb{Z}
\end{aligned}
\label{S_bd_beta}
\end{equation}
which is trivial by the boundary conditions~\eqref{eq: bx}. Similarly, the total boundary term at the (combined) lower junction $\gamma$ is
\begin{equation}
\begin{aligned}
    S_{bdy,\gamma} &= S(\gamma)+S(\delta) + \left. S_{global}\right|_{\gamma \delta}+S(\delta') \\
    &= {iN\over 2\pi}\int_{\Sigma_{\gamma\delta}} (A_L+A_I'+A_I)\wedge (A_I-A_R)\in2\pi i\mathbb{Z}\,,
\end{aligned}
\label{S_bd_gamma}
\end{equation}
which is also trivial by the boundary conditions.

Therefore, we are left with a decoupled $(\mathcal{Z}_N)_N$ theory supported on the interval $[\gamma,\beta]$ in Eq.~\eqref{eq: bl1}.  By comparing the action given in \eqref{eq: bl1} with \eqref{eq: ZNN}, we identify that the boundary condition at $\gamma$ corresponds to $\mathcal{A}_-$, while the two boundary conditions (1) and (2) at $\beta$ correspond to $\mathcal{A}_-$ and $\mathcal{A}_+$, respectively.

More concretely, for the boundary condition (1) the upper and lower junctions have the same boundary conditions for the gauge fields $A_L+A_I'+A_R$ and $A_I-A_R$ in the $(\mathcal{Z}_N)_N$ action in \eqref{eq: bl1}, which are written in Table~\ref{tab:1}. Following the discussion in Section~\ref{sec: 4d method}, reducing the interval $[\gamma,\beta]$ results in a 2d $\mathbb{Z}_N$ gauge theory.
\begin{table}[H]
    \centering
    \begin{tabular}{c|cc}
    \hline
         & $A_L+A_I'+A_R$ & $A_I-A_R$  \\
         \hline
         $\alpha,\beta',\beta$ & $\mathbb{Z}_N$ & pure $U(1)$ gauge\\
         $\gamma,\delta',\delta \, (1)$ & $\mathbb{Z}_N$ & pure $U(1)$ gauge\\
         \hline
    \end{tabular}
    \caption{Boundary conditions for the gauge fields in Eq. \eqref{eq: bl1} for the case (1) of Eq.~\eqref{eq: bx}. Here, by $\mathbb{Z}_N$ we mean that the corresponding gauge field $a$ has the boundary condition $Na = \text{pure $U(1)$ gauge}$ and that $e^{i\oint a}$ is a $\mathbb{Z}_N$ line.}
    \label{tab:1}
\end{table}

Under the boundary condition (2), on the other hand, the gauge fields have different boundary conditions at the upper and lower junctions, as written in Table~\ref{tab:2}. In this case, we only find $\mathbb{Z}_2$ untwisted local operators arising from the Wilson lines $e^{i \frac{Nk}{2} \oint (a_1 + a_2)}$ ($k \in \mathbb{Z}_2$). Hence, reducing the interval $[\gamma,\beta]$ results in a $\bz_2$ gauge theory. 
\begin{table}[H]
    \centering
    \begin{tabular}{c|cc}
    \hline
         & $2(A_L+A_I'+A_R)+(A_I-A_R)$ & $A_I-A_R$  \\
         \hline
         $\alpha,\beta',\beta$ & $\mathbb{Z}_{N\over 2}$ & pure $U(1)$ gauge\\
         $\gamma,\delta',\delta \, (2)$ & pure $U(1)$ gauge & $\mathbb{Z}_{N\over 2}$\\
         \hline
    \end{tabular}
    \caption{Boundary conditions for the gauge fields in \eqref{eq: bl1} for the case (2) of Eq.~\eqref{eq: bx}.}
    % Here $\mathbb{Z}_{N\over 2}$ denotes an unconstrained $\mathbb{Z}_{N\over 2}$ gauge field on the boundary.
    \label{tab:2}
\end{table}

% \sk{Need an argument something like "Hence, the only non-trivial untwisted local operator in the dimensionally reduced 2d theory is from the Wilson line of charge $(N/2,N/2)$.} 

In summary,
\begin{equation}
    (\mathcal{Z}_N)_N \quad \overset{\text{reducing interval}}{\longrightarrow} \quad \begin{cases}
        (1) : &\mathbb{Z}_N \text{ gauge theory}\\
        (2): & \text{$\bz_2$ gauge theory}\,\,\,.
    \end{cases}
\label{ZN_reduction}
\end{equation}

Overall, we obtained that the boundary terms do not contribute to this associator 1-morphism (see Eqs. \eqref{S_bd_beta} and \eqref{S_bd_gamma}), and hence it is given only by the 2d theory obtained by reducing the $(\mathcal{Z}_N)_N$ theory on the interval $[\gamma,\beta]$, which we found in \eqref{ZN_reduction}. That is, 
\begin{equation}
\left[ F^{D_3}_{D_3,D_3,\overline{D}_3} \right]_{\mathbb{1},U(1)_N\overline{D}_3} = \mathbb{Z}_N \text{ gauge theory} \quad \text{or}\quad \text{$\bz_2$ gauge theory} 
\label{F_L}
\end{equation}
where the two possibilities correspond to the two options for the boundary conditions at the junction $U(1)_N\times U(1)_{-N}\to \bi$.\footnote{There are only two possible topological boundary conditions for the $U(1)_N\times U(1)_{-N}$ theory since we have restricted here to $N=2\times(\textrm{odd prime})$.}   

\paragraph{Other associators} We compute all the other associators involving the triality defect $D_3$ and its orientation reversal $\overline{D}_3$ in Appendix~\ref{app: triality associators}.  The results are listed in Table~\ref{tab: triality associators}.
\begin{table}[H]
    \centering
    \renewcommand{\arraystretch}{1.5}
    \begin{tabular}{c|c}
    \hline
         & $F$-symbol associator  \\
         \hline
       $D_3\times D_3\times D_3$  & $\mathbb{Z}_N$ condensation defect \\
       \hline
       $D_3\times \overline{D}_3\times D_3$ & Trivial theory\\
       \hline
       $D_3\times D_3\times \overline{D}_3$ & \multirow{2}{*}{$\mathbb{Z}_N$ gauge theory or $\bz_2$ gauge theory}\\
       $\overline{D}_3 \times D_3\times D_3$ & \\
       \hline
    \end{tabular}
    \caption{Associators corresponding to various combinations of the $D_3$ and $\overline{D}_3$ defects, for $N=2\times(\textrm{odd prime})$ and with trivial Witt elements (i.e. the identity defect) selected at the fusion junctions. The two different possibilities for the associator of the last two combinations correspond to the two types of topological boundary conditions of the $U(1)_N\times U(1)_{-N}$ theory that appears in the fusion process.  The other possible four associators are related to those in the table by orientation reversal. }
    \label{tab: triality associators}
\end{table}
For general even values of $N$, we expect to have more topological boundary conditions at the morphism junction $U(1)_N\times U(1)_{-N} \to \bi$, corresponding to condensing nonanomalous $\bz_{2k}$ lines for $k|{N\over 2}$, which results in $\bz_{2k}$ gauge theories for the $F$-symbol in the last row of Table~\ref{tab: triality associators}.

In the next section, we introduce a group-theoretical construction for the triality defects for certain odd values of $N$ and compute the above associators, which agree with the ones in Table~\ref{tab: triality associators} except for the $\bz_2$ gauge theory in the last column which becomes the trivial theory.\footnote{The $\bz_2$ gauge theory in these associators results from the dimensional reduction of the $\bz_N$ gauge theory with twist $N$, $(\mathcal{Z}_N)_N$, on the interval.  Due to the non-trivial twist $N$, this theory has an anomaly for even $N$, and hence the 2d theory obtained after the dimensional reduction cannot be trivial.  On the other hand, it is not anomalous for odd $N$ and indeed $(\mathcal{Z}_N)_N\simeq (\mathcal{Z}_N)_0$ for such $N$, which allows for a trivial TQFT after the dimensional reduction.}

Lastly, in Appendix~\ref{app: duality}, we also provide an example for the computation of a duality defect associator, $[F_{D_2,\overline{D}_2,D_2}^{D_2}]_{\mathbb{1},\mathbb{1}}$, using the Lagrangian method and the group-theoretical approach and list the results for the other duality associators, which agree between the two methods.

%-----------------new section

\section{Triality defects in Maxwell theory using group theoretical construction}
\label{Sec_Group}
In this section, we describe a complementary computation of the associativity map for triality defects that admit certain group-theoretical constructions for odd $N$. Recall that the fusion rules are
\begin{equation}
    D_{3}\times D_{3}=SU(N)_{-1}\overline{D}_{3}\;\,,\;\,\overline{D}_{3}\times\overline{D}_{3}=SU(N)_{1}D_{3}\;\,,\;\,\overline{D}_{3}\times D_{3}=D_{3}\times\overline{D}_{3}=C_{0} ~.
\end{equation}
Here, CS theory $SU(N)_{\pm 1}$ is a bosonic TQFT containing $\mathbb{Z}_N$ Abelian anyons. Denoting the generating line as $a$, the topological spin $h(a^k)$ of $a^k$ and the braiding phase $B(a^k,a^{k'})$ are \footnote{This can be directly computed using \cite{DiFrancesco:1997nk}, or use the fact that $SU(N)_{1}$ is the minimal TQFT $\mathcal{A}^{N,N-1}$ \cite{Hsin:2018vcg,Choi:2022jqy} and $\frac{k(N-k)}{2N} = \frac{N-1}{2N}k^2 \mod 1$.}
\begin{equation}\label{eq:SU(N)m1}
    h(a^k) = \pm \frac{k(N-k)}{2N} \mod 1 ~, \quad B(a^k,a^{k'}) = e^{\mp\frac{2\pi i}{N}k k'} ~, \quad \text{for} \quad SU(N)_{\pm 1} ~.
\end{equation}
In the case of the spin theory, there is an additional transparent fermion $\psi$. $SU(N)_{-1}\boxtimes \mathbb{Z}_2^\psi$ now contains $\mathbb{Z}_{2N} = \mathbb{Z}_N^a \times \mathbb{Z}_2^\psi$ Abelian anyons (as $N$ is odd), and is generated by the anyon $b = a \psi$. The topological spin and the braiding phases are
\begin{equation}
    h(b^k) = \frac{k^2}{2N} ~, \quad B(b^k,b^{k'}) = e^{\frac{2\pi i}{N} kk'} ~,
\end{equation}
which matches the spectrum of the $U(1)_N$ CS theory. This relation is known as the level-rank duality \cite{Hsin:2016blu}. For the following, we will assume that we work with bosonic theory for simplicity; although the result holds for the spin theory as well since the additional fermion does not have any essential effect due to its transparent nature.

\subsection{Group-theoretical construction of triality defects}
As first pointed out in \cite{Kaidi:2021xfk}, a non-invertible duality defect in 4d theory can be constructed from gauging a certain 1-form symmetry in a 2-group symmetry with a certain mixed anomaly. In \cite{Choi:2022zal}, a group-theoretical construction for certain duality and triality defects is provided. Let us first explain in detail the triality defects. Consider a 4d QFT $\mathcal{X}$ with $\mathbb{Z}_N$ (where $N$ is odd) 1-form symmetry, and its partition function is invariant under the following order-$3$ twisted gauging
\begin{equation}
    Z_{\mathcal{X}}[A^{(2)}] = \sum_{a^{(2)} \in H^2(X_4,\mathbb{Z}_N)} Z_{\mathcal{X}}[a^{(2)}] \exp\left(\frac{2\pi i}{N} \frac{N+1}{2} \int a^{(2)} \cup a^{(2)} + \frac{2\pi i}{N} \int a^{(2)} \cup A^{(2)}  \right) ~.
\end{equation}
One can define a new theory $\widetilde{\mathcal{X}}$ from $\mathcal{X}$ by twisted gauging the 1-form symmetry
\begin{equation}
    Z_{\widetilde{\mathcal{X}}}[A^{(2)}] = \sum_{a^{(2)} \in H^2(X_4,\mathbb{Z}_N)} Z_{\mathcal{X}}[a^{(2)}] \exp\left(\frac{2\pi i}{N} p' \int a^{(2)} \cup a^{(2)} + \frac{2\pi i}{N}\int a^{(2)}\cup A^{(2)} \right) ~.
\end{equation}
When the odd $N$ is such that the following equation of $p'$ admits solutions
\begin{equation}\label{eq:tri_pp_eq}
    (2p')^2 - 2p' + 1 = 0 \mod N ~,
\end{equation}
the partition function of $\widetilde{\mathcal{X}}$ satisfies
\begin{equation}\label{eq:tpi}
    Z_{\widetilde{\mathcal{X}}}[A^{(2)}] = Z_{\widetilde{\mathcal{X}}}[-(2p')^{-1} A^{(2)}] \exp\left(\frac{2\pi i}{N}\left(-\frac{N+1}{2} (2p')^{-1}\right)\int A^{(2)}\cup A^{(2)} \right) ~,
\end{equation}
%
% \sk{from here}
where $(2p')^{-1}$ denotes the mod-$N$ inverse of $(2p')$. By \eqref{eq:tri_pp_eq}, we can choose $(2p')^{-1} = 1 - 2p'$ and we will also choose $(2p')$ to be odd such that $(2p')^{-1} = 1 - 2p'$ is even.

Using $(-2p')^3=1$ from \eqref{eq:tri_pp_eq}, one can see that the triality twisted gauging in $\mathcal{X}$ is mapped to an invertible $\bz_3^{(0)}$ automorphism of the dual 1-form symmetry $\bz_N^{(1)}$ in $\widetilde{\mathcal{X}}$. Together, they form a non-Abelian 2-group symmetry $\mathbb{Z}_N^{(1)}\rtimes \mathbb{Z}_3^{(0)}$. The additional phase indicates the mixed anomaly between the $\bz_3^{(0)}$ and $\bz_N^{(1)}$ symmetries. Due to this anomaly, the $\bz_3^{(0)}$-symmetry operator $\bd_3$ is not gauge invariant after gauging the dual 1-form symmetry to regain the theory $\mathcal{X}$. To acquire a gauge invariant codim-1 operator, one can stack a TQFT coupled to the gauge field $a^{(2)}$ of the dual symmetry to cancel the anomaly. There is a canonical choice known as the minimal TQFT.

Let us recall some basic properties of the minimal TQFT $\mathcal{A}^{N,p}$ in the special case where $N$ is odd and $p$ is even and coprime with $N$ \cite{Hsin:2018vcg}. The non-spin version of this TQFT contains $\mathbb{Z}_N$ Abelian anyons $a^s$ (where $a^N = 1$) with the topological spin
\begin{equation}
    h(a^s) = \frac{p s^2}{2N} \mod 1 ~.
\end{equation}
The braiding between any two generic Abelian anyons $a_1$ and $a_2$ is
\begin{equation}
    B(a_1,a_2) = e^{2\pi i [h(a_1 a_2) - h(a_1) - h(a_2)]} ~.
\end{equation}
The minimal TQFT $\mathcal{A}^{N,p}$ can be used to cancel the anomaly of the 1-form symmetry captured by the 4d SPT \cite{Hsin:2018vcg,Choi:2022jqy}
\begin{equation}
    -\frac{2\pi i}{N} \frac{N+1}{2} p \int B^{(2)} \cup B^{(2)} ~.
\end{equation}
Therefore, $D_3 = \mathbb{D}_3 \mathcal{A}^{N,(2p')^{-1}}\left[a^{(2)}\right]$\footnote{Notice that we also use the letter $a$ to denote Abelian anyons. Whenever we use $a^{(n)}$ to refer to gauge fields, we will always add the superscript ${(n)}$ to denote its degree so the readers can distinguish between $n$-form gauge fields $a^{(n)}$ and anyons $a$'s.} with an even choice of $(2p')^{-1}$ is gauge invariant in $\mathcal{X}$, but becomes non-invertible because of $\mathcal{A}^{N,(2p')^{-1}}[a^{(2)}]$.

The orientation reversal of $\overline{D}_3$ is 
\begin{equation}\label{eq:D3bgt}
    \overline{D}_3 = \mathcal{A}^{N,-(2p')^{-1}}[a^{(2)}]\overline{\mathbb{D}}_3 ~,
\end{equation} 
where the anomaly of the minimal TQFT is flipped because of the orientation reversal. 

It is instructive to check the fusion rule explicitly to highlight two subtleties compared to the simple case in \cite{Kaidi:2021xfk}. First, the non-Abelian 2-group here does not commute with the minimal TQFT coupled to $a^{(2)}$:
\begin{equation}
    \mathbb{D}_3\mathcal{A}^{N,p}[a^{(2)}] = \mathcal{A}^{N,p}[-(2p')^{-1} a^{(2)}]\mathbb{D}_3 ~.
\end{equation}
Notice that here we interpret the action of $\mathbb{D}_3$ as changing how the bulk field $a^{(2)}$ couples to the anyons in the minimal TQFT $\mathcal{A}^{N,p}$, and $\mathbb{D}_3$ itself does not act on the lines in $\mathcal{A}^{N,p}$. Second, for generic minimal TQFTs, it is not easy to manipulate the Lagrangian to prove various relations, and we will demonstrate how to do so by directly analyzing the anyon spectrum. 

For instance, let's consider the fusion of $D_3 \times D_3$. We find
\begin{equation}\label{eq:gtD3D3f}
\begin{aligned}
    D_3 \times D_3 &= \mathbb{D}_3 \mathcal{A}^{N,(2p')^{-1}}_{a_1}[a^{(2)}] \mathbb{D}_3 \mathcal{A}^{N,(2p')^{-1}}_{a_2}[a^{(2)}] \\
    &= \overline{\mathbb{D}}_3 \mathcal{A}^{N,(2p')^{-1}}_{a_1}[-(2p')a^{(2)}] \mathcal{A}^{N,(2p')^{-1}}_{a_2}[a^{(2)}] \\
    &= \left[SU(N)_{-1}\right]_{a_1 a_2^{2p'}} \, \overline{\mathbb{D}}_3 \mathcal{A}_{a_1^{-2p'} a_2}^{N,-N+1}[a^{(2)}] \equiv SU(N)_{-1} \overline{D}_3 ~.
\end{aligned}
\end{equation}
Here, the subscript under each TQFT denotes its generating line. In the second equal sign, we used
\begin{equation}\label{eq:tcr}
    \mathbb{D}_3\mathcal{A}^{N,(2p')^{-1}}[a^{(2)}] = \mathcal{A}^{N,(2p')^{-1}}[-(2p')^{-1} a^{(2)}]\mathbb{D}_3 ~,
\end{equation}
followed from the action of $\mathbb{Z}_3^{(0)}$ on $\mathbb{Z}_N^{(1)}$. To show the third step, one can consider the anyon spectrum of $\mathcal{A}^{N,(2p')^{-1}}_{a_1}[-(2p')a^{(2)}] \mathcal{A}^{N,(2p')^{-1}}_{a_2}[a^{(2)}]$. We can redefine the generating lines as
\begin{equation}
    b_1 = a_1^{-(2p')} a_2 ~, \quad b_2 = a_1 a_2^{(2p')} ~.
\end{equation}
Furthermore, because $b_1$ and $b_2$ braid trivially with each other, the theory is factorized into two theories generated by $b_i$, respectively. First, the bulk 2-form gauge field $a^{(2)}$ couples to the $\mathbb{Z}_N$ line $b_1$ which has topological spin 
\begin{equation}
\begin{aligned}
    h(a_1^{-2p'} a_2) &= \frac{(2p')^{-1}}{2N}\left((-2p')^2+1\right) \mod 1\\
    &= \frac{(2p')^{-1}}{2N} \left(2p'\right) \mod 1 \\
    &= \frac{-N+1}{2N} \mod 1
\end{aligned}
\end{equation}
where in the last step we used $(2p')^{-1} (2p') = -N +1 \mod 2N$ (as $(2p')^{-1} (2p')$ is even by our choice). Thus, we see that the theory generated by $b_1$ is the minimal TQFT $\mathcal{A}^{N,-N+1}_{b_1}[a^{(2)}]$. To see that $\overline{\mathbb{D}}_3 \mathcal{A}^{N,-N+1}_{b_1}[a^{(2)}]$ is indeed $\overline{D}_3$ given in \eqref{eq:D3bgt}, one uses first the commutation relation \eqref{eq:tcr} and then the certain identity of minimal TQFT
\begin{equation}
    \overline{\mathbb{D}}_3 \mathcal{A}^{N,-N+1}_{b_1}[a^{(2)}] = \mathcal{A}^{N,-N+1}_{b_1}[-(2p')a^{(2)}] \overline{\mathbb{D}}_3 = \mathcal{A}^{N,-(2p')^{-1}}_{c}[a^{(2)}] \overline{\mathbb{D}}_3 ~,
\end{equation}
where $c = b_1^{-2p'}$. The last step can be shown as follows: $a^{(2)}$ couples to the line $b_1^{-2p'}$ with topological spin
\begin{equation}
    h(c) = h(b_1^{-2p'}) = \frac{-(2p')^{-1}}{2N} \mod 1 ~,
\end{equation}
thus if we consider a change of variable $c = b_1^{-2p'} = a_1^{-(2p')^{-1}}a_2^{-2p'}$, we see that the minimal TQFT can be rewritten as $\mathcal{A}^{N,-(2p')^{-1}}_{c}[a^{(2)}]$.

On the other hand, the anyons $b_2^k$ form $\mathbb{Z}_{N}$ and their topological spins are 
\begin{equation}
    h(b_2^k) = \frac{-N+1}{2N} k^2 \mod 1 = - \frac{k(N-k)}{2N} \mod 1 ~,
\end{equation}
thus they give rise to the theory $SU(N)_{-1}$ \eqref{eq:SU(N)m1}. Combining the two discussions, we recover the fusion rule \eqref{eq:gtD3D3f}.

\

As another example, let's consider $D_3 \times \overline{D}_3$:
\begin{equation}
    D_3 \times \overline{D}_3 = \mathbb{D}_3 \mathcal{A}^{N,(2p')^{-1}}_{a_1}[a^{(2)}] \mathcal{A}^{N,-(2p')^{-1}}_{a_2}[a^{(2)}]\overline{\mathbb{D}}_3 = \mathbb{D}_3 \mathcal{C}_0 \overline{\mathbb{D}}_3 = \mathcal{C}_0 ~.
\end{equation}
The key step here is to show that $\mathcal{A}^{N,(2p')^{-1}}_{a_1}[a^{(2)}] \mathcal{A}^{N,-(2p')^{-1}}_{a_2}[a^{(2)}]$ is the condensation defect $\mathcal{C}_0$. We can reparameterize the $\mathbb{Z}_N \times \mathbb{Z}_N$ anyons with the following two generators
\begin{equation}
    b_1 = (a_1 a_2^{-1})^{-\frac{N+1}{2} p'} ~, \quad b_2 = a_1 a_2 ~.
\end{equation}
The new generating lines $b_i$ are $\mathbb{Z}_N$ bosons and have mutual braiding
\begin{equation}
    B(b_1,b_2) = e^{-\frac{2\pi i}{N}} ~.
\end{equation}
This implies that $b_i$'s generate a $\mathbb{Z}_N$ gauge theory with no twist and $b_1$ and $b_2$ correspond to the electric line and the magnetic line respectively. Thus, we can write down an action for it
\begin{equation}\label{eq:gtcd}
    S = \frac{2\pi i}{N} \int b^{(1)} \cup \delta \hat{b}^{(1)} + \hat{b}^{(1)} \cup a^{(2)} ~,
\end{equation}
where the coupling $\hat{b}^{(1)} \cup a^{(2)}$ comes from the fact that $a^{(2)}$ couples to the line $b_2$. This is nothing but a condensation defect \cite{Kaidi:2021xfk}, as
\begin{equation}
\begin{aligned}
    \int [Db^{(1)}][D\hat{b}^{(1)}] e^{\frac{2\pi i}{N} \int b^{(1)} \cup \delta \hat{b}^{(1)} + \hat{b}^{(1)} \cup a^{(2)}} =& \frac{1}{\sqrt{|H^1(X_3,\mathbb{Z}_N)|}}\sum_{\hat{b}^{(1)}\in H^1(X_3,\mathbb{Z}_N)} e^{\frac{2\pi i}{N}\int b^{(1)}\cup a^{(2)}} \\
    =& \frac{1}{\sqrt{|H^1(X_3,\mathbb{Z}_N)|}} \sum_{\sigma \in H_2(X_3,\mathbb{Z}_N)} e^{\frac{2\pi i}{N}\oint_\sigma a^{(2)}} \equiv \mathcal{C}_0 ~,
\end{aligned}
\end{equation}
where we first integrate out $b^{(1)}$ to set $\hat{b}^{(1)}$ flat and then re-express the sum using the Poincar\'e dual $\sigma \in H_2(X_3,\mathbb{Z}_N)$ of $\hat{b}^{(1)} \in H^1(X_3,\mathbb{Z}_N)$. 

It is interesting to notice that the magnetic line $e^{\frac{2\pi i}{N}\oint_\gamma \hat{b}^{(1)}}$ is topological and corresponds to higher quantum symmetry. The electric line $e^{\frac{2\pi i}{N} \oint_\gamma b^{(1)}}$, however, is no longer a genuine line operator as $b^{(1)}$ due to the modified gauge transformation $b^{(1)}\rightarrow b^{(1)} - \lambda^{(1)}_a$ under the gauge transformation of $a^{(2)}\rightarrow a^{(2)} + \delta \lambda_a^{(1)}$. This leads to two interesting configurations. First, the bulk 1-form symmetry operator $e^{\frac{2\pi i}{N}\int a^{(2)}}$ can terminate topologically on $\mathcal{C}_0$ (as the 1-form symmetry is gauged on $\mathcal{C}_0$). This is realized by dressing $e^{-\frac{2\pi i}{N} \oint_\gamma b^{(1)}}$ on the intersection of $e^{\frac{2\pi i}{N}\int a^{(2)}}$ with $\mathcal{C}_0$ to ensure the gauge invariance. Second, a non-genuine line operator $L(\partial \sigma)$ bounding the 1-form symmetry operator $e^{\frac{2\pi i}{N}\int_{\sigma} a^{(2)}}$ (so that the combined $L(\partial \sigma)e^{\frac{2\pi i}{N}\int_{\sigma} a^{(2)}}$ is gauge invariant) in the bulk will become a genuine line when pushing to $\mathcal{C}_0$ because the 1-form symmetry is gauged. This is described by the gauge invariant line $L(\partial \sigma)e^{-\frac{2\pi i}{N}\oint_{\partial \sigma} b^{(1)}}$. Furthermore, $L(\partial \sigma)e^{-\frac{2\pi i}{N}\oint_{\partial \sigma} b^{(1)}}$ is charged under the higher quantum symmetry generated by the magnetic line. Comparing with the description of the Lagrangian condensation defect in Maxwell theory, we see that the higher quantum symmetry $e^{\frac{2\pi i}{N}\oint_\gamma \hat{b}^{(1)}}$ is matched by $e^{i\oint A_L - A_R}$ there; and the genuine line operator $e^{i\oint a}$ on the condensation defect in the Lagrangian description is an example of the bulk non-genuine line operator which becomes genuine on the condensation defect.

\subsection{Overview of anyon condensations and fusion of domain walls}
% \sk{Instead of citing the paper, maybe we can just refer to Section 2.2 of this paper.} \zs{Sounds good.}
Following the general discussion in Section \ref{sec: higher category}, to compute the associator for the three defects $\mathcal{D}_1,\mathcal{D}_2,\mathcal{D}_3$ (where $\mathcal{D}_i$ are triality defects dressed with corresponding minimal TQFT), we need to evaluate the following diagram
\begin{equation}
    \begin{tikzpicture}[baseline={([yshift=-.5ex]current bounding box.center)},vertex/.style={anchor=base,circle,fill=black!25,minimum size=18pt,inner sep=2pt},scale=1]
        \draw[line width = 0.4mm, ->-=0.5] (0,1.5)--(0,2.5);
        \node[above] at (0,2.5) {\footnotesize $\lls\mathcal{D}_1 \otimes (\mathcal{D}_2 \otimes \mathcal{D}_3)\rrs_0$}; %a(bc)
        \draw[line width = 0.4mm, ->-=0.5] (-1.5,0)--(0,1.5); %a
        \node[left] at (-1.5,0) {\footnotesize $\mathcal{D}_1$};
        \draw[line width = 0.4mm, ->-=0.5] (0.75,0.75)--(0,1.5);
        \node[right] at (0.25,1.25) {\footnotesize $\lls \mathcal{D}_2 \otimes \mathcal{D}_3\rrs_0$}; %bc
        \draw[line width = 0.4mm, ->-=0.5] (1.5,0)--(0.75,0.75); 
        \draw[line width = 0.4mm, ->-=0.5] (-0.75,-0.75)--(0.75,0.75);
        \node[left] at (-0.1,0) {\footnotesize $\mathcal{D}_2$}; %b
        \draw[line width = 0.4mm, ->-=0.5] (0,-1.5)--(1.5,0);
        \node[right] at (1.5,0) {\footnotesize $\mathcal{D}_3$}; %c
        \draw[line width = 0.4mm, ->-=0.5] (0,-1.5)--(-0.75,-0.75);
        \node[left] at (-0.25,-1.25) {\footnotesize $\lls\mathcal{D}_1 \otimes \mathcal{D}_2\rrs_0$}; %ab
        \draw[line width = 0.4mm, ->-=0.5] (-0.75,-0.75)--(-1.5,0);
        \draw[line width = 0.4mm, ->-=0.5] (0,-2.5)--(0,-1.5);
        \node[below] at (0,-2.5) {\footnotesize $\lls(\mathcal{D}_1 \otimes \mathcal{D}_2) \otimes \mathcal{D}_3\rrs_0$}; %(ab)c
        \filldraw[red] (0,1.5) circle (2pt);
        \filldraw[red] (0.75,0.75) circle (2pt);
        \filldraw[blue] (-0.75,-0.75) circle (2pt);
        \filldraw[blue] (0,-1.5) circle (2pt);
    \end{tikzpicture} ~.
\end{equation}
Fusing the defects in the middle, the final result will take the form
\begin{equation}
\begin{tikzpicture}[baseline={([yshift=-.5ex]current bounding box.center)},vertex/.style={anchor=base,circle,fill=black!25,minimum size=18pt,inner sep=2pt},scale=1]
    \draw[line width = 0.4mm, ->-=0.5] (0,-2) -- (0,-1);
    \draw[line width = 0.4mm, ->-=0.5] (0,-1) -- (0,+1);
    \draw[line width = 0.4mm, ->-=0.5] (0,+1) -- (0,+2);
    \filldraw[black] (0,-1) circle (2pt);
    \filldraw[black] (0,+1) circle (2pt);
    \node[below] at (0,-2) {\footnotesize $\mathbb{D}^{x} \, \mathcal{T}_-[a^{(2)}]$};
    \node[above] at (0,+2) {\footnotesize $\mathbb{D}^{x} \, \mathcal{T}_+[a^{(2)}]$};
    \node[right] at (0,0) {\footnotesize $\mathbb{D}^{x} \, \mathcal{T}[a^{(2)}] \equiv \mathcal{D}_1 \times \mathcal{D}_2 \times \mathcal{D}_3$};
    \node[right] at (0,-1) {\footnotesize $\mathcal{I}_-$};
    \node[right] at (0,+1) {\footnotesize $\mathcal{I}_+$};
\end{tikzpicture} \implies \begin{tikzpicture}[baseline={([yshift=-.5ex]current bounding box.center)},vertex/.style={anchor=base,circle,fill=black!25,minimum size=18pt,inner sep=2pt},scale=1]
    \draw[line width = 0.4mm, ->-=0.5] (0,-2) -- (0,0);
    \draw[line width = 0.4mm, ->-=0.5] (0,0) -- (0,+2);
    \filldraw[black] (0,0) circle (2pt);
    \node[right] at (0,0) {\footnotesize $\mathcal{I}_{+-}[a^{(2)}] \equiv \mathbf{F}_{\mathcal{D}_1,\mathcal{D}_2,\mathcal{D}_3}$};
    \node[below] at (0,-2) {\footnotesize $\mathbb{D}^{x} \, \mathcal{T}_-[a^{(2)}]$};
    \node[above] at (0,+2) {\footnotesize $\mathbb{D}^{x} \, \mathcal{T}_+[a^{(2)}]$};
\end{tikzpicture} ~,
\end{equation}
where $x = 0,1,2$ depending $\mathcal{D}_i$, $\mathcal{I}_\pm$ are the topological interfaces between $\mathcal{T}$ and $\mathcal{T}_\pm$ respectively. The $F$-symbol here is then acquired by fusing the two interfaces $\mathcal{I}_+$ and $\mathcal{I}_-$. 

The simplification of computing associators in this group-theoretical construction is that everything can be phrased in the language of 3d TQFT, i.e. unitary modular tensor categories (UMTC). Below, we will first include a summary of the results, and will include a detailed description on how to use anyon condensation theory to acquire this in Appendix \ref{app:ac_dw}. 

First, the TQFT $\mathcal{T}$ is described by a UMTC and the interfaces $\mathcal{I}_\pm, \mathcal{T}_\pm$ are acquired by anyon condensations. Following \cite{Copetti:2023mcq}, we use $\mathbb{A}_\pm$ to denote the group of condensed anyons. Then, we want to describe the topological line operator contents in the following configuration
\begin{equation}\label{eq:conf_1}
\begin{aligned}
    \begin{tikzpicture}[baseline={([yshift=-.5ex]current bounding box.center)},vertex/.style={anchor=base,circle,fill=black!25,minimum size=18pt,inner sep=2pt},scale=1]
        \draw[grey, thick] (-2.5,-1) -- (-1.5,-1);
        \draw[grey, thick] (-2.5,+1) -- (-1.5,+1);
        \draw[blue, thick] (-1.5,-1) -- (-1.5,+1);
        \draw[blue, thick] (+1.5,-1) -- (+1.5,+1);
        \draw[grey, thick] (+2.5,-1) -- (+1.5,-1);
        \draw[grey, thick] (+2.5,+1) -- (+1.5,+1);
        \filldraw[blue, opacity = 0.1] (-1.5,-1) -- (-1.5,+1) -- (1.5,+1) -- (1.5,-1) -- (-1.5,-1);
        \node[black] at (-2,0) {\footnotesize $\mathcal{T}_-$};
        \node[black] at (+2,0) {\footnotesize $\mathcal{T}_+$};
        \node[black, below] at (-1.5,-1) {\footnotesize $\mathcal{I}_-$};
        \node[black, below] at (+1.5,-1) {\footnotesize $\mathcal{I}_+$};
        \node[black] at (0,0) {\footnotesize $\mathcal{T}$};
    \end{tikzpicture} ~,
\end{aligned}
\end{equation}
and they are summarized as follows in Table~\ref{tab:anyon_spec}.
\begin{table}[H]
    \centering
    \begin{tabular}{|c|c|l|}
    \hline  
        & Line Operators & \makecell[c]{Fusion rules} \\
    \hline
        $\mathcal{T}_+$ & $L_b^+: b \in \mathcal{T}/\mathbb{A}_+$ and $B(b,a) = 1 \,\, \forall a\in \mathbb{A}_+$ & $\mathcal{T}_+ \boxtimes \mathcal{T}_+ \rightarrow \mathcal{T}_+: L_b^+ \times L_{b'}^+ = L_{bb'}^+$ \\ 
    \hline
        $\mathcal{T}_-$ & $L_b^-: b \in \mathcal{T}/\mathbb{A}_-$ and $B(b,a) = 1 \,\, \forall a\in \mathbb{A}_-$ & $\mathcal{T}_- \boxtimes \mathcal{T}_- \rightarrow \mathcal{T}_-: L_b^- \times L_{b'}^- = L_{bb'}^-$\\
    \hline
        \multirow{3}{*}{$\mathcal{I}_+$} & \multirow{3}{*}{$\ell^+_b, \quad b \in \mathcal{T}/\mathbb{A}_+$} & $  \mathcal{T} \boxtimes \mathcal{I}_+ \rightarrow \mathcal{I}_+: b \otimes \ell^+_{b'} = \ell_{bb'}^+$ \\ \cline{3-3}
        & & $\mathcal{I}_+ \boxtimes \mathcal{I}_+ \rightarrow \mathcal{I}_+: \ell^+_b \otimes \ell^+_{b'} = \ell_{bb'}^+$  \\ \cline{3-3}
        & & $\mathcal{I}_+ \boxtimes \mathcal{T}_+ \rightarrow \mathcal{I}_+: \ell^+_b \otimes L^+_{b'} = \ell^+_{bb'}$\\ \cline{3-3}
    \hline
        \multirow{3}{*}{$\mathcal{I}_-$} & \multirow{3}{*}{$\ell^-_b, \quad b \in \mathcal{T}/\mathbb{A}_-$} & $\mathcal{I}_- \boxtimes \mathcal{T} \rightarrow \mathcal{I}_-: \ell^-_{b} \otimes b' = \ell_{bb'}^-$ \\ \cline{3-3}
        & & $\mathcal{I}_- \boxtimes \mathcal{I}_- \rightarrow \mathcal{I}_-: \ell^-_b \otimes \ell^-_{b'} = \ell_{bb'}^-$ \\ \cline{3-3}
        & & $\mathcal{T}_- \boxtimes \mathcal{I}_- \rightarrow \mathcal{I}_-: L^-_b \otimes \ell^-_{b'} = \ell^-_{bb'}$\\ \cline{3-3}
    \hline
    \end{tabular}
    \caption{The operator contents and fusion rules in \eqref{eq:conf_1}. When list the operator contents, by $b \in \mathcal{T}/\mathbb{A}_+$ in $\mathcal{I}_+$ for instance, we mean the line $\ell_b^+$ is the same line as $\ell_{b'}^+$ if $b^{-1}b' \in \mathbb{A}_+$. Notice that while we only list the fusion rules, the line operators in $\mathcal{T}_\pm$ actually form a modular tensor category respectively and the braiding structure is inherent from $\mathcal{T}$. The domain walls $\mathcal{I}_{\pm}$ are fusion categories, but they also carry actions of the theories on both sides, as captured by the additional fusion rules given above.}
    \label{tab:anyon_spec}
\end{table}

To acquire the $F$-symbol, we consider fusing the domain walls $\mathcal{I}_+$ and $\mathcal{I}_-$ which again can be computed using the anyon condensation theory. The outcome is a domain wall $\mathcal{I}_{-+}$ between $\mathcal{T}_-$ and $\mathcal{T}_+$, which we identify as the $F$-symbol. On the fused domain wall, besides the topological line operators, generically it may also admit local operators, which link non-trivially with the topological line operators on the wall. The result is summarized as follows in Table~\ref{tab:anyon_spec2} and details can be found in Appendix \ref{app:ac_dw}.
\begin{table}[H]
    \centering
    \begin{tabular}{|c|c|c|}
    \hline
       \multirow{3}{*}{Line operators} & \multirow{3}{*}{$\ell^{-+}_{b}, \quad b \in \mathbb{A}_- \text{\textbackslash} \mathcal{T} / \mathbb{A}_+$} & $\mathcal{I}_{-+} \boxtimes \mathcal{I}_{-+} \rightarrow \mathcal{I}_{-+}: \ell^{-+}_b\otimes \ell^{-+}_{b'} = \ell^{-+}_{bb'}$ \\ \cline{3-3} 
       & & $\mathcal{I}_{-+} \boxtimes \mathcal{T}_+ \rightarrow \mathcal{I}_{-+}: L_{b}^- \otimes \ell^{-+}_{b'} = \ell^{-+}_{bb'}$ \\ \cline{3-3}
       & & $\mathcal{T}_{-} \boxtimes \mathcal{I}_{-+} \rightarrow \mathcal{I}_{-+}: \ell^{-+}_{b} \otimes L_{b'}^+ = \ell^{-+}_{bb'}$ \\ \cline{3-3}
    \hline
       Local operators  & $V_{a}, \quad a \in \mathbb{A}_+ \cap \mathbb{A}_-$ & $B(V_{a},\ell^{+-}_b) = B(a,b)$ \\
    \hline
    \end{tabular}
    \caption{The topological operator contents on the fused wall $\mathcal{I}_{-+}$. Generically, there is non-trivial braiding between the local operators and the line operators inherent from the braiding in $\mathcal{T}$.}
    \label{tab:anyon_spec2}
\end{table}

Finally, one also needs to remember to keep track if the resulting 2d interface $\mathcal{I}_{-+}$ couples to the bulk gauge field $a^{(2)}$. As we will see, this will change the physical meaning of the 2d interface.

\subsection{Local fusion junctions}
In this subsection, we list the local fusion junctions where the defects are given by the simplest representatives in the equivalence classes. We will demonstrate the construction by an example, and list all other results afterwards.

Recall that $D_3 \times \overline{D}_3 = \mathcal{C}_0$, however, because the condensation defect $\mathcal{C}_0$ is in the same equivalence class as the identity operator, we can construct a local fusion junction
\begin{equation}\label{eq:D3D3bgt}
    \begin{tikzpicture}[scale=0.8,baseline={([yshift=-.5ex]current bounding box.center)},vertex/.style={anchor=base,
    circle,fill=black!25,minimum size=18pt,inner sep=2pt},scale=0.50]
    \draw[thick, black] (-2,-2) -- (0,0);
    \draw[thick, black] (+2,-2) -- (0,0);
    \draw[thick, black] (0,0) -- (0,2);
    \draw[thick, black, -stealth] (-2,-2) -- (-1,-1);
    \draw[thick, black, -stealth] (+2,-2) -- (1,-1);
    \draw[thick, black, -stealth] (0,0) -- (0,1);
    \filldraw[thick, black] (0,0) circle (3pt);
    \node[black, below] at (-2,-2) {$D_3$};
    \node[black, below] at (2,-2) {$\overline{D}_3$};
    \node[black, above] at (0,2) {$\mathbb{1}$};
\end{tikzpicture} ~,
\end{equation}
via anyon condensations. First, recall that we have shown that $\mathbb{Z}_N$-gauge theory coupled to the bulk gauge field $a^{(2)}$ leads to the condensation defect, and in the action,
\begin{equation}
    S = \frac{2\pi i}{N} \int \left(b^{(1)} \cup \delta \hat{b}^{(1)} + \hat{b}^{(1)} \cup a^{(2)} \right)~,
\end{equation}
we see that generically only the $\hat{b}^{(1)}$-Wilson line $e^{i\oint_\gamma \hat{b}^{(1)}}$ is topological from the equation of motion, thus is identified as the dual 1-form $\mathbb{Z}_N$ quantum symmetry on the condensation defect. The canonical topological boundary of any condensation defect can be acquired by half-space gauging (condensing) the dual quantum symmetry. Therefore, the following interface $\mathcal{I}$ between $\mathcal{C}_0$ and $\mathbb{1}$
\begin{equation}
    \begin{tikzpicture}[baseline={([yshift=-.5ex]current bounding box.center)},vertex/.style={anchor=base,circle,fill=black!25,minimum size=18pt,inner sep=2pt},scale=1]
    \draw[line width = 0.4mm, ->-=0.5] (0,0) -- (0,+1);
    \draw[line width = 0.4mm, ->-=0.5] (0,+1) -- (0,+2);
    \filldraw[black] (0,+1) circle (2pt);
    \node[below] at (0,0) {\footnotesize $\mathcal{C}_0 = (\mathcal{Z}_N)_0[a^{(2)}]$};
    \node[above] at (0,+2) {\footnotesize $\mathbb{1}$};
    \node[right] at (0,+1) {\footnotesize $\mathcal{I}$};
\end{tikzpicture} ~,
\end{equation}
is acquired by condensing the anyon coupled to the bulk gauge field $a^{(2)}$. Applying this to the local junction \eqref{eq:D3D3bgt}, in the global fusion
\begin{equation}
    D_3 \times \overline{D}_3  = \mathcal{A}^{N,(2p')^{-1}}_{a_1}[-(2p')a^{(2)}] \mathcal{A}^{N,-(2p')^{-1}}_{a_2}[-(2p')a^{(2)}] = (\mathcal{Z}_7)_0[-(2p')a^{(2)}] \equiv \mathcal{C}_0 ~,
\end{equation}
the bulk gauge field $a^{(2)}$ couples to the anyon $b_2 = (a_1 a_2)^{-2p'}$. We thus conclude that the condensed anyon $\mathbb{A} = \langle  (a_1 a_2)^{-2p'}\rangle = \langle a_1 a_2\rangle$.

The other local fusion junctions can be constructed similarly, and we summarize the results in Table~\ref{tab:triality_local_junction}.
\begin{table}[]
    \centering
    \begin{tabular}{|c|c|c|}
    \hline
    Local Junctions & Global Fusion & Condensed Anyons $\mathbb{A}$\\
    \hline
    \begin{tikzpicture}[scale=0.8,baseline={([yshift=-.5ex]current bounding box.center)},vertex/.style={anchor=base,
    circle,fill=black!25,minimum size=18pt,inner sep=2pt},scale=0.50]
    \draw[thick, black] (-2,-2) -- (0,0);
    \draw[thick, black] (+2,-2) -- (0,0);
    \draw[thick, black] (0,0) -- (0,2);
    \draw[thick, black, -stealth] (-2,-2) -- (-1,-1);
    \draw[thick, black, -stealth] (+2,-2) -- (1,-1);
    \draw[thick, black, -stealth] (0,0) -- (0,1);
    \filldraw[thick, black] (0,0) circle (3pt);
    \node[black, below] at (-2,-2) {$D_3$};
    \node[black, below] at (2,-2) {$\overline{D}_3$};
    \node[black, above] at (0,2) {$\mathbb{1}$};
    \end{tikzpicture} & \makecell[l]{\footnotesize $\quad [D_3]_{a_1} \times [\overline{D}_3]_{a_2}$ \\\footnotesize  $= \mathcal{A}^{N,(2p')^{-1}}_{a_1}[-(2p')a^{(2)}]\mathcal{A}^{N,-(2p')^{-1}}_{a_2}[-(2p')a^{(2)}]$} & $\langle a_1 a_2 \rangle $ \\ 
    \hline
    \begin{tikzpicture}[scale=0.8,baseline={([yshift=-.5ex]current bounding box.center)},vertex/.style={anchor=base,
    circle,fill=black!25,minimum size=18pt,inner sep=2pt},scale=0.50]
    \draw[thick, black] (-2,-2) -- (0,0);
    \draw[thick, black] (+2,-2) -- (0,0);
    \draw[thick, black] (0,0) -- (0,2);
    \draw[thick, black, -stealth] (-2,-2) -- (-1,-1);
    \draw[thick, black, -stealth] (+2,-2) -- (1,-1);
    \draw[thick, black, -stealth] (0,0) -- (0,1);
    \filldraw[thick, black] (0,0) circle (3pt);
    \node[black, below] at (-2,-2) {$\overline{D}_3$};
    \node[black, below] at (2,-2) {$D_3$};
    \node[black, above] at (0,2) {$\mathbb{1}$};
    \end{tikzpicture} & {\footnotesize $ [\overline{D}_3]_{a_1} \times [D_3]_{a_2} = \mathcal{A}^{N,-(2p')^{-1}}_{a_1}[a^{(2)}]\mathcal{A}^{N,(2p')^{-1}}_{a_2}[a^{(2)}]$} & $\langle a_1 a_2 \rangle $ \\
    \hline
    \begin{tikzpicture}[scale=0.8,baseline={([yshift=-.5ex]current bounding box.center)},vertex/.style={anchor=base,
    circle,fill=black!25,minimum size=18pt,inner sep=2pt},scale=0.50]
    \draw[thick, black] (-2,-2) -- (0,0);
    \draw[thick, black] (+2,-2) -- (0,0);
    \draw[thick, black] (0,0) -- (0,2);
    \draw[thick, black, -stealth] (-2,-2) -- (-1,-1);
    \draw[thick, black, -stealth] (+2,-2) -- (1,-1);
    \draw[thick, black, -stealth] (0,0) -- (0,1);
    \filldraw[thick, black] (0,0) circle (3pt);
    \node[black, below] at (-2,-2) {\scriptsize $D_3$};
    \node[black, below] at (2,-2) {\scriptsize $D_3$};
    \node[black, above] at (0,2) {\scriptsize $SU(N)_{-1} \overline{D}_3 $};
    \end{tikzpicture} & {\footnotesize $\quad [D_3]_{a_1} \times [D_3]_{a_2} = [SU(N)_{-1}]_{a_1 a_2^{2p'}} [\overline{D}_3]_{a_1^{-(2p')^{-1}}a_2^{-2p'}}$} & {$\emptyset$} \\
    \hline
    \begin{tikzpicture}[scale=0.8,baseline={([yshift=-.5ex]current bounding box.center)},vertex/.style={anchor=base,
    circle,fill=black!25,minimum size=18pt,inner sep=2pt},scale=0.50]
    \draw[thick, black] (-2,-2) -- (0,0);
    \draw[thick, black] (+2,-2) -- (0,0);
    \draw[thick, black] (0,0) -- (0,2);
    \draw[thick, black, -stealth] (-2,-2) -- (-1,-1);
    \draw[thick, black, -stealth] (+2,-2) -- (1,-1);
    \draw[thick, black, -stealth] (0,0) -- (0,1);
    \filldraw[thick, black] (0,0) circle (3pt);
    \node[black, below] at (-2,-2) {\scriptsize $D_3$};
    \node[black, below] at (2,-2) {\scriptsize $D_3 SU(N)_{1}$};
    \node[black, above] at (0,2) {\scriptsize $\overline{D}_3$};
    \end{tikzpicture} & \makecell[l]{\footnotesize $\quad [D_3]_{a_1} \times [D_3]_{a_2} [SU(N)_{1}]_{a_3}$ \\ \footnotesize $ = [\overline{D}_3]_{a_1^{-(2p')^{-1}}a_2^{-2p'}} [SU(N)_{-1}]_{a_1 a_2^{2p'}} [SU(N)_{1}]_{a_3}$} & \makecell[l]{$\quad \, \langle a_1 a_2^{2p'} a_3\rangle$ \\ or $\langle a_1 a_2^{2p'} a_3^{-1}\rangle$} \\
    \hline
    \begin{tikzpicture}[scale=0.8,baseline={([yshift=-.5ex]current bounding box.center)},vertex/.style={anchor=base,
    circle,fill=black!25,minimum size=18pt,inner sep=2pt},scale=0.50]
    \draw[thick, black] (-2,-2) -- (0,0);
    \draw[thick, black] (+2,-2) -- (0,0);
    \draw[thick, black] (0,0) -- (0,2);
    \draw[thick, black, -stealth] (-2,-2) -- (-1,-1);
    \draw[thick, black, -stealth] (+2,-2) -- (1,-1);
    \draw[thick, black, -stealth] (0,0) -- (0,1);
    \filldraw[thick, black] (0,0) circle (3pt);
    \node[black, below] at (-2,-2) {\scriptsize $\overline{D}_3$};
    \node[black, below] at (2,-2) {\scriptsize $\overline{D}_3$};
    \node[black, above] at (0,2) {\scriptsize $SU(N)_1 D_3$};
    \end{tikzpicture} & {\footnotesize $\quad [\overline{D}_3]_{a_1} \times [\overline{D}_3]_{a_2} = [D_3]_{a_1^{-2p'} a_2^{-(2p')^{-1}}} [SU(N)_{1}]_{a_1^{2p'}a_2}$}  & {$\emptyset$}\\
    \hline
    \begin{tikzpicture}[scale=0.8,baseline={([yshift=-.5ex]current bounding box.center)},vertex/.style={anchor=base,
    circle,fill=black!25,minimum size=18pt,inner sep=2pt},scale=0.50]
    \draw[thick, black] (-2,-2) -- (0,0);
    \draw[thick, black] (+2,-2) -- (0,0);
    \draw[thick, black] (0,0) -- (0,2);
    \draw[thick, black, -stealth] (-2,-2) -- (-1,-1);
    \draw[thick, black, -stealth] (+2,-2) -- (1,-1);
    \draw[thick, black, -stealth] (0,0) -- (0,1);
    \filldraw[thick, black] (0,0) circle (3pt);
    \node[black, below] at (-2,-2) {\scriptsize $\overline{D}_3 $};
    \node[black, below] at (2,-2) {\scriptsize $\overline{D}_3 SU(N)_{-1}$};
    \node[black, above] at (0,2) {\scriptsize $D_3$};
    \end{tikzpicture} & \makecell[l]{\footnotesize $\quad [\overline{D}_3]_{a_1}  \times [\overline{D}_3]_{a_2} [SU(N)_{-1}]_{a_3}$ \\\footnotesize $= [D_3]_{a_1^{-2p'} a_2^{-(2p')^{-1}}} [SU(N)_{1}]_{a_1^{2p'}a_2} [SU(N)_{-1}]_{a_3}$} &\makecell[l]{$\quad \, \langle a_1^{2p'} a_2 a_3\rangle$ \\ or $\langle a_1^{2p'} a_2 a_3^{-1}\rangle$} \\
    \hline
    \end{tabular}
    \caption{The list of local fusion junctions that requires anyon condensations to construct. We use  the subscript $a$ in $[D_3]_a$ and $[\overline{D}_3]_a$ to denote the generating anyons in the stacked minimal TQFT and the presentation of $\overline{D}_3$ in \eqref{eq:D3bgt} is used. We also the junctions where no anyon is condensed to make the change of anyon generators explicit. In the $4$-th and the last junctions, we use the anyon condensation to remove $[SU(N)_1]_a \times [SU(N)_{-1}]_{b}$. Since $N$ is odd, it is straightforward to check this is the same as a $\mathbb{Z}_N$-gauge theory with no twist (where $ab$ generates the electric lines and $ab^{-1}$ generates the magnetic lines). Generically there will be more than two choices of condensable anyons we provided depending on $N$; here, we only list the two which exists for all $N$.}
    \label{tab:triality_local_junction}
\end{table}

\subsection{Associators for triality defects}\label{app:tri_asso}
With the above tools we can now compute the associators of the triality defect admitting the above group-theoretical construction. For simplicity, we will consider the simplest special case where $N = 7$, and it's straightforward to generalize our approach for generic allowed $N$. We will choose 
\begin{equation}
    2p' = 3 ~, \quad (2p')^{-1} = 1 - 2p' = -2 ~, \quad -(2p')^{-1} = 2 ~,
\end{equation}
and the triality defects are given by
\begin{equation}
    D_3 = \mathbb{D}_3 \mathcal{A}^{7,-2}[a^{(2)}] ~, \quad \overline{D}_3 = \mathcal{A}^{7,2}[a^{(2)}] \overline{\mathbb{D}}_3 ~.
\end{equation}
In the following, we will demonstrate the computation through the following four examples:
\begin{equation}
    D_3\times D_3\times D_3 ~, \quad D_3\times \overline{D}_3\times D_3 ~, \quad \overline{D}_3\times D_3\times D_3 ~, \quad D_3\times D_3 \times \overline{D}_3 ~,
\end{equation}
and the other four are related to the above by orientation reversal.

\subsubsection*{$\mathbf{\left[F_{D_3,{D}_3,D_3}^{D_3}\right]_{SU(7)_{-1} \overline{D}_3,SU(7)_{-1} \overline{D}_3}}$}

Let's consider the $F$-symbol for $D_3 \times D_3 \times D_3$. Evaluating the global fusion in two different ways, we find
\begin{equation}
\begin{aligned}
    & (D_3 \times D_3) \times D_3 = SU(7)_{-1} \overline{D}_3 \times D_3 = SU(7)_{-1}  \mathcal{C}_0 \rightarrow SU(7)_{-1}  ~, \\
    & D_3 \times (D_3 \times D_3) = D_3 \times (SU(7)_{-1}  \overline{D}_3) = SU(7)_{-1} \mathcal{C}_0 \rightarrow SU(7)_{-1}  ~.
\end{aligned}
\end{equation}
Thus, the corresponding diagram is 
\begin{equation}
    \begin{tikzpicture}[baseline={([yshift=-.5ex]current bounding box.center)},vertex/.style={anchor=base,circle,fill=black!25,minimum size=18pt,inner sep=2pt},scale=1]
            \draw[line width = 0.4mm, ->-=0.5] (0,1.5)--(0,3);
            \node[above] at (0,3) {\footnotesize $SU(7)_{-1}$}; %a(bc)
            \draw[line width = 0.4mm, ->-=0.5] (-1.5,0)--(0,1.5); %a
            \node[left] at (-1.5,0) {\footnotesize $D_3$};
            \draw[line width = 0.4mm, ->-=0.5] (0.75,0.75)--(0,1.5);
            \node[right] at (0.25,1.25) {\footnotesize $SU(7)_{-1} \overline{D}_3$}; %bc
            \draw[line width = 0.4mm, ->-=0.5] (1.5,0)--(0.75,0.75); 
            \draw[line width = 0.4mm, ->-=0.5] (-0.75,-0.75)--(0.75,0.75);
            \node[left] at (-0.1,0) {\footnotesize $D_3$}; %b
            \draw[line width = 0.4mm, ->-=0.5] (0,-1.5)--(1.5,0);
            \node[right] at (1.5,0) {\footnotesize $D_3$}; %c
            \draw[line width = 0.4mm, ->-=0.5] (0,-1.5)--(-0.75,-0.75);
            \node[left] at (-0.25,-1.25) {\footnotesize $SU(7)_{-1} \overline{D}_3$}; %ab
            \draw[line width = 0.4mm, ->-=0.5] (-0.75,-0.75)--(-1.5,0);
            \draw[line width = 0.4mm, ->-=0.5] (0,-3)--(0,-1.5);
            \node[below] at (0,-3) {\footnotesize $SU(7)_{-1}$}; %(ab)c

	\end{tikzpicture} \longrightarrow \begin{tikzpicture}[baseline={([yshift=-.5ex]current bounding box.center)},vertex/.style={anchor=base,circle,fill=black!25,minimum size=18pt,inner sep=2pt},scale=1]
    \draw[line width = 0.4mm, ->-=0.5] (0,-2) -- (0,-1);
    \draw[line width = 0.4mm, ->-=0.5] (0,-1) -- (0,+1);
    \draw[line width = 0.4mm, ->-=0.5] (0,+1) -- (0,+2);
    \filldraw[black] (0,-1) circle (2pt);
    \filldraw[black] (0,+1) circle (2pt);
    \node[below] at (0,-2) {\footnotesize $SU(7)_{-1}$};
    \node[above] at (0,+2) {\footnotesize $SU(7)_{-1}$};
    \node[right] at (0,0) {\footnotesize $\mathcal{T}[a^{(2)}] \equiv \mathcal{A}^{7,-2}[2a^{(2)}] \mathcal{A}^{7,-2}[4a^{(2)}] \mathcal{A}^{7,-2}[a^{(2)}]$};
    \node[right] at (0,-1) {\footnotesize $\mathcal{I}_-$};
    \node[right] at (0,+1) {\footnotesize $\mathcal{I}_+$};
\end{tikzpicture} \longrightarrow \begin{tikzpicture}[baseline={([yshift=-.5ex]current bounding box.center)},vertex/.style={anchor=base,circle,fill=black!25,minimum size=18pt,inner sep=2pt},scale=1]
    \draw[line width = 0.4mm, ->-=0.5] (0,-2) -- (0,0);
    \draw[line width = 0.4mm, ->-=0.5] (0,0) -- (0,+2);
    \filldraw[black] (0,0) circle (2pt);
    \node[right] at (0,0) {\footnotesize $\mathcal{I}_{-+}[a^{(2)}]$};
    \node[below] at (0,-2) {\footnotesize $SU(7)_{-1}$};
    \node[above] at (0,+2) {\footnotesize $SU(7)_{-1}$};
\end{tikzpicture} ~.
\end{equation}
Here, the theory $\mathcal{T}$ is extracted using \eqref{eq:tcr}
\begin{equation}
\begin{aligned}
    D_3 \times D_3 \times D_3 &= \mathbb{D}_3 \mathcal{A}^{7,-2}_{a_1}[a^{(2)}] \mathbb{D}_3 \mathcal{A}^{7,-2}_{a_2}[a^{(2)}] \mathbb{D}_3 \mathcal{A}^{7,-2}_{a_3}[a^{(2)}] \\
    &= \mathbb{D}_3 \mathcal{A}^{7,-2}_{a_1}[a^{(2)}]\overline{D}_3 \mathcal{A}^{7,-2}_{a_2}[4a^{(2)}] \mathcal{A}^{7,-2}_{a_3}[a^{(2)}] \\
    &= \mathcal{A}^{7,-2}_{a_1}[2a^{(2)}] \mathcal{A}^{7,-2}_{a_2}[4a^{(2)}] \mathcal{A}^{7,-2}_{a_3}[a^{(2)}]  ~.
\end{aligned}
\end{equation}
To find that the condensed anyons $\mathbb{A}_+$, let's follow the fusion tree
\begin{equation}
     D_3 \times (D_3 \times D_3) = D_3 \times (SU(N)_{-1} \overline{D}_3) = SU(N)_{-1}  \mathcal{C}_0 \rightarrow SU(N)_{-1} ~.
\end{equation}
In the first junction, $(D_3 \times D_3) = U(1)_N \mathcal{A}^{7,2}[a^{(2)}] \overline{\mathbb{D}}_3$, as discussed previously, the minimal TQFT $\mathcal{A}^{7,2}[a^{(2)}]$ is generated by the anyon $a_2^4 a_3$ to which the gauge field $a^{(2)}$ couples. No anyon condensation happens here. Next, the second junction is simply the local junction between $D_3 \times \overline{D}_3$ and the $\mathbb{1}$, and as shown previously, the anyon that couples to the gauge field $a^{(2)}$ is condensed on this interface; therefore $\mathbb{A}_+ = \langle a_1^2 a_2^4 a_3 \rangle$. By the similar analysis, one can show that $\mathbb{A}_- = \langle a_1^2 a_2^4 a_3\rangle$.

Using the result in Table~\ref{tab:anyon_spec2}, we see that both $\mathcal{T}_\pm$ are $SU(7)_{-1}$'s generated by $L_{a_2 a_3^3}^\pm$ respectively. The fused interface $\mathcal{I}_{-+}$ contains topological line operators factorized into $\mathbb{Z}_{7} \times \mathbb{Z}_7$, where one $\mathbb{Z}_{7}$ factor corresponds to the unconfined lines in $SU(7)_{-1}$ and the other $\mathbb{Z}_7$ factor corresponds to topological lines confined on the interface. By checking the fusion rule between the $\mathcal{T}_\pm = SU(7)_{-1}$ and the $\mathbb{Z}_{7}$ unconfined sector, we see that the unconfined $\mathbb{Z}_{7}$ sector forms the identity interface between $\mathcal{T}_\pm$ theories. 

Furthermore, the interface also contains $\mathbb{Z}_7$ topological vertex operators labeled by $V_a$ where $a \in \mathbb{A}_+ \cap \mathbb{A}_- = \langle a_1^2 a_2^4 a_3\rangle$. From the braiding relation between the vertex operators and the confined lines, we see that they form a 2d $\mathbb{Z}_7$ TQFT on the interface. However, this TQFT actually couples to the bulk gauge field $a^{(2)}$ through the vertex operator, thus an effective action is given by
\begin{equation}
    S = \frac{2\pi i}{7} \int \left(b^{(1)} \cup \delta \phi^{(0)} + \phi^{(0)} a^{(2)}\right) ~. \label{eq: ea}
\end{equation}
By the similar multiplication where we integrate out $b^{(1)} \in C^1(M_2,\mathbb{Z}_7)$, we see that this is nothing but a 2d condensation defect of the bulk $\mathbb{Z}_7^{(1)}$ symmetry (which is actually just a projector). Hence, we conclude in this case that the $F$-symbol interface is given by the identity interface of the $SU(7)_{-1}$ TQFT stacked with a 2d condensation defect of the bulk $\mathbb{Z}_7^{(1)}$ symmetry.

Let us briefly comment on the operator spectrum. First, $e^{\frac{2\pi i k}{N}\phi^{(0)}}$ are $\mathbb{Z}_N$ topological local operators (arising from the higher quantum symmetry line $e^{\frac{2\pi i k}{N}\int a^{(1)}}$ stretching between two boundaries) and they can be interpreted as the 1-form higher quantum symmetry on the 2d condensation defect. The gauge field $b^{(1)}$ again has the modified gauge transformation $b^{(1)} \rightarrow b^{(1)} - \lambda_a^{(1)}$ under the $a^{(2)}$ gauge transformation $a^{(2)} \rightarrow a^{(2)} + \delta \lambda_a^{(1)}$. Therefore, $e^{\frac{2\pi i}{N}\oint b^{(1)}}$ plays the same role as the 3d condensation defect, namely to ensure the non-genuine line bounding the bulk 1-form $\mathbb{Z}_N$ symmetry becomes genuine on the 2d condensation defect. Notice that this matches straightforwardly with the same discussion in the Lagrangian formalism in Appendix \ref{app:tri_asso}.

To summarize, we find
\begin{equation}
    \mathbf{\left[F_{D_3,{D}_3,D_3}^{D_3}\right]_{SU(7)_{-1} \overline{D}_3,SU(7)_{-1} \overline{D}_3}} = \text{2d condensation defect of $\mathbb{Z}_7$ 1-form symmetry} ~.
\end{equation}

\subsubsection*{$\mathbf{\left[F_{D_3,\overline{D}_3,D_3}^{D_3}\right]_{\bi,\bi}}$}
As another example, one can consider $D_3 \times \overline{D}_3 \times D_3$. In this case, the diagram can be summarized as
\begin{equation}
    \begin{tikzpicture}[baseline={([yshift=-.5ex]current bounding box.center)},vertex/.style={anchor=base,circle,fill=black!25,minimum size=18pt,inner sep=2pt},scale=1]
            \draw[line width = 0.4mm, ->-=0.5] (0,1.5)--(0,3);
            \node[above] at (0,3) {\footnotesize $D_3$}; %a(bc)
            \draw[line width = 0.4mm, ->-=0.5] (-1.5,0)--(0,1.5); %a
            \node[left] at (-1.5,0) {\footnotesize $D_3$};
            \draw[dashed, line width = 0.4mm, ->-=0.5] (0.75,0.75)--(0,1.5);
            \node[right] at (0.25,1.25) {\footnotesize $\mathbb{1}$}; %bc
            \draw[line width = 0.4mm, ->-=0.5] (1.5,0)--(0.75,0.75); 
            \draw[line width = 0.4mm, ->-=0.5] (-0.75,-0.75)--(0.75,0.75);
            \node[left] at (-0.1,0) {\footnotesize $\overline{D}_3$}; %b
            \draw[line width = 0.4mm, ->-=0.5] (0,-1.5)--(1.5,0);
            \node[right] at (1.5,0) {\footnotesize $D_3$}; %c
            \draw[dashed, line width = 0.4mm, ->-=0.5] (0,-1.5)--(-0.75,-0.75);
            \node[left] at (-0.25,-1.25) {\footnotesize $\mathbb{1}$}; %ab
            \draw[line width = 0.4mm, ->-=0.5] (-0.75,-0.75)--(-1.5,0);
            \draw[line width = 0.4mm, ->-=0.5] (0,-3)--(0,-1.5);
            \node[below] at (0,-3) {\footnotesize $D_3$}; %(ab)c
	\end{tikzpicture} \longrightarrow \begin{tikzpicture}[baseline={([yshift=-.5ex]current bounding box.center)},vertex/.style={anchor=base,circle,fill=black!25,minimum size=18pt,inner sep=2pt},scale=1]
    \draw[line width = 0.4mm, ->-=0.5] (0,-2) -- (0,-1);
    \draw[line width = 0.4mm, ->-=0.5] (0,-1) -- (0,+1);
    \draw[line width = 0.4mm, ->-=0.5] (0,+1) -- (0,+2);
    \filldraw[black] (0,-1) circle (2pt);
    \filldraw[black] (0,+1) circle (2pt);
    \node[below] at (1.3,-2) {\footnotesize $D_3 = \mathbb{D}_3 \mathcal{A}^{7,-2}[a^{(2)}]$};
    \node[above] at (1.3,+2) {\footnotesize $D_3 = \mathbb{D}_3 \mathcal{A}^{7,-2}[a^{(2)}]$};
    \node[right] at (0,0) {\footnotesize $\mathbb{D}_3 \mathcal{T}[a^{(2)}] \equiv \mathbb{D}_3 \mathcal{A}^{7,-2}_{a_1}[a^{(2)}] \mathcal{A}^{7,2}_{a_2}[a^{(2)}] \mathcal{A}^{7,-2}_{a_3}[a^{(2)}]$};
    \node[right] at (0,-1) {\footnotesize $\mathcal{I}_-$};
    \node[right] at (0,+1) {\footnotesize $\mathcal{I}_+$};
\end{tikzpicture} \longrightarrow \begin{tikzpicture}[baseline={([yshift=-.5ex]current bounding box.center)},vertex/.style={anchor=base,circle,fill=black!25,minimum size=18pt,inner sep=2pt},scale=1]
    \draw[line width = 0.4mm, ->-=0.5] (0,-2) -- (0,0);
    \draw[line width = 0.4mm, ->-=0.5] (0,0) -- (0,+2);
    \filldraw[black] (0,0) circle (2pt);
    \node[right] at (0,0) {\footnotesize $\mathcal{I}_{-+}[a^{(2)}]$};
    \node[below] at (0,-2) {\footnotesize $\mathbb{D}_3 \mathcal{A}^{7,-2}[a^{(2)}]$};
    \node[above] at (0,+2) {\footnotesize $\mathbb{D}_3 \mathcal{A}^{7,-2}[a^{(2)}]$};
\end{tikzpicture} ~.
\end{equation}
The theory $\mathcal{T} = \mathcal{A}^{7,-2}_{a_1}[a^{(2)}] \mathcal{A}^{7,2}_{a_2}[4a^{(2)}] \mathcal{A}^{7,-2}_{a_3}[a^{(2)}]$ is extracted similarly using \eqref{eq:tcr} with the presentation of $\overline{D}_3 = \mathcal{A}^{7,-2}[a^{(2)}] \overline{\mathbb{D}}_3$. From Table~\ref{tab:triality_local_junction}, we find
\begin{equation}
    \mathbb{A}_- = \langle a_1 a_2 \rangle ~, \quad \mathbb{A}_+ = \langle a_2 a_3 \rangle ~.
\end{equation}

In this case, the fused domain wall $\mathcal{I}_{-+}$ contains only topological lines labeled by $\mathbb{Z}_{7}$ and no local operators as $\mathbb{A}_+ \cap \mathbb{A}_-$ is trivial. The theory $\mathcal{T}_+ = \mathcal{A}^{7,-2}[a^{(2)}]$ is generated by the anyon $L^+_{a_1}$ and the theory $\mathcal{T}_- = \mathcal{A}^{7,-2}[a^{(2)}]$ is generated by the anyon $L^-_{a_3}$. Because both $L^+_{a_1}$ and $L^-_{a_3}$ couple to the bulk gauge field $a^{(2)}$ respectively, it is natural to identify the two lines. On the other hand, pushing $L^+_{a_1}$ and $L^-_{a_3}$ to the interface, they become $\ell^{-+}_{a_1}$ and $\ell^{-+}_{a_3}$ respectively. But because $a_1 = a_1 (a_1 a_2)^{-1} (a_2 a_3)$, we see that
\begin{equation}
    \ell^{-+}_{a_1} = \ell^{-+}_{a_1 (a_1 a_2)^{-1} (a_2 a_3)} = \ell^{-+}_{a_3} ~,
\end{equation}
therefore pushing $L^+_{a_1}$ and $L^-_{a_3}$ leads to the identical line on the interface. Thus, we conclude that the $F$-symbol interface here is the identity interface between triality defect $D_3$ and $D_3$.

\subsubsection*{$\mathbf{\left[F_{\overline{D}_3,{D}_3,D_3}^{D_3}\right]_{\bi,SU(7)_{-1}\overline{D}_3}}$} 
This component of the associator relates the following two fusion channels:
\begin{equation}
\begin{aligned}
    \overline{D}_3\times (D_3\times D_3) &= \overline{D}_3\times (SU(7)_{-1} \, \overline{D}_3)  = (SU(7)_1 \times SU(7)_{-1}) D_3  \to D_3 ~, \\
    (\overline{D}_3\times D_3) \times D_3 &= C_0 \times D_3 \to D_3 ~,
\end{aligned}
\end{equation}
and the corresponding diagram can be worked out by referring to Table~\ref{tab:triality_local_junction}:
\begin{equation}
    \begin{tikzpicture}[baseline={([yshift=-.5ex]current bounding box.center)},vertex/.style={anchor=base,circle,fill=black!25,minimum size=18pt,inner sep=2pt},scale=1]
            \draw[line width = 0.4mm, ->-=0.5] (0,1.5)--(0,3);
            \node[above] at (0,3) {\footnotesize $[D_3]_{L^+_{a_1^4 a_2^4 a_3}}$}; %a(bc)
            \draw[line width = 0.4mm, ->-=0.5] (-1.5,0)--(0,1.5); %a
            \node[left] at (-1.5,0) {\footnotesize $[\overline{D}_3]_{a_1}$};
            \draw[line width = 0.4mm, ->-=0.5] (0.75,0.75)--(0,1.5);
            \node[right] at (0.25,1.25) {\footnotesize $[SU(7)_{-1}]_{a_2 a_3^3} [\overline{D}_3]_{a_2^2 a_3^{-3}}$}; %bc
            \draw[line width = 0.4mm, ->-=0.5] (1.5,0)--(0.75,0.75); 
            \draw[line width = 0.4mm, ->-=0.5] (-0.75,-0.75)--(0.75,0.75);
            \node[left] at (-0.1,0) {\footnotesize $[D_3]_{a_2}$}; %b
            \draw[line width = 0.4mm, ->-=0.5] (0,-1.5)--(1.5,0);
            \node[right] at (1.5,0) {\footnotesize $[D_3]_{a_3}$}; %c
            \draw[dashed, line width = 0.4mm, ->-=0.5] (0,-1.5)--(-0.75,-0.75);
            \node[left] at (-0.25,-1.25) {\footnotesize $\mathbb{1}$}; %ab
            \draw[line width = 0.4mm, ->-=0.5] (-0.75,-0.75)--(-1.5,0);
            \draw[line width = 0.4mm, ->-=0.5] (0,-3)--(0,-1.5);
            \node[below] at (0,-3) {\footnotesize $[D_3]_{L^-_{a_3}}$}; %(ab)c

            \node[red, left] at (0,1.9) {\footnotesize $\mathbb{A}_{\text{I}} = \langle a_1 a_2 \rangle $};
            \node[red, left] at (0,1.5) {\footnotesize $\mathbb{A}_{\text{II}} = \langle a_1^3 a_2 a_3 \rangle $};
            \filldraw[red] (0,1.5) circle (2pt);
            \filldraw[blue] (-0.75,-0.75) circle (2pt);
            \node[blue, left] at (-0.75,-0.75) {\footnotesize $\mathbb{A} = \langle a_1 a_2\rangle$};
	\end{tikzpicture} \longrightarrow \begin{tikzpicture}[baseline={([yshift=-.5ex]current bounding box.center)},vertex/.style={anchor=base,circle,fill=black!25,minimum size=18pt,inner sep=2pt},scale=1]
    \draw[line width = 0.4mm, ->-=0.5] (0,-2) -- (0,-1);
    \draw[line width = 0.4mm, ->-=0.5] (0,-1) -- (0,+1);
    \draw[line width = 0.4mm, ->-=0.5] (0,+1) -- (0,+2);
    \filldraw[black] (0,-1) circle (2pt);
    \filldraw[black] (0,+1) circle (2pt);
    \node[below] at (1.3,-2) {\footnotesize $D_3 = \mathbb{D}_3 \mathcal{A}^{7,-2}[a^{(2)}]$};
    \node[above] at (1.3,+2) {\footnotesize $D_3 = \mathbb{D}_3 \mathcal{A}^{7,-2}[a^{(2)}]$};
    \node[right] at (0,-1) {\footnotesize $\mathcal{I}_-: \mathbb{A}_-= \langle a_1 a_2\rangle$};
    \node[right] at (0,+1) {\footnotesize $\mathcal{I}_+: \mathbb{A}_{+,\text{I}} = \langle a_1 a_2 \rangle ~, \quad  \mathbb{A}_{+,\text{II}} = \langle a_1^3 a_2 a_3 \rangle$};
\end{tikzpicture} ~,
\end{equation}
from which we conclude
\begin{equation}
    \mathbb{A}_- = \langle a_1 a_2 \rangle ~, \quad \mathbb{A}_{+,\text{I}} = \langle a_1 a_2 \rangle ~, \quad \mathbb{A}_{+,\text{II}} = \langle a_1^3 a_2 a_3 \rangle ~,
\end{equation}
where for $\mathbb{A}_+$ we have two choices. For the choice of $\ba_{+,\text{I}}$, the interface $\mathcal{I}_{-+}$ contains vertex operators labeled by $\mathbb{A}_{+,\text{I}}\cap \mathbb{A}_- = \langle a_1 a_2 \rangle \simeq \mathbb{Z}_7$ and confined $\mathbb{Z}_7$ topological lines generated by $\ell^{-+}_{a_1 a_2^{-1}}$. Together they form a decoupled 2d $\mathbb{Z}_7$-gauge theory. One can further check from the fusion rules between unconfined lines and the bulk lines that the interface $\mathcal{I}_{-+,\text{I}}$ is the identity interface in the minimal TQFT $\mathcal{A}^{7,-2}[a^{(2)}]$ plus the decoupled 2d $\mathbb{Z}_7$ gauge theory.

For the choice of $\ba_{+,\text{II}}$, there are no confined lines and no vertex operators on the interface $\mathcal{I}_{-+}$. And it is straightforward to check that the $\mathcal{I}_{-+,\text{II}}$ is simply the identity interface in the minimal TQFT $\mathcal{A}^{7,-2}[a^{(2)}]$.

In summary,
\begin{equation}
    \mathbf{\left[F_{\overline{D}_3,{D}_3,D_3}^{D_3}\right]_{\bi,SU(7)_{-1}\overline{D}_3}} = \text{decoupled $\bz_7$ gauge theory \quad or \quad trivial TQFT} ~.
\end{equation}

\subsubsection*{$\mathbf{\left[F_{ D_3,{D}_3,\overline{D}_3}^{D_3}\right]_{SU(7)_{-1}\overline{D}_3,\bi}}$}
% \zs{I am thinking about writing in the same way:}
This component of the associator relates the two following fusions.
\begin{equation}
\begin{aligned}
    D_3\times (D_3\times \overline{D}_3) &= D_3\times C_0 \to D_3 ~, \\
    (D_3\times D_3)\times \overline{D}_3 &= SU(7)_{-1} \, \overline{D}_3\times \overline{D}_3 = SU(7)_{-1} \times SU(7)_{1}\, {D}_3 \to D_3  ~,
\end{aligned}
\end{equation}
and the corresponding diagram can be worked out by referring to Table~\ref{tab:triality_local_junction}:
\begin{equation}
    \begin{tikzpicture}[baseline={([yshift=-.5ex]current bounding box.center)},vertex/.style={anchor=base,circle,fill=black!25,minimum size=18pt,inner sep=2pt},scale=1]
            \draw[line width = 0.4mm, ->-=0.5] (0,1.5)--(0,3);
            \node[above] at (0,3) {\footnotesize $[D_3]_{L^+_{a_1}}$}; %a(bc)
            \draw[line width = 0.4mm, ->-=0.5] (-1.5,0)--(0,1.5); %a
            \node[left] at (-1.5,0) {\footnotesize $[D_3]_{a_1}$};
            \draw[dashed, line width = 0.4mm, ->-=0.5] (0.75,0.75)--(0,1.5);
            \node[right] at (0.25,1.25) {\footnotesize $\mathbb{1}$}; %bc
            \draw[line width = 0.4mm, ->-=0.5] (1.5,0)--(0.75,0.75); 
            \draw[line width = 0.4mm, ->-=0.5] (-0.75,-0.75)--(0.75,0.75);
            \node[left] at (-0.1,0) {\footnotesize $[D_3]_{a_2}$}; %b
            \draw[line width = 0.4mm, ->-=0.5] (0,-1.5)--(1.5,0);
            \node[right] at (1.5,0) {\footnotesize $[\overline{D}_3]_{a_3}$}; %c
            \draw[line width = 0.4mm, ->-=0.5] (0,-1.5)--(-0.75,-0.75);
            \node[left] at (-0.25,-1.25) {\footnotesize $[\overline{D}_3]_{a_1^2 a_2^{-3}} [SU(7)_{-1}]_{a_1 a_2^3}$}; %ab
            \draw[line width = 0.4mm, ->-=0.5] (-0.75,-0.75)--(-1.5,0);
            \draw[line width = 0.4mm, ->-=0.5] (0,-3)--(0,-1.5);
            \node[below] at (0,-3) {\footnotesize $[D_3]_{L^-_{a_1 a_2^2 a_3^2}}$}; %(ab)c
            \node[red, right] at (0,-1.9) {\footnotesize $\mathbb{A}_{\text{II}} = \langle a_2 a_3 \rangle $};
            \node[red, right] at (0,-1.5) {\footnotesize $\mathbb{A}_{\text{I}} = \langle a_1^{-2} a_2^2 a_3 \rangle $};
            \filldraw[red] (0,-1.5) circle (2pt);
            \filldraw[blue] (+0.75,+0.75) circle (2pt);
            \node[blue, right] at (+0.75,+0.75) {\footnotesize $\mathbb{A} = \langle a_2 a_3\rangle$};
	\end{tikzpicture} \longrightarrow \begin{tikzpicture}[baseline={([yshift=-.5ex]current bounding box.center)},vertex/.style={anchor=base,circle,fill=black!25,minimum size=18pt,inner sep=2pt},scale=1]
    \draw[line width = 0.4mm, ->-=0.5] (0,-2) -- (0,-1);
    \draw[line width = 0.4mm, ->-=0.5] (0,-1) -- (0,+1);
    \draw[line width = 0.4mm, ->-=0.5] (0,+1) -- (0,+2);
    \filldraw[black] (0,-1) circle (2pt);
    \filldraw[black] (0,+1) circle (2pt);
    \node[below] at (1.3,-2) {\footnotesize $D_3 = \mathbb{D}_3 \mathcal{A}^{7,-2}[a^{(2)}]$};
    \node[above] at (1.3,+2) {\footnotesize $D_3 = \mathbb{D}_3 \mathcal{A}^{7,-2}[a^{(2)}]$};
    \node[right] at (0,-1) {\footnotesize $\mathcal{I}_-: \mathbb{A}_{-,\text{I}} = \langle a_2 a_3 \rangle ~, \quad  \mathbb{A}_{-,\text{II}} = \langle a_1^{-2} a_2^2 a_3 \rangle$};
    \node[right] at (0,+1) {\footnotesize $\mathcal{I}_+: \mathbb{A}_+ = \langle a_1 a_2\rangle$};
\end{tikzpicture} ~,
\end{equation}
from which we conclude
\begin{equation}
    \mathbb{A}_{-,\text{I}} = \langle a_2 a_3 \rangle ~, \quad \mathbb{A}_{-,\text{II}} = \langle a_1^{-2} a_2^2 a_3 \rangle ~, \quad \mathbb{A}_{+,\text{II}} = \langle a_2 a_3 \rangle ~,
\end{equation}
where $\mathbb{A}_{-}$ has two choices. For the choice of $\ba_{-,\text{I}}$, the interface $\mathcal{I}_{-+}$ is the identity interface in the minimal TQFT $\mathcal{A}^{7,-2}[a^{(2)}]$ together with a decoupled 2d $\mathbb{Z}_7$ gauge theory. And for the choice of $\ba_{-,\text{II}}$, $\mathcal{I}_{-+,\text{II}}$ is simply the identity interface in the minimal TQFT $\mathcal{A}^{7,-2}[a^{(2)}]$. Hence, we find
\begin{equation}
    \mathbf{\left[F_{ D_3,{D}_3,\overline{D}_3}^{D_3}\right]_{SU(7)_{-1}\overline{D}_3,\bi}} = \text{decoupled $\bz_7$ gauge theory \quad or \quad trivial TQFT} ~.
\end{equation}

\section*{Acknowledgments}

We thank Per Kraus, Ryan Thorngren and Michele Del Zotto for valuable discussions. S.K. is supported by Mani L. Bhaumik Institute for Theoretical Physics. Z.S. is supported by the Simons Collaboration on Global Categorical Symmetries. O.S. is supported by Simons Foundation grant 994310 (Simons Collaboration
on Confinement and QCD Strings).

\appendix

\section{Fusion junctions and associators in the 2d compact boson theory}
\subsection{Local fusion junctions} \label{app: local fusion junctions}
In this subsection, we analyze the $T\times T \to \eta^k$ junction using the junction data for the fusion $T\times \eta^{-k} \to T$, and then complete our analysis of the local fusion junctions by analyzing the $\eta^k \times T = T$ and $\eta^m\times \eta^n = \eta^{m+n}$ junctions.

\paragraph{$T\times T\to \eta^k$ using $T\times \eta^{-k}\to T$} One can begin with the local fusion junction for $T\times \eta^{-k}\to T$ and flip the direction of the $\eta^k$ defect along $L_2$ and similarly also that of the $T$ defect along $t>0$, as described in Figure~\ref{fig: alt T-T fusion}.
\begin{figure}[H]
	\centering
	\begin{tikzpicture}
		\begin{scope}[very thick,decoration={
				markings,
				mark=at position 0.5 with {\arrow{>}}}
			] 
			\draw[postaction={decorate}] (-1.73*0.8,-2*0.8)--(0,-1*0.8) node[at start,left] (l) {$T$} node[midway,left] () {$L_1~$};
			\draw[postaction={decorate}] (1.73*0.8,-2*0.8)--(0,-1*0.8) node[at start,right] () {$\eta^{-k}$} node[midway,right] () {$~L_2$};
			\draw[postaction={decorate}] (0,-1*0.8)--(0,1*0.8) node[at end,right] () {$T$};
			\node at (-1*0.8,-0.5*0.8) () {$\phi_3$};
			\node at (1*0.8,-0.5*0.8) () {$\phi_1$};
			\node at (0,-1.8*0.8) () {$\phi_2$};
			\node[right] at (0,-0.9*0.8) () {$0$};
			\filldraw[blue] (0,-1*0.8) circle (2pt) node[right] () {};
            \node at (2.5*0.8,0) (f) {=};
		\end{scope}
		\begin{scope}[shift={($(f.east)+(2,0)$)}, very thick,decoration={
				markings,
				mark=at position 0.5 with {\arrow{>}}}
			] 
			\draw[postaction={decorate}] (-1.73*0.8,-2*0.8)--(0,-1*0.8) node[at start,left] (l) {$T$} node[midway,left] () {$L_1~$};
			\draw[postaction={decorate}] (0,-1*0.8)--(1.73*0.8,-2*0.8) node[at end,right] () {$\eta^{k}$} node[midway,right] () {$~L_2$};
			\draw[postaction={decorate}] (0,-1*0.8)--(0,1*0.8) node[at end,right] () {$T$};
			\node at (-1*0.8,-0.5*0.8) () {$\phi_3$};
			\node at (1*0.8,-0.5*0.8) () {$\phi_1$};
			\node at (0,-1.8*0.8) () {$\phi_2$};
			\node[right] at (0,-0.9*0.8) () {$0$};
			\filldraw[blue] (0,-1*0.8) circle (2pt) node[right] () {};
            \node at (2.5*0.8,0) (ff) {=};
            \end{scope}
		\begin{scope}[shift={($(ff.east)+(2,0)$)}, very thick,decoration={
				markings,
				mark=at position 0.5 with {\arrow{>}}}
			] 
	\draw[postaction={decorate}] (-1.73*0.8,-2*0.8)--(0,-1*0.8) node[at start,left] (l) {$T$} node[midway,left] () {$L_1~$};
			\draw[postaction={decorate}] (0,-1*0.8)--(1.73*0.8,-2*0.8) node[at end,right] () {$\eta^{k}$} node[midway,right] () {$~L_2$};
			\draw[postaction={decorate}] (0,1*0.8)--(0,-1*0.8) node[at start,right] () {$T$};
			\node at (-1*0.8,-0.5*0.8) () {$\phi_3$};
			\node at (1*0.8,-0.5*0.8) () {$\phi_1$};
			\node at (0,-1.8*0.8) () {$\phi_2$};
			\node[right] at (0,-0.9*0.8) () {$0$};
			\filldraw[blue] (0,-1*0.8) circle (2pt) node[right] () {};
            \node at (3*0.8,0) () {$-\left. {iN\over 2\pi}\phi_1\phi_3 \right|^0_{\infty}$};
            \end{scope}
	\end{tikzpicture}
	\caption{The local fusion junction Hom($T\times T,\eta^k$), obtained by flipping the directions of $\eta^{-k}$ on $L_2$ and $T$ on $t>0$.}
	\label{fig: alt T-T fusion}
\end{figure}
Flipping the direction of $\eta^k$ corresponds to taking its inverse, i.e. $S_{\eta^{-k}}(L_2) = S_{\eta^k}(-L_2)$, because
\begin{equation}
\begin{aligned}
\begin{tikzpicture}[baseline=($(G1.base)!.5!(G2.base)$)]
    \begin{scope}[very thick,decoration={
            markings,
            mark=at position 0.5 with {\arrow{>}}}
        ] 
        \draw[postaction={decorate}] (1,0)--(0,1) node[at start,right] (G1) {$\eta^{-k}$} node[midway, left] () {$\phi_2~$} node[midway, right] () {$~\phi_1$} node[at end] (G2) {};
        \filldraw[blue] (0,1) circle (2pt) node[above,blue] () {$0$};
    \end{scope}
\end{tikzpicture} = {i\over 2\pi}\int_{L_2} \dr\varphi \left(\phi_2 - \phi_1 + {2\pi \eta\over N}\right) = {i\over 2\pi}\int_{-L_2} \dr\varphi\left(\phi_1 - \phi_2 - {2\pi\eta\over N} \right) = \begin{tikzpicture}[baseline=($(G1.base)!.5!(G2.base)$)]
    \begin{scope}[very thick,decoration={
            markings,
            mark=at position 0.5 with {\arrow{>}}}
        ] 
        \draw[postaction={decorate}] (0,1)--(1,0) node[at end,right] (G1) {$\eta^k$} node[midway, left] () {$\phi_2~$} node[midway, right] () {$~\phi_1$} node[at start] (G2) {};
        \filldraw[blue] (0,1) circle (2pt) node[above,blue] () {$0$};
    \end{scope}
\end{tikzpicture}
\end{aligned} \label{eq: ah}
\end{equation}
However, flipping the direction of the $T$-defect requires a boundary term since
\begin{equation}
\begin{aligned}
\begin{tikzpicture}[baseline=($(G1.base)!.5!(G2.base)$)]
    \begin{scope}[very thick,decoration={
            markings,
            mark=at position 0.5 with {\arrow{>}}}
        ] 
        \draw[postaction={decorate}] (0,-0.5)--(0,0.5) node[at start,right] (G1) {} node[midway, right] () {$T$} node[at end] (G2) {};
        \filldraw[blue] (0,-0.5) circle (2pt) node[below,blue] () {$0$};
%        \filldraw[blue] (0,0.5) circle (2pt) node[above,blue] () {$L$};
        \node at (0.7,-0.2) () {$\phi_1$};
        \node at (-0.5,-0.2) () {$\phi_3$};
    \end{scope}
\end{tikzpicture}
 &= {iN\over 2\pi}\int^\infty_0 \phi_3 \dr\phi_1={iN \over 2\pi}\int^0_\infty \phi_1 \dr\phi_3  - \left. {iN\over 2\pi}\phi_1\phi_3 \right|^0_{\infty}  =  \begin{tikzpicture}[baseline=($(G1.base)!.5!(G2.base)$)]
    \begin{scope}[very thick,decoration={
            markings,
            mark=at position 0.5 with {\arrow{>}}}
        ] 
        \draw[postaction={decorate}] (0,0.5)--(0,-0.5) node[at end,right] (G1) {} node[midway, right] () {$T$} node[at start] (G2) {};
        \filldraw[blue] (0,-0.5) circle (2pt) node[below,blue] () {$0$};
        %\filldraw[blue] (0,0.5) circle (2pt) node[above,blue] () {$L$};
        \node at (0.7,-0.2) () {$\phi_1$};
        \node at (-0.5,-0.2) () {$\phi_3$};
    \end{scope}
\end{tikzpicture} - \left. {iN\over 2\pi}\phi_1\phi_3 \right|^0_{\infty}\,.
\end{aligned} \label{eq: ai}
\end{equation}
Combining the results from $T\times \eta^{-k}= T$~\eqref{eq: aj} with a suitable relabeling of the fields and the extra boundary term~\eqref{eq: ai} needed for flipping the direction of $T$, one reproduces the boundary data~\eqref{eq: ak}.

\begin{figure}[H]
	\centering
	\begin{tikzpicture}
		\begin{scope}[very thick,decoration={
				markings,
				mark=at position 0.5 with {\arrow{>}}}
			] 
			\draw[postaction={decorate}] (-1.73,-2)--(0,-1) node[at start,left] (l) {$\eta^k,\varphi$} node[midway,left] () {$L_1~$};
			\draw[postaction={decorate}] (1.73,-2)--(0,-1) node[at start,right] () {$T$} node[midway,right] () {$~L_2$};
			\draw[postaction={decorate}] (0,-1)--(0,1) node[at end,right] () {$T$};
			\node at (-1,-0.5) () {$\phi_1$};
			\node at (1,-0.5) () {$\phi_2$};
			\node at (0,-1.5) () {$\phi_3$};
			\node[right] at (0,-0.9) () {$0$};
			\filldraw[blue] (0,-1) circle (2pt) node[right] () {};
            \node at (3,0) (f) {};
		\end{scope}
		\begin{scope}[shift={($(f.east)+(3,0)$)}, very thick,decoration={
				markings,
				mark=at position 0.5 with {\arrow{>}}}
			] 
	\draw[postaction={decorate}] (-1.73,-2)--(0,-1) node[at start,left] (l) {$\eta^m,\varphi_1$} node[midway,left] () {$L_1~$};
			\draw[postaction={decorate}] (1.73,-2)--(0,-1) node[at start,right] () {$\eta^n,\varphi_2$} node[midway,right] () {$~L_2$};
			\draw[postaction={decorate}] (0,-1)--(0,1) node[at end,right] () {$\eta^{m+n},\varphi$};
			\node at (-1,-0.5) () {$\phi_1$};
			\node at (1,-0.5) () {$\phi_2$};
			\node at (0,-1.5) () {$\phi_3$};
			\node[right] at (0,-0.9) () {$0$};
			\filldraw[blue] (0,-1) circle (2pt) node[right] () {};
		\end{scope}
	\end{tikzpicture}
	\caption{The local fusion junctions Hom($\eta^k\times T,T$) (left) and Hom($\eta^m\times \eta^n,\eta^{m+n}$) (right).}
	\label{fig: other fusion junctions}
\end{figure}
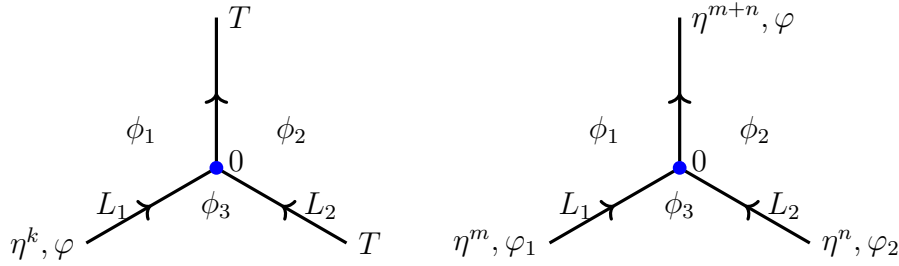

\paragraph{$\eta^k\times T = T$}
Consider fusing the upper part of the two defects $\eta^k$ and $T$ to obtain the local fusion junction in Figure~\ref{fig: other fusion junctions} (left).  
The defect action in the fused part at $t>0$ is given by
\begin{equation}
\begin{aligned}
S =& {i\over 2\pi}\int_{L_1} \dr\varphi \left( \phi_1- \phi_3 + {2\pi k \over N} \right) + {iN\over 2\pi}\int_{L_2}\phi_3 \dr\phi_2  \\ 
& + {i\over 2\pi}\int_{t>0} \left[N \phi_1 \dr\phi_2 + \left(\phi_1-\phi_3+{2\pi k \over N}\right)\dr \left(\varphi - N \phi_2\right) \right] + \left. ik \phi_2\right|^\infty_0
\end{aligned}
\end{equation}
where the first term on the second line corresponds to the $T$-duality defect as expected from the fusion outcome, the second term is a trivial decoupled TQFT and the last term is a boundary term.  The trivial TQFT in the current form is such that its Dirichlet boundary conditions are consistent with the relation between the bulk fields across the $\eta^k$ defect, $\left.\phi_1-\phi_3+{2\pi k \over N}\right|_{L_1} =0$ mod $2\pi$.  The Dirichlet boundary conditions and the boundary term at this junction can be summarized as
\begin{equation}
\begin{aligned}
    \text{Hom}(\eta^k\times T, T): \quad &\left.  \phi_1-\phi_3+{2\pi k \over N}\right|_{t=0} = 0\,, \quad \left. \varphi-N\phi_2\right|_{t=0} = 0 \quad \text{mod $2\pi$}\\
    &\quad S_{bdry} = -ik\phi_2(0)\,.
\end{aligned} \label{eq: eta-T=T}
\end{equation}

\paragraph{$\eta^m\times \eta^n = \eta^{m+n}$} Following the same steps for the local fusion junction $\eta^m\times \eta^n = \eta^{m+n}$ in Figure~\ref{fig: other fusion junctions} (right), the action along the fused segment takes the form 
\begin{equation}
\begin{aligned}
S=& {i\over 2\pi} \int_{t>0} \left[ \dr\varphi_1 \left( \phi_1 - \phi_3 + {2\pi m \over N}  \right) + \dr\varphi_2 \left( \phi_3- \phi_2 + {2\pi n \over N} \right) \right]\\
=& {i\over 2\pi} \int_{t>0} \left[ \dr\varphi_1 \left( \phi_1 - \phi_2 +{2\pi(m+n) \over N} \right) + \left( \phi_3 - \phi_2 + {2\pi n \over N} \right)\dr(\varphi_2 - \varphi_1) \right]\,.
\end{aligned}
\end{equation}
The first term on the second line is the action for the expected $\eta^{m+n}$ defect and the second term is a trivial decoupled TQFT, whose Dirichlet boundary condition is consistent with the relation between the bulk fields on the defect.  Hence, there is no boundary term but the boundary condition is given by
\begin{equation}
\begin{aligned}
    \text{Hom}(\eta^m\times \eta^n, \eta^{m+n}): \quad & \left. \phi_3-\phi_2+{2\pi n \over N}\right|_{t=0}=0\,, \quad \left. \varphi \right|_{t=0} = \left. \varphi_2 \right|_{t=0}= \left. \varphi_1 \right|_{t=0} \quad \text{mod $2\pi$}\\
    &\quad S_{bdry} = 0\,.
\end{aligned}
\end{equation}

% ------------------ associator section ------------

\subsection{$F$-symbols} \label{app: F of compact bosons}
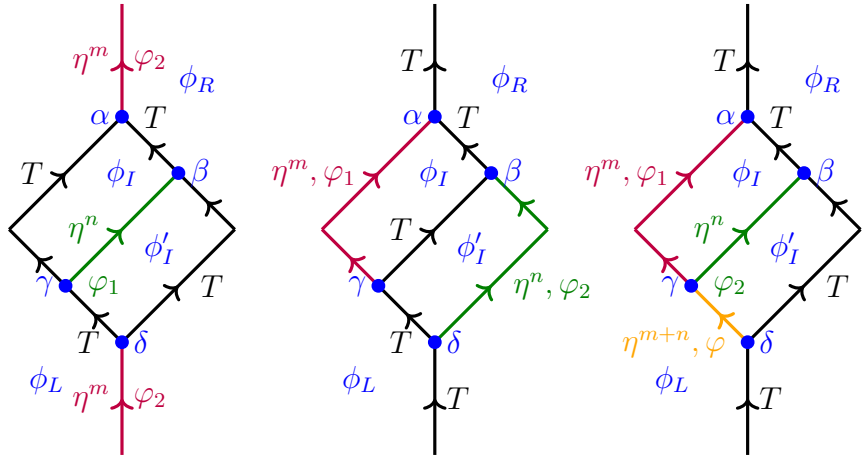
\begin{figure}[H]
	\centering
	\begin{tikzpicture}
		\begin{scope}[very thick,decoration={
				markings,
				mark=at position 0.5 with {\arrow{>}}}
			]
            \draw[postaction={decorate},purple] (0,1.5)--(0,3) node[midway,left] () {$\eta^m$} node[midway,right] () {$\varphi_2$}; %a(bc)
            \draw[postaction={decorate}] (-1.5,0)--(0,1.5) node[midway,left] () {$T~$}; %a
            \draw[postaction={decorate}] (0.75,0.75)--(0,1.5) node[at end,right] () {$~T$}; %bc
            \draw[postaction={decorate}] (1.5,0)--(0.75,0.75); 
            \draw[postaction={decorate},Green] (-0.75,-0.75)--(0.75,0.75) node[midway,left] () {$\eta^n~$} node[at start,right] () {$~\varphi_1$}; %b
            \draw[postaction={decorate}] (0,-1.5)--(1.5,0) node[midway,right] () {$~T$}; %c
            \draw[postaction={decorate}] (0,-1.5)--(-0.75,-0.75) node[at start,left] () {$T~$}; %ab
            \draw[postaction={decorate}] (-0.75,-0.75)--(-1.5,0);
            \draw[postaction={decorate},purple] (0,-3)--(0,-1.5) node[midway,left] () {$\eta^m$} node[midway,right] () {$\varphi_2$}; %(ab)c
            			
	\node[blue] at (-1,-2) () {$\phi_L$};
        \node[blue] at (1,2) () {$\phi_R$};
        \node[blue] at (0.5,-0.25) () {$\phi_I'$};
        \node[blue] at (0,0.75) () {$\phi_I$};
			
            \filldraw[blue] (0,1.5) circle (2pt) node[left,blue] () {$\alpha$};
            \filldraw[blue] (0.75,0.75) circle (2pt) node[right,blue] () {$\beta$};
            \filldraw[blue] (-0.75,-0.75) circle (2pt) node[left,blue] () {$\gamma$};
            \filldraw[blue] (0,-1.5) circle (2pt) node[right,blue] () {$\delta$};
        \node at (2,0) (f) {};
        \end{scope}
        \begin{scope}[shift={($(f.east)+(2,0)$)},very thick,decoration={
				markings,
				mark=at position 0.5 with {\arrow{>}}}
			]
            \draw[postaction={decorate}] (0,1.5)--(0,3) node[midway,left] () {$T$}; %a(bc)
            \draw[postaction={decorate},purple] (-1.5,0)--(0,1.5) node[midway,left] () {$\eta^m,\varphi_1~$}; %a
            \draw[postaction={decorate}] (0.75,0.75)--(0,1.5) node[at end,right] () {$~T$}; %bc
            \draw[postaction={decorate},Green] (1.5,0)--(0.75,0.75); 
            \draw[postaction={decorate}] (-0.75,-0.75)--(0.75,0.75) node[midway,left] () {$T~$}; %b
            \draw[postaction={decorate},Green] (0,-1.5)--(1.5,0) node[midway,right] () {$~\eta^n,\varphi_2$}; %c
            \draw[postaction={decorate}] (0,-1.5)--(-0.75,-0.75) node[at start,left] () {$T~$}; %ab
            \draw[postaction={decorate},purple] (-0.75,-0.75)--(-1.5,0);
            \draw[postaction={decorate}] (0,-3)--(0,-1.5) node[midway,right] () {$T$}; %(ab)c
            			
	\node[blue] at (-1,-2) () {$\phi_L$};
        \node[blue] at (1,2) () {$\phi_R$};
        \node[blue] at (0.5,-0.25) () {$\phi_I'$};
        \node[blue] at (0,0.75) () {$\phi_I$};
			
            \filldraw[blue] (0,1.5) circle (2pt) node[left,blue] () {$\alpha$};
            \filldraw[blue] (0.75,0.75) circle (2pt) node[right,blue] () {$\beta$};
            \filldraw[blue] (-0.75,-0.75) circle (2pt) node[left,blue] () {$\gamma$};
            \filldraw[blue] (0,-1.5) circle (2pt) node[right,blue] () {$\delta$};
        \node at (2,0) (ff) {};
        \end{scope}
        \begin{scope}[shift={($(ff.east)+(2,0)$)},very thick,decoration={
				markings,
				mark=at position 0.5 with {\arrow{>}}}
			]
            \draw[postaction={decorate}] (0,1.5)--(0,3) node[midway,left] () {$T$}; %a(bc)
            \draw[postaction={decorate},purple] (-1.5,0)--(0,1.5) node[midway,left] () {$\eta^m,\varphi_1~$}; %a
            \draw[postaction={decorate}] (0.75,0.75)--(0,1.5) node[at end,right] () {$~T$}; %bc
            \draw[postaction={decorate}] (1.5,0)--(0.75,0.75); 
            \draw[postaction={decorate},Green] (-0.75,-0.75)--(0.75,0.75) node[midway,left] () {$\eta^n~$} node[at start, right] () {$~\varphi_2$}; %b
            \draw[postaction={decorate}] (0,-1.5)--(1.5,0) node[midway,right] () {$~T$}; %c
            \draw[postaction={decorate},Orange] (0,-1.5)--(-0.75,-0.75) node[at start,left] () {$\eta^{m+n},\varphi~$}; %ab
            \draw[postaction={decorate},purple] (-0.75,-0.75)--(-1.5,0);
            \draw[postaction={decorate}] (0,-3)--(0,-1.5) node[midway,right] () {$T$}; %(ab)c
            			
	\node[blue] at (-1,-2) () {$\phi_L$};
        \node[blue] at (1,2) () {$\phi_R$};
        \node[blue] at (0.5,-0.25) () {$\phi_I'$};
        \node[blue] at (0,0.75) () {$\phi_I$};
			
            \filldraw[blue] (0,1.5) circle (2pt) node[left,blue] () {$\alpha$};
            \filldraw[blue] (0.75,0.75) circle (2pt) node[right,blue] () {$\beta$};
            \filldraw[blue] (-0.75,-0.75) circle (2pt) node[left,blue] () {$\gamma$};
            \filldraw[blue] (0,-1.5) circle (2pt) node[right,blue] () {$\delta$};

        \node at (2,0) (fff) {};
        \end{scope}
        
	\end{tikzpicture}
	\caption{The diagrams for $F^{\eta^m}_{T,\eta^n,T},F^T_{\eta^m,T,\eta^n}$, and $F^T_{\eta^m,\eta^n,T}$.}
	\label{fig: F-symbols first three}
\end{figure}

In this subsection, we compute all the $F$-symbols in the 2d compact boson theory except for $F^T_{T,T,T}$.  In Figure~\ref{fig: F-symbols first three}, we collected the diagrams for the first three $F$-symbols.

\paragraph{$F^{\eta^m}_{T,\eta^n,T}$}
Using the results from the previous subsection, the boundary conditions and boundary terms at each junction are
\begin{equation}
    \begin{aligned}
        \alpha: \quad & \varphi_2 - N\phi_I =0\,,\quad \phi_L-\phi_R+{2\pi m\over N} = 0\,, \quad S(\alpha) = -{iN\over 2\pi}\phi_I\left( \phi_R - {2\pi m \over N} \right), \\
        \beta: \quad & \phi_I - \phi_I' +{2\pi n\over N} = 0\,, \quad \varphi_1 - N\phi_R =0\,, \quad S(\beta) = -in \phi_R(\beta), \\
        \gamma: \quad & \phi_I - \phi_I' + {2\pi n \over N} = 0\,, \quad \varphi_1 - N\phi_L = 0\,, \quad S(\gamma) = {iN\over 2\pi}\phi_L\left( \phi_I - \phi_I' +{2\pi n \over N} \right), \\
        \delta: \quad & \varphi_2 - N\phi_I' =0\,,\quad \phi_L-\phi_R+{2\pi m\over N} = 0\,, \quad S(\delta) = {iN\over 2\pi}\phi_I'\left( \phi_R + {2\pi\eta_2\over N} \right).
    \end{aligned} \label{eq: za}
\end{equation}
The result of the global fusion of the three defects, $T,\eta^n,T$ in the middle is
\begin{equation}
    \begin{aligned}
    S=&{iN\over 2\pi} \int^\beta_\gamma \dr\phi_I\left( \phi_L-\phi_R + {2\pi m \over N} \right) + {i\over 2\pi} \int^\beta_\gamma \left( \phi_I- \phi_I' + {2\pi n \over N} \right)\dr \left(\varphi_1 - N\phi_R \right)\\
& + \underbrace{\left[ -im \phi_I+ {iN\over 2\pi} \phi_R\phi_I+ in \phi_R \right]^\beta_\gamma}_{S_{bdry;global}}
    \end{aligned}
\end{equation}
where the first term is the expected $\eta^m$ defect, the second term is a trivial decoupled TQFT, and the last term is a collection of boundary terms.  Both decoupled TQFT fields $\varphi_1-N\phi_R$ and $\phi_I-\phi_I'+{2\pi n \over N}$ have trivial Dirichlet boundary conditions from the boundary conditions~\eqref{eq: za}, and hence the decoupled TQFT does not contribute to the $F$-symbol after reducing the interval.  Adding the boundary terms at the upper junction,
\begin{equation}
    S(\alpha) + S(\beta) + \left.S_{bdry;global}\right|_\beta = 0
\end{equation}
and at the lower junction,
\begin{equation}
\begin{aligned}
    S(\gamma) + S(\delta) + \left. S_{bdry;global}\right|_\delta &= {iN\over 2\pi}\left( \phi_L - \phi_R +{2\pi m \over N} \right)\left(\phi_I -\phi_I'+{2\pi n \over N}\right) - {2\pi i mn\over N} \\
    &= -{2\pi i mn\over N} \quad \text{mod $2\pi$}\,.
\end{aligned}
\end{equation}
Hence, the $F$-symbol is
\begin{equation}
    F^{\eta^m}_{T,\eta^n,T} = \exp\left({{2\pi i mn\over N}}\right)\,.
\end{equation}

\paragraph{$F^T_{\eta^m,T,\eta^n}$} Using the results from the previous subsection, the boundary conditions (modulo $2\pi$) and the boundary terms at each junction are
\begin{equation}
\begin{aligned}
\alpha: \quad & \phi_L - \phi_I + {2\pi m \over N}=0, \quad \varphi_1 - N\phi_R =0, \quad S(\alpha) = -im\phi_R(\alpha),\\
\beta: \quad & \varphi_2 - N\phi_I = 0, \quad \phi_I' - \phi_R + {2\pi n \over N} =0, \quad S(\beta) = -{iN\over 2\pi}\phi_I\left( \phi_I'-\phi_R +{2\pi n \over N} \right),\\
\gamma: \quad & \phi_L - \phi_I + {2\pi m \over N} =0, \quad \varphi_1 - N\phi_I' =0, \quad S(\gamma) = im \phi_I'(\gamma),\\
\delta: \quad & \varphi_2 -N\phi_I = 0, \quad \phi_I'-\phi_R + {2\pi n \over N} =0, \quad S(\delta) = {iN\over 2\pi}\phi_L\left( \phi_I'-\phi_R + {2\pi n \over N} \right)
\end{aligned}\label{eq: zb}
\end{equation}
and the result of the global fusion of $\eta^m,T,\eta^n$ in the middle is
\begin{equation}
    \begin{aligned}
        S=& {iN \over 2\pi}\int^\beta_\gamma \phi_L\dr\phi_R +\left[ {i\over 2\pi}\left( \phi_I' - \phi_R + {2\pi n \over N} \right)\varphi_2 + im \phi_I' \right]^\beta_\gamma \\
&+{i\over 2\pi}\int^\beta_\gamma \left( \phi_L- \phi_I+{2\pi m\over N} \right) \dr (\varphi_1 - N\phi_I') - {i\over 2\pi}\int^\beta_\gamma \left( \varphi_2 - N\phi_L \right)\dr\left( \phi_I' - \phi_R + {2\pi n \over N} \right) .
    \end{aligned}
\end{equation}
The trivial TQFT fields have trivial Dirichlet boundary conditions and do not contribute to the $F$-symbol.  Adding the boundary terms at the upper junction,
\begin{equation}
\begin{aligned}
    S(\alpha) + S(\beta) + \left.S_{bdry;global}\right|_\beta &= {i \over 2\pi}(\varphi_2 - N\phi_I)\left( \phi_I'-\phi_R + {2\pi n \over N} \right) +  im\left( \phi_I' - \phi_R  \right)  \\
    &= -{2\pi i mn\over N} 
\end{aligned}
\end{equation}
and at the lower junction
\begin{equation}
\begin{aligned}
    S(\gamma)+S(\delta) + \left. S_{bdry;global}\right|_\delta &= {i\over 2\pi}\left( \phi_I' - \phi_R + {2\pi n \over N} \right)(N\phi_L - \varphi_2) = 0\,.
\end{aligned}
\end{equation}
Hence,
\begin{equation}
    F_{\eta^m,T,\eta^n}^T = \exp\left({2\pi i mn\over N} \right) ~.
\end{equation}

\paragraph{$F^T_{\eta^m,\eta^n,T}$} The boundary conditions and boundary terms at each junction are
\begin{equation}
    \begin{aligned}
        \alpha: &\quad \phi_L - \phi_I + {2\pi m \over N} =0, \quad \varphi_1 - N\phi_R =0, \quad S(\alpha) = -im \phi_R(\alpha) ~, \\
        \beta: &\quad \phi_I-\phi_I' + {2\pi n \over N} =0, \quad \varphi_2 - N\phi_R =0, \quad S(\beta) = -in \phi_R(\beta) ~, \\
        \delta: &\quad \phi_L - \phi_I' + {2\pi (m+n) \over N} =0, \quad \varphi - N\phi_R =0, \quad S(\delta) = i(m+n)\phi_R(\delta) ~, \\
        \gamma: &\quad \varphi = \varphi_1=\varphi_2 , \quad \phi_I - \phi_I' + {2\pi n \over N}=0, \quad S(\gamma) = 0 ~.
    \end{aligned}
\end{equation}
The global fusion of $\eta^m,\eta^n,T$ in the middle is
\begin{equation}
    \begin{aligned}
        S=& {iN\over 2\pi}\int^\beta_\gamma \phi_L \dr\phi_R + \left. i(m+n)\phi_R\right|^\beta_\gamma \\
        & + {i\over 2\pi}\int^\beta_\gamma \left( \phi_L - \phi_I +{2\pi m \over N} \right)\dr(\varphi_1-\varphi_2) +\left(\phi_L - \phi_I'+{2\pi (m+n)\over N} \right)\dr(\varphi_2-N\phi_R) ~,
    \end{aligned}
\end{equation}
and the trivial decoupled TQFT fields have the trivial Dirichlet boundary conditions. The sums of the boundary terms at the upper and lower junctions vanish, and thus the $F$-symbol is trivial,
\begin{equation}
    F^{T}_{\eta^m,\eta^n,T} = 1\,.
\end{equation}

\begin{figure}[H]
	\centering
	\begin{tikzpicture}
    \begin{scope}[very thick,decoration={
				markings,
				mark=at position 0.5 with {\arrow{>}}}
			]
            \draw[postaction={decorate}] (0,1.5)--(0,3) node[midway,left] () {$T$}; %a(bc)
            \draw[postaction={decorate}] (-1.5,0)--(0,1.5) node[midway,left] () {$T~$}; %a
            \draw[postaction={decorate},Orange] (0.75,0.75)--(0,1.5) node[at end,right] () {$~\eta^{m+n},\varphi$}; %bc
            \draw[postaction={decorate},Green] (1.5,0)--(0.75,0.75); 
            \draw[postaction={decorate},purple] (-0.75,-0.75)--(0.75,0.75) node[midway,left] () {$\eta^m~$} node[at start, right] () {$~\varphi_1$}; %b
            \draw[postaction={decorate},Green] (0,-1.5)--(1.5,0) node[midway,right] () {$~\eta^n,\varphi_2$}; %c
            \draw[postaction={decorate}] (0,-1.5)--(-0.75,-0.75) node[at start,left] () {$T~$}; %ab
            \draw[postaction={decorate}] (-0.75,-0.75)--(-1.5,0);
            \draw[postaction={decorate}] (0,-3)--(0,-1.5) node[midway,right] () {$T$}; %(ab)c
            			
	\node[blue] at (-1,-2) () {$\phi_L$};
        \node[blue] at (1,2) () {$\phi_R$};
        \node[blue] at (0.5,-0.25) () {$\phi_I'$};
        \node[blue] at (0,0.75) () {$\phi_I$};
			
            \filldraw[blue] (0,1.5) circle (2pt) node[left,blue] () {$\alpha$};
            \filldraw[blue] (0.75,0.75) circle (2pt) node[right,blue] () {$\beta$};
            \filldraw[blue] (-0.75,-0.75) circle (2pt) node[left,blue] () {$\gamma$};
            \filldraw[blue] (0,-1.5) circle (2pt) node[right,blue] () {$\delta$};

            \node at (2,0) (fff) {};
        \end{scope}
		\begin{scope}[shift={($(fff.east)+(2,0)$)},very thick,decoration={
				markings,
				mark=at position 0.5 with {\arrow{>}}}
			]
            \draw[postaction={decorate},purple] (0,1.5)--(0,3) node[midway,left] () {$\eta^m$} node[midway,right] () {$\varphi_1$}; %a(bc)
            \draw[postaction={decorate}] (-1.5,0)--(0,1.5) node[midway,left] () {$T~$}; %a
            \draw[postaction={decorate}] (0.75,0.75)--(0,1.5) node[at end,right] () {$~T$}; %bc
            \draw[postaction={decorate},Green] (1.5,0)--(0.75,0.75); 
            \draw[postaction={decorate}] (-0.75,-0.75)--(0.75,0.75) node[midway,left] () {$T~$}; %b
            \draw[postaction={decorate},Green] (0,-1.5)--(1.5,0) node[midway,right] () {$~\eta^n,\varphi_2$}; %c
            \draw[postaction={decorate},Orange] (0,-1.5)--(-0.75,-0.75) node[at start,left] () {$\eta^{m-n},\varphi~$}; %ab
            \draw[postaction={decorate}] (-0.75,-0.75)--(-1.5,0);
            \draw[postaction={decorate},purple] (0,-3)--(0,-1.5) node[midway,left] () {$\eta^m$} node[midway,right] () {$\varphi_1$}; %(ab)c
            			
	\node[blue] at (-1,-2) () {$\phi_L$};
        \node[blue] at (1,2) () {$\phi_R$};
        \node[blue] at (0.5,-0.25) () {$\phi_I'$};
        \node[blue] at (0,0.75) () {$\phi_I$};
			
            \filldraw[blue] (0,1.5) circle (2pt) node[left,blue] () {$\alpha$};
            \filldraw[blue] (0.75,0.75) circle (2pt) node[right,blue] () {$\beta$};
            \filldraw[blue] (-0.75,-0.75) circle (2pt) node[left,blue] () {$\gamma$};
            \filldraw[blue] (0,-1.5) circle (2pt) node[right,blue] () {$\delta$};
        \node at (2,0) (f) {};
        \end{scope}
        \begin{scope}[shift={($(f.east)+(2,0)$)},very thick,decoration={
				markings,
				mark=at position 0.5 with {\arrow{>}}}
			]
            \draw[postaction={decorate},purple] (0,1.5)--(0,3) node[midway,left] () {$\eta^m$} node[midway,right] () {$\varphi_1$}; %a(bc)
            \draw[postaction={decorate},Green] (-1.5,0)--(0,1.5) node[midway,left] () {$\eta^n,\varphi_2~$}; %a
            \draw[postaction={decorate},Orange] (0.75,0.75)--(0,1.5) node[at end,right] () {$~\eta^{m-n},\varphi$}; %bc
            \draw[postaction={decorate}] (1.5,0)--(0.75,0.75); 
            \draw[postaction={decorate}] (-0.75,-0.75)--(0.75,0.75) node[midway,left] () {$T~$}; %b
            \draw[postaction={decorate}] (0,-1.5)--(1.5,0) node[midway,right] () {$~T$}; %c
            \draw[postaction={decorate}] (0,-1.5)--(-0.75,-0.75) node[at start,left] () {$T~$}; %ab
            \draw[postaction={decorate},Green] (-0.75,-0.75)--(-1.5,0);
            \draw[postaction={decorate},purple] (0,-3)--(0,-1.5) node[midway,left] () {$\eta^m$} node[midway,right] () {$\varphi_1$}; %(ab)c
            			
	\node[blue] at (-1,-2) () {$\phi_L$};
        \node[blue] at (1,2) () {$\phi_R$};
        \node[blue] at (0.5,-0.25) () {$\phi_I'$};
        \node[blue] at (0,0.75) () {$\phi_I$};
			
            \filldraw[blue] (0,1.5) circle (2pt) node[left,blue] () {$\alpha$};
            \filldraw[blue] (0.75,0.75) circle (2pt) node[right,blue] () {$\beta$};
            \filldraw[blue] (-0.75,-0.75) circle (2pt) node[left,blue] () {$\gamma$};
            \filldraw[blue] (0,-1.5) circle (2pt) node[right,blue] () {$\delta$};
        \node at (2,0) (ff) {};
        \end{scope}     
	\end{tikzpicture}
	\caption{The diagrams for $F^T_{T,\eta^m,\eta^n},F^{\eta^m}_{T,T,\eta^n}$ and $F^{\eta^m}_{\eta^n,T,T}$.}
	\label{fig: F-symbols first four}
\end{figure}
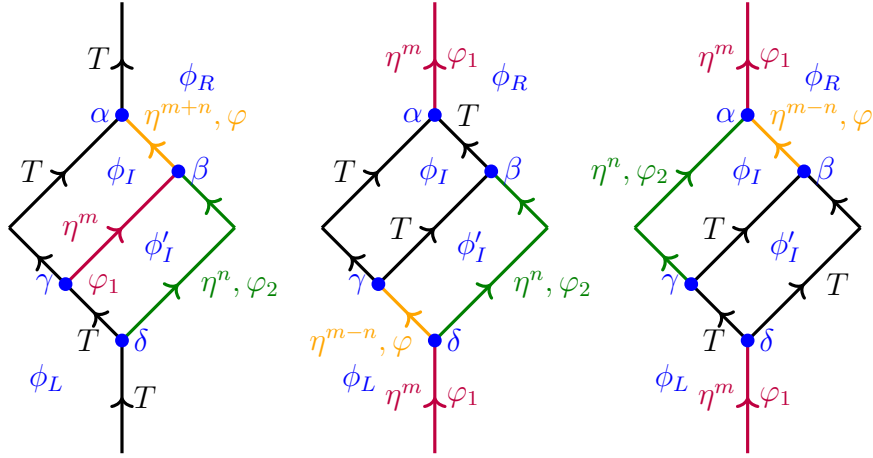

\paragraph{$F^T_{T,\eta^m,\eta^n}$} The boundary conditions (mod $2\pi$) and boundary terms at each junction are
\begin{equation}
    \begin{aligned}
        \alpha: &\quad \phi_I - \phi_R +{2\pi(m+n)\over N} = 0, \quad \varphi-N\phi_L = 0, \quad S(\alpha) = -{iN\over 2\pi}\phi_L\left(\phi_I - \phi_R +{2\pi (m+n)\over N} \right) ~, \\
        \beta:&\quad \varphi = \varphi_1 = \varphi_2 , \quad \phi_I'-\phi_R + {2\pi n\over N} = 0, \quad S(\beta) = 0 ~,\\
        \gamma: &\quad \phi_I-\phi_I'+{2\pi m \over N}= 0, \quad \varphi_1-N\phi_L =0, \quad S(\gamma) = {iN\over 2\pi}\phi_L\left( \phi_I - \phi_I' +{2\pi m \over N} \right) ~, \\
        \delta: &\quad \phi_I'-\phi_R+{2\pi n\over N}=0,\quad \varphi_2-N\phi_L=0, \quad S(\delta) = {iN\over 2\pi}\phi_L\left( \phi_I' - \phi_R +{2\pi n \over N} \right) ~,
    \end{aligned}
\end{equation}
and the global fusion of the three defects $\eta^m,\eta^n,T$ is
\begin{equation}
    \begin{aligned}
        S = &{iN\over 2\pi}\int^\beta_\gamma \phi_L\dr\phi_R +  \left[ {i\over 2\pi}\varphi_1 \left(\phi_I-\phi_R+{2\pi(m+n) \over N} \right)  \right]^\beta_\gamma \\
        &\quad +{i\over 2\pi}\int^\beta_\gamma \left[(N\phi_L - \varphi_1)\dr(\phi_I-\phi_R+{2\pi (m+n)\over N}) + \left(\phi_I'-\phi_R-{2\pi\eta_2 \over N}\right) \dr(\varphi_2 -\varphi_1) \right]\,.
    \end{aligned}
\end{equation}
All trivial decoupled TQFT fields have the trivial Dirichlet boundary conditions and so the two decoupled TQFTs do not contribute to the $F$-symbol. The sum  of the boundary terms at the upper junction is given by
\begin{equation}
    S(\alpha)+S(\beta)+\left.S_{bdry;global}\right|_\beta = {i\over 2\pi}(\varphi_1-N\phi_L)\left(\phi_I-\phi_R+{2\pi(m+n) \over N} \right)  = 0\quad \text{mod $2\pi$}
\end{equation}
and at the lower junction by
\begin{equation}
    S(\gamma)+S(\delta)+\left. S_{bdry;global}\right|_\gamma = {i\over 2\pi}(N\phi_L - \varphi_1)\left(\phi_I-\phi_R+{2\pi(m+n) \over N} \right)  = 0\quad \text{mod $2\pi$}\,.
\end{equation}
Hence, the $F$-symbol is trivial,
\begin{equation}
    F^{T}_{T,\eta^m,\eta^n} = 1\,.
\end{equation}

\paragraph{$F^{\eta^m}_{T,T,\eta^n}$} The boundary conditions (mod $2\pi$) and boundary terms at each junction are
\begin{equation}
    \begin{aligned}
        \alpha: &\quad \varphi_1 - N\phi_I=0,\quad \phi_L-\phi_R+{2\pi m\over N} = 0, \quad S(\alpha) = -{iN\over 2\pi}\phi_I\left(\phi_R -{2\pi m \over N}\right), \\
        \beta: &\quad \varphi_2-N\phi_I =0, \quad \phi_I'-\phi_R +{2\pi n\over N} = 0, \quad S(\beta) = -{iN\over 2\pi}\phi_I\left(\phi_I'-\phi_R + {2\pi n\over N} \right),\\
        \gamma: &\quad \varphi - N\phi_I=0, \quad \phi_L-\phi_I'+{2\pi (m-n)\over N} = 0, \quad S(\gamma) = {iN\over 2\pi} \phi_I\left( \phi_I' -{2\pi(m-n)\over N} \right),\\
        \delta: &\quad \varphi=\varphi_1= \varphi_2, \quad \phi_I'-\phi_R + {2\pi n \over N}=0,\quad \phi_L-\phi_R+{2\pi m\over N} = 0, \quad S(\delta) = 0 ~,
    \end{aligned}
\end{equation}
and the global fusion of the three defects $T,T,\eta^n$ in the middle is
\begin{equation}
\begin{aligned}
    S =& {iN\over 2\pi}\int^\beta_\gamma \left(\phi_L - \phi_R +{2\pi m \over N}\right)\dr\phi_I  + {i\over 2\pi}\int^\beta_\gamma (\phi_I'-\phi_R+{2\pi n \over N})\dr(\varphi_2-N\phi_I) \\
    &+{iN\over 2\pi}\left[ \phi_I\left(\phi_I' - {2\pi(m-n)\over N}\right) \right]^\beta_\gamma \,.
\end{aligned}
\end{equation}
The trivial decoupled TQFT fields have trivial Dirichlet boundary conditions.  The sum of the boundary terms at the upper junction vanishes,
\begin{equation}
    S(\alpha) + S(\beta) + \left. S_{bdry;global}\right|_\beta = 0 ~,
\end{equation}
and the sum at the lower junction vanishes as well,
\begin{equation}
    S(\gamma) + S(\delta) + \left. S_{bdry;global}\right|_\gamma = 0\,.
\end{equation}
Hence, the $F$-symbol is trivial:
\begin{equation}
    F^{\eta^m}_{T,T,\eta^n} = 1\,.
\end{equation}

\paragraph{$F^{\eta^m}_{\eta^n,T,T}$} The boundary conditions (mod $2\pi$) and boundary terms at each junction are
\begin{equation}
    \begin{aligned}
        \alpha: &\quad \varphi_1=\varphi_2 = \varphi, \quad \phi_L-\phi_R+{2\pi m \over N} = 0, \quad \phi_L-\phi_I+{2\pi n\over N}=0, \quad S(\alpha) = 0,\\
        \beta: &\quad \varphi-N\phi_I' = 0, \quad \phi_I - \phi_R +{2\pi(m-n)\over N}=0, \quad S(\beta) = -{iN\over 2\pi}\phi_I'\left( \phi_R - {2\pi(m-n)\over N}\right),\\
        \gamma: &\quad \phi_L - \phi_I+{2\pi n\over N} = 0, \quad \varphi_2 - N\phi_I' = 0, \quad S(\gamma) = in\phi_I'(\gamma),\\
        \delta: &\quad \varphi_1-N\phi_I' = 0, \quad \phi_L-\phi_R+{2\pi m \over N} = 0, \quad S(\delta) = {iN\over 2\pi}\phi_I'\left( \phi_R - {2\pi m \over N}\right)
    \end{aligned}
\end{equation}
and the global fusion of the three defects $\eta^n,T,T$ in the middle is
\begin{equation}
    \begin{aligned}
        S=&{i\over 2\pi}\int^\beta_\gamma \dr \varphi_1 \left(\phi_L-\phi_R + {2\pi m\over N}\right)+ \left[ {iN\over 2\pi}\phi_I'\left(\phi_R -{2\pi (m-n)\over N}\right) \right]^\beta_\gamma \\
        &+ {i\over 2\pi}\int^\beta_\gamma \left(\phi_L-\phi_R + {2\pi m\over N}\right)\dr (N\phi_I'-\varphi_1)  +{i\over 2\pi}\int^\beta_\gamma \left(\phi_L -\phi_I + {2\pi n\over N} \right)\dr(\varphi_2-N\phi_I')\,.
    \end{aligned}
\end{equation}
All decoupled trivial TQFT fields have trivial Dirichlet boundary conditions, and the sum of the boundary terms at both the upper and lower junctions vanishes,
\begin{equation}
    S(\alpha) + S(\beta) + \left. S_{bdry;global}\right|_\beta = 0\,,\quad S(\gamma) + S(\delta) + \left. S_{bdry;global}\right|_\gamma = 0\,.
\end{equation}
Hence, the $F$-symbol is trivial,
\begin{equation}
    F^{\eta^m}_{\eta^n,T,T} = 1\,.
\end{equation}

% -------------- New Section ------------------
\section{Comments on alternative formalisms on the local junctions}\label{app:altj}
The topological boundary/interface of a topological defect (or a TFT) can be described either in terms of boundary conditions or by introducing boundary degrees of freedom. In Section \ref{sec: 4dtri}, we take the first approach, and in this Appendix, we want to present the alternative description commonly used in the literature \cite{Roumpedakis:2022aik,Bah:2025oxi} and discuss their relations in the concrete example of the 3d $\mathbb{Z}_N$ gauge theory written in terms of the CS action. Notice that there are no new results here, and we simply hope to clarify a few subtleties in various formalisms.

Let's first recall the topological boundary condition for $\mathbb{Z}_N$ gauge theory given by the CS action
\begin{equation}
    S = \frac{i N}{2\pi} \int_{V} a\wedge db ~.
\end{equation}
For generic $N > 1$, there always exist two boundary conditions 
\begin{equation}
\begin{aligned}
    \text{(I)} &: \quad a|_{\partial V} = d\phi_a ~, \quad Nb |_{\partial V} = d\phi_b ~, \\
    \text{(II)} &: Na|_{\partial V} = d\phi_a ~, \quad \quad b|_{\partial V} = d\phi_b ~,
\end{aligned}
\end{equation}
and we will restrict our discussion to these two for simplicity.

In the boundary condition (I), the gauge transformations are given by
\begin{equation}
    a \rightarrow a + d\lambda_a ~, \quad b\rightarrow b + d\lambda_b ~, \quad \phi_a \rightarrow \phi_a + \lambda_a ~, \quad \phi_b \rightarrow \phi_b + N\lambda_b ~.
\end{equation}
It is straightforward to check that the action is gauge invariant and is consistent with the variation principle. In this boundary condition, the $a$-Wilson line is allowed to terminate on the boundary, described by the gauge-invariant operator
\begin{equation}
    \exp\left(-i \phi_a (p) + i \int_\gamma a\right) ~, \quad \gamma \cap \partial V = \{p\} ~,
\end{equation}
where $\gamma$ is some path in the bulk $V$ that terminates at the boundary point $p \in \partial V$. On the boundary, the charge $N$ $b$-Wilson line is identified with the identity line as there exists a local junction between the two. In other words, the line $e^{i N \int b}$ can terminate on the vertex operator $e^{-i\phi_b}$ in a gauge-invariant way\footnote{Notice that similar story happens in the bulk, except the role of $e^{i\phi_b}$ is played by the monopole operator of $a$ \cite{Kapustin:2010hk}.}. Hence, $e^{i\oint_\gamma b}$ generates $\mathbb{Z}_N$ symmetry on the boundary for the boundary condition (I). Combined, we see the boundary condition (I) yields the correct operator spectrum for the $e^{i\oint a}$-condensed boundary. It is straightforward to repeat the same analysis for the boundary condition (II) and see it describes the $e^{i\oint b}$-condensed boundary correctly.

\

The same boundary conditions can also be described by introducing boundary scalars and coupling them to the bulk field, while leaving the bulk fields completely free in the action on the boundary. The boundary coupling as well as the gauge transformations are given by
\begin{equation}
\begin{aligned}
    \text{(I)} &: S_{\partial V} = \frac{iN}{2\pi} \int_{\partial V} \phi db ~, \\
    & \quad a\rightarrow a + d\lambda_a ~, \quad b\rightarrow b + d\lambda_b ~, \quad \phi \rightarrow \phi - \lambda_a ~, \\
    \text{(II)} &: S_{\partial V} = \frac{iN}{2\pi} \int_{\partial V} \phi da + a \wedge b ~, \\
    & \quad a\rightarrow a + d\lambda_a ~, \quad b\rightarrow b + d\lambda_b ~, \quad \phi \rightarrow \phi - \lambda_b ~.
\end{aligned}
\end{equation}
It is straightforward to check that both boundary conditions are well-defined and yield the correct operator spectrum. In the boundary condition (I), $e^{i\int a}$ can terminate on the $e^{-i\phi}$ in a gauge-invariant way on the boundary. At the same time, the equation of motion of $\phi$ implies that $e^{iN\oint b} = \mathbb{1}$ thus $e^{i\oint b}$ generates $\mathbb{Z}_N$ symmetry on the boundary\footnote{To see that $e^{iN\int b}$ can terminate on the boundary, we notice that there exists a boundary operator $\mathcal{O}(x)$, defined by removing the point $x$ from $\partial V$ and require the compact scalar $\phi$ to have unit winding around $x$ when inserting in the path integral. $e^{iN\int b}$ can terminate gauge-invariantly on $\mathcal{O}(x)$.}. Similarly, in the boundary condition (II), $e^{i\int b}$ can terminate on the boundary while $e^{i\oint a}$ generates $\mathbb{Z}_N$-symmetry instead.

The advantage of this formalism is that one can manifestly produce an action in terms of continuous fields in the interval reduction. For instance, let's consider placing the theory on $[0,1]_t \times M_2$. Let us first consider choosing boundary condition (I) for both $t = 0$ and $t = 1$ boundaries. In this case, the full action is given by
\begin{equation}
    \frac{i N}{2\pi} \int_{[0,1]_t \times M_2} a\wedge db - \frac{iN}{2\pi} \int_{M_2|_{t=0}} \phi \, db + \frac{iN}{2\pi} \int_{M_2|_{t=1}} \widetilde{\phi} \, db ~, 
\end{equation}
with the gauge transformation
\begin{equation}
    a \rightarrow a+d\lambda_a ~, \quad b\rightarrow b + d\lambda_b ~, \quad \phi \rightarrow \phi - \lambda_a ~, \quad \widetilde{\phi} \rightarrow \phi - \lambda_a ~.
\end{equation}
Reducing along the interval by fusing two boundaries leads to the 2d action
\begin{equation}
    \frac{iN}{2\pi} \int_{M_2} (\widetilde{\phi} - \phi) \, db \equiv \frac{iN}{2\pi} \int_{M_2} \Phi \, db ~,
\end{equation}
which becomes the standard 2d $\mathbb{Z}_N$ gauge theory upon introducing the gauge invariant compact scalar $\Phi = \widetilde{\phi} - \phi$, thus reproduces the well-known result.

On the other hand, one can also consider choosing boundary condition (I) for the $t = 0$ boundary and boundary condition (II) for the $t = 1$ boundary. In this case, the full action is given by
\begin{equation}
    \frac{i N}{2\pi} \int_{[0,1]_t \times M_2} a\wedge db - \frac{iN}{2\pi} \int_{M_2|_{t=0}} (\phi \, da + a \wedge b) + \frac{iN}{2\pi} \int_{M_2|_{t=1}} \widetilde{\phi}\, db ~, 
\end{equation}
with the gauge transformation
\begin{equation}\label{eq:DNgt}
    a \rightarrow a+d\lambda_a ~, \quad b\rightarrow b + d\lambda_b ~, \quad \phi \rightarrow \phi - \lambda_b ~, \quad \widetilde{\phi} \rightarrow \phi - \lambda_a ~.
\end{equation}
Upon the reduction, we find the 2d action
\begin{equation}
    \frac{iN}{2\pi} \int_{M_2} - a \wedge b - \phi \,da + \widetilde{\phi} \,db ~,
\end{equation}
with the gauge transformation \eqref{eq:DNgt}. Despite having a seemingly non-trivial action, the 2d theory is actually trivial as there are no gauge-invariant operators which do not commute upon canonical quantization, also matching the well-known result.

% -------------- New Section ------------------

\section{Triality fusion junctions and associators in 4d Maxwell theory}\label{app: triality}

\subsection{Triality fusion junctions}\label{app: triality junctions}
The triality and the condensation defects form a closed non-invertible fusion algebra.  In this section, we focus on the associator for the fusion of three triality defects (not of a condensation defect) for even $N$ and analyze the following relevant local fusion and morphism junctions,
\begin{equation}
    \begin{gathered}
C_0\to \mathbb{1}~\eqref{eq: bj}, \quad C_{N\over 2}\to \mathbb{1}~\eqref{eq: bp}, \quad U(1)_N\times U(1)_{-N}\to \bi~\eqref{eq: bi}, \\
D_3\times \bi=D_3~\eqref{eq: bo}, \quad \bi\times D_3 =D_3~\eqref{eq: bk}, \quad \bi\times \overline{D}_3=\overline{D}_3~\eqref{eq: wb}, \quad \overline{D}_3\times \bi =\overline{D}_3~\eqref{eq: wc}\\
D_3\times D_3=U(1)_N\overline{D}_3~\eqref{eq: bf}, \quad \overline{D}_3\times \overline{D}_3=U(1)_{-N}D_3~\eqref{eq: bff},\\
\overline{D}_3\times D_3=C_0~\eqref{eq: wd}, \quad D_3\times \overline{D}_3=C_{N\over 2}~\eqref{eq: we} \,.
    \end{gathered} \label{eq: wa}
\end{equation}
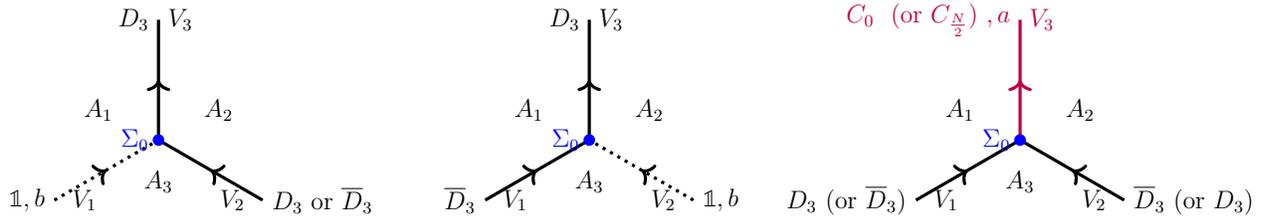
\begin{figure}[H]
	\centering
	\begin{tikzpicture}
    \begin{scope}[scale=0.8,every node/.style={scale=0.8}]
		\begin{scope}[very thick,decoration={
				markings,
				mark=at position 0.5 with {\arrow{>}}}
			] 
			\draw[postaction={decorate},dotted] (-1.73,-2)--(0,-1) node[at start,left] (l) {$\mathbb{1},b$} node[at start,right] () {$~V_1$};
			\draw[postaction={decorate}] (1.73,-2)--(0,-1) node[at start,right] () {${D}_3$ or $\overline{D}_3$} node[at start,left] () {$V_2~$};
			\draw[postaction={decorate}] (0,-1)--(0,1) node[at end,left] () {${D}_3$} node[at end,right] () {$V_3$};
			\node at (-1,-0.5) () {$A_1$};
			\node at (1,-0.5) () {$A_2$};
			\node at (0,-1.7) () {$A_3$};
        
			\filldraw[blue] (0,-1) circle (2pt) node[left,blue] () {$\Sigma_0$};
            \node at (3,0) (f) {};
		\end{scope}
        \begin{scope}[shift={($(f.east)+(4,0)$)},very thick,decoration={
				markings,
				mark=at position 0.5 with {\arrow{>}}}
			] 
			\draw[postaction={decorate}] (-1.73,-2)--(0,-1) node[at start,left] (l) {$\overline{D}_3$} node[at start,right] () {$~V_1$};
			\draw[postaction={decorate},dotted] (1.73,-2)--(0,-1) node[at start,right] () {$\bi,b$} node[at start,left] () {$V_2~$};
			\draw[postaction={decorate}] (0,-1)--(0,1) node[at end,left] () {${D}_3$} node[at end,right] () {$V_3$};
			\node at (-1,-0.5) () {$A_1$};
			\node at (1,-0.5) () {$A_2$};
			\node at (0,-1.7) () {$A_3$};

			\filldraw[blue] (0,-1) circle (2pt) node[left,blue] () {$\Sigma_0$};
            \node at (3,0) (ff) {};
		\end{scope}
        \begin{scope}[shift={($(ff.east)+(4,0)$)},very thick,decoration={
				markings,
				mark=at position 0.5 with {\arrow{>}}}
			] 
			\draw[postaction={decorate}] (-1.73,-2)--(0,-1) node[at start,left] (l) {$D_3$ (or $\overline{D}_3$)} node[at start,right] () {$~V_1$};
			\draw[postaction={decorate}] (1.73,-2)--(0,-1) node[at start,right] () {$\overline{D}_3$ (or $D_3$)} node[at start,left] () {$V_2~$};
			\draw[postaction={decorate},purple] (0,-1)--(0,1) node[at end,left] () {$C_0~~(\text{or } C_{N\over 2})~,a$} node[at end,right] () {$V_3$};
			\node at (-1,-0.5) () {$A_1$};
			\node at (1,-0.5) () {$A_2$};
			\node at (0,-1.7) () {$A_3$};

			\filldraw[blue] (0,-1) circle (2pt) node[left,blue] () {$\Sigma_0$};
		\end{scope}
        \end{scope}
	\end{tikzpicture}
	\caption{Fusion junctions on the second line of Eq.~\eqref{eq: wa}; $b$ is the defect field on $V_2$ for the identity defect.}
	\label{fig: triality fusion jns}
\end{figure}
\paragraph{$\bi \times D_3 = D_3$} Consider the local fusion junction for $\mathbb{1}\times D_3\to D_3$, the first diagram of Figure~\ref{fig: triality fusion jns}.  The action along the fused upper part $V_3$ is
\begin{equation}
\begin{aligned}
    S&= \frac{i}{2\pi } \int_{V_3} \left[ \dr b (A_1 - A_3) + N A_3 \dr A_2 + \frac{N}{2}A_3 \dr A_3 \right] \\
    &= \frac{iN}{2\pi} \int_{V_3} \left( A_1 \dr A_2 + \frac{1}{2}A_1\dr A_1  \right) +{iN\over4\pi}\int_{\Sigma_0} A_1\wedge A_3\\
		&\qquad +  \frac{i}{2\pi} \int_{V_3} (A_1 - A_3) \, \dr \left(b-\frac{N}{2}A_3 - \frac{N}{2}A_1-NA_2 \right)
\end{aligned}
\end{equation}
where the first term on the last line describes the triality defect $D_3$ as expected from the fusion algebra, second term is the boundary term and the last term is a trivial TQFT.  Since the fields $A_3$ and $b$ above $\Sigma_0$ are supported only along the $D_3$ defect, (by shifting $A_3\to A_3+A_1$ and $b\to b+{N\over 2}A_3 + {N\over 2} A_1 + NA_2$, the TQFT is decoupled from the bulk.  Thus, we impose the boundary conditions~\eqref{eq: decoupled TQFT bc}.  In summary,
\begin{equation}
\begin{aligned}
    \text{Hom}(\bi\times D_3, D_3): \quad &\left. A_1-A_3 ~, \quad  b-\frac{N}{2}A_3 - \frac{N}{2}A_1-NA_2\right|_{\Sigma_0} = \text{pure $U(1)$ gauge} ~, \\
    & S_{bdry} = {iN\over4\pi}\int_{\Sigma_0} A_1\wedge A_3 ~.
\end{aligned} \label{eq: bk}
\end{equation}

\paragraph{$\bi \times \overline{D}_3 = \overline{D}_3$} We again use the first diagram in Figure~\ref{fig: triality fusion jns} with the action along the $V_3$ segment,
\begin{equation}
    \begin{aligned}
        S &= {i\over 2\pi}\int_{V_3} \left[\dr b (A_1-A_3) - NA_3 \dr A_2 - {N\over 2}A_2\dr A_2 \right] \\
        &=\underbrace{-{iN\over 2\pi}\int_{V_3} \left(A_1 \dr A_2+{1\over 2}A_2\dr A_2\right)}_{\overline{D}_3} + \underbrace{{i\over 2\pi}\int_{V_3}(A_1-A_3)\dr (b+NA_2) }_{\text{Decoupled trivial TQFT}} ~.
    \end{aligned} 
\end{equation}
\begin{equation}
\text{Hom}(\bi \times \overline{D}_3, \overline{D}_3):\quad    \left. A_1-A_3, \quad  b+NA_2\right|_{\Sigma_0} = \text{pure $U(1)$ gauge}, \quad S_{bdry} = 0 ~. \label{eq: wb}
\end{equation}

\paragraph{$\overline{D}_3 \times \bi = \overline{D}_3$} This corresponds to the second diagram in Figure~\ref{fig: triality fusion jns} with the action along the $V_3$ segment given by 
\begin{equation}
    \begin{aligned}
S &= {i\over 2\pi}\int_{V_3} \left[-N A_1 \dr A_3 - {N\over 2}A_3 \dr A_3 + \dr b \wedge (A_3-A_2)\right]\\
&= \underbrace{-{iN\over 2\pi}\int_{V_3} \left(A_1 \dr A_2 + {1\over 2}A_2\dr A_2\right)}_{\overline{D}_3} + \underbrace{\frac{i}{2\pi} \int_{V_3} (A_3 - A_2) d \left[b-NA_1 -{N\over 2}(A_2+A_3) \right] }_{\text{Decoupled trivial TQFT}} \\
    & \quad \underbrace{-\frac{i}{2\pi} \int_{\Sigma_0} NA_1 \wedge (A_3 - A_2) + \frac{N}{2} A_2 \wedge A_3}_{S_{bdry}} ~.
    \end{aligned}
\end{equation}
Imposing the boundary conditions~\eqref{eq: decoupled TQFT bc} for the decoupled TQFT, we have
\begin{equation}
\begin{aligned}
\text{Hom}(\overline{D}_3 \times \bi , \overline{D}_3): \quad &\left. b-NA_1-{N\over 2}(A_2+A_3), \quad A_2-A_3\right|_{\Sigma_0} = \text{pure $U(1)$ gauge} ~, \\
&S_{bdry} = -\frac{i}{2\pi} \int_{\Sigma_0} NA_1 \wedge (A_3 - A_2) + \frac{N}{2} A_2 \wedge A_3 ~.
\end{aligned} \label{eq: wc}
\end{equation}

\paragraph{$\overline{D}_3\times D_3=C_0$} This junction corresponds to the third diagram of Figure~\ref{fig: triality fusion jns}. The action along the fused part $V_3$ is
\begin{equation}
    \begin{aligned}
        S &= {iN \over 2\pi}\int_{V_3} \left( -A_1 \dr A_3 - {1\over 2}A_3 \dr A_3 +A_3\dr A_2 + {1\over 2}A_3\dr A_3 \right)\\
        &= \underbrace{-{iN \over 2\pi}\int_{V_3}\dr A_3 \,(A_1-A_2)}_{C_0} - \underbrace{{iN\over 2\pi}\int_{\Sigma_0} A_2 \wedge A_3}_{S_{bdry}} ~.
    \end{aligned}
\end{equation}
No particular boundary condition is imposed, hence 
\begin{equation}
 \text{Hom}(\overline{D}_3\times D_3,C_0): \text{no b.c.}, \quad S_{bdry} = -{iN\over 2\pi}\int_{\Sigma_0} A_2 \wedge A_3 ~. \label{eq: wd}
\end{equation}
Attaching a morphism junction between $C_0$ and the identity (with the identity defect field being $b$) using Eq.~\eqref{eq: bj}, we get 
\begin{equation}
 \text{Hom}(\overline{D}_3\times D_3,\bi): \left. b+NA_3,\quad A_1-A_2\right|=\text{pure $U(1)$ gauge}, \quad S_{bdry} = -{iN\over 2\pi}\int_{\Sigma_0} A_2 \wedge A_3 ~. \label{eq: wda}
\end{equation}

\paragraph{$D_3\times \overline{D}_3 = C_{N\over 2}$} This junction corresponds to the third diagram of Figure~\ref{fig: triality fusion jns}, with the action along the fused segment $V_3$ given by
\begin{equation}
    \begin{aligned}
S &= {iN\over 2\pi}\int_{V_3} \left(A_1 \dr A_3 + {1\over 2}A_1\dr A_1  -A_3 \dr A_2- {1\over 2}A_2\dr A_2 \right)\\
&= \underbrace{{iN\over 2\pi}\int_{V_3} \left[ (A_1-A_2)\dr (A_3+A_2) +{1\over 2}(A_1-A_2)\dr(A_1-A_2) \right]}_{C_{N\over 2}} +\underbrace{{iN\over 2\pi}\int_{\Sigma_0}\left(A_2\wedge A_3 - {1\over 2}A_1\wedge A_2 \right)}_{S_{bdry}} .
    \end{aligned}
\end{equation}
There is no particular boundary condition imposed here, hence 
\begin{equation}
 \text{Hom}(D_3\times \overline{D}_3 , C_{N\over 2}): \quad   \text{No b.c.}, \quad S_{bdry} = {iN\over 2\pi}\int_{\Sigma_0}\left(A_2\wedge A_3 - {1\over 2}A_1\wedge A_2 \right) \,.
\label{eq: we}
\end{equation}
Attaching a morphism junction between $C_0$ and the identity (with the identity defect field being $b$) using Eq.~\eqref{eq: bp}, we have 
\begin{equation}
\begin{aligned}
    \text{Hom}(D_3\times \overline{D}_3 ,\bi): \quad   &N(A_2+A_3)-b, \quad A_1-A_2 = \text{pure $U(1)$ gauge},\\
    & S_{bdry} = {iN\over 2\pi}\int_{\Sigma_0}\left(A_2\wedge A_3 - {1\over 2}A_1\wedge A_2 \right) ~. \label{eq: wea}
\end{aligned}
\end{equation}

% ---- new subsection ----

\subsection{Triality defect associator} \label{app: triality associators}
\begin{figure}[H]
	\centering
	\begin{tikzpicture}
		\begin{scope}[very thick,decoration={
				markings,
				mark=at position 0.5 with {\arrow{>}}}
			]
        % a(bc)
            \draw[postaction={decorate},dotted,Purple] (0,2.25)--(0,3) node[midway,left] () {$\mathbb{1},b$}; % a(bc)
        
            \draw[postaction={decorate}] (0,1.5)--(0,2.25) node[midway,left] () {$C_{N\over 2}$}; % a(bc)

        % bc
             \draw[postaction={decorate},Green,rounded corners] (0.75,0.75)--(0.9,0.9)--(0.15,1.65)--(0.15,3) node[at end,right] () {$U(1)_N,a_1$}; % bc 
            \draw[postaction={decorate},purple] (0.75,0.75)--(0,1.5) node[at end,right] () {$~\overline{D}_3$}; % bc 

        % a
            \draw[postaction={decorate}] (-0.75,-0.75)--(-1.5,0)--(0,1.5) node[midway,left] () {${D}_3~$}; % a

        % b           
            \draw[postaction={decorate}] (-0.75,-0.75)--(0.75,0.75) node[midway,left] () {$D_3~$}; % b

        % c
        \draw[postaction={decorate}] (0,-1.5)--(1.5,0) node[midway,right] () {$~D_3$}--(0.75,0.75); % c

        % ab            
            \draw[postaction={decorate},purple] (0,-1.5)--(-0.75,-0.75) node[at start,left] () {$\overline{D}_3~$}; %ab
            
            \draw[postaction={decorate},orange] (-0.15,-3)[rounded corners]--(-0.15,-1.65) node[at start,left] () {$U(1)_N,a_2$}-- (-0.9,-0.9)[rounded corners]--(-0.75,-0.75) ; % ab

        % (ab)c    
            \draw[postaction={decorate}] (0,-3)[dotted,Purple]--(0,-1.5) node[midway,right] () {$\bi,b$};%(ab)c
            % \draw[postaction={decorate}](0,-2.25)--(0,-1.5) node[midway,right] () {$C_0$}; %(ab)c
            			
	\node[blue] at (-1,-2) () {$A_L$};
        \node[blue] at (1,2) () {$A_R$};
        \node[blue] at (0.5,-0.25) () {$A_I'$};
        \node[blue] at (-0.15,0.6) () {$A_I$};

            \filldraw[blue] (0,2.25) circle (2pt) node[right,blue] () {$\alpha'$};
            \filldraw[blue] (0,1.5) circle (2pt) node[left,blue] () {$\alpha$};
            \filldraw[blue] (0.75,0.75) circle (2pt) node[right,blue] () {$\beta$};
            \filldraw[blue] (-0.75,-0.75) circle (2pt) node[left,blue] () {$\gamma$};
            \filldraw[blue] (0,-1.5) circle (2pt) node[right,blue] () {$\delta$};
            %\filldraw[blue] (0,-2.25) circle (2pt) node[right,blue] () {$\delta'$};
    \node at (3,0) (f) {};
        \end{scope}
    \begin{scope}[shift={($(f.east)+(2,0)$)},very thick,decoration={
				markings,
				mark=at position 0.5 with {\arrow{>}}}
			]
        % a(bc)
            \draw[postaction={decorate}] (0,1.5)--(0,3) node[midway,left] () {$D_3$}; % a(bc)

        % bc
            \draw[postaction={decorate},dotted,Purple] (0.75,0.75)--(0,1.5) node[at end,right] () {$~\bi,b$}; % bc 
            % \draw[] (0.75,0.75)--(0.6,0.9) node[at end,right] () {$~C_0$} ; % bc

        % a
            \draw[postaction={decorate}] (-0.75,-0.75)--(-1.5,0)--(0,1.5) node[midway,left] () {${D}_3~$}; % a

        % b           
            \draw[postaction={decorate},purple] (-0.75,-0.75)--(0.75,0.75) node[midway,left] () {$\overline{D}_3~$}; % b

        % c
        \draw[postaction={decorate}] (0,-1.5)--(1.5,0) node[midway,right] () {$~D_3$}--(0.75,0.75); % c

        % ab            
            \draw[postaction={decorate},dotted,Purple] (0,-1.5)--(-0.75,-0.75) node[at start,left] () {$\bi,b'~$}; %ab

        % (ab)c    
            \draw[postaction={decorate}] (0,-3)--(0,-1.5) node[midway,right] () {$D_3$}; % (ab)c
            			
	\node[blue] at (-1,-2) () {$A_L$};
        \node[blue] at (1,2) () {$A_R$};
        \node[blue] at (0.5,-0.25) () {$A_I'$};
        \node[blue] at (-0.15,0.6) () {$A_I$};

            \filldraw[blue] (0,1.5) circle (2pt) node[left,blue] () {$\alpha$};
            \filldraw[blue] (0.75,0.75) circle (2pt) node[right,blue] () {$\beta$};
            \filldraw[blue] (-0.75,-0.75) circle (2pt) node[left,blue] () {$\gamma$};
            \filldraw[blue] (0,-1.5) circle (2pt) node[right,blue] () {$\delta$};

    \node at (3,0) (ff) {};
    \end{scope}
    \begin{scope}[shift={($(ff.east)+(2,0)$)},very thick,decoration={
				markings,
				mark=at position 0.5 with {\arrow{>}}}
			]
        % bc
            \draw[postaction={decorate},purple] (0.75,0.75)--(0,1.5) node[midway,left] () {$\overline{D}_3$}; % bc 
            \draw[postaction={decorate},rounded corners,orange] (0.75,0.75)--(0.9,0.9)--(0.2,1.65)--(0.2,2.25) node[at start,right] () {$~U(1)_N,a_1$}; % bc 

        % a(bc)
            \draw[postaction={decorate}] (0,1.5)--(0,3) node[at end,left] () {$D_3$}; % a(bc)
            \draw[postaction={decorate},rounded corners, Green] (0,1.5)--(0.2,1.65)--(0.2,2.25) node[at end,right] () {$U(1)_{-N},a_2$}; % a(bc)

        % a
            \draw[postaction={decorate},purple] (-0.75,-0.75)--(-1.5,0)--(0,1.5) node[midway,left] () {$\overline{D}_3~$}; % a

        % b           
            \draw[postaction={decorate}] (-0.75,-0.75)--(0.75,0.75) node[midway,left] () {$ D_3~$}; % b

        % c
        \draw[postaction={decorate}] (0,-1.5)--(1.5,0) node[midway,right] () {$~ D_3$}--(0.75,0.75); % c

        % ab            
            \draw[postaction={decorate},dotted,Purple] (0,-1.5)--(-0.75,-0.75) node[at start, left] () {$\bi,b$}; %ab

        % (ab)c    
            \draw[postaction={decorate}] (0,-3)--(0,-1.5) node[near start,right] () {$D_3$}; % (ab)c
            			
	\node[blue] at (-1,-2) () {$A_L$};
        \node[blue] at (1.5,1) () {$A_R$};
        \node[blue] at (0.5,-0.25) () {$A_I'$};
        \node[blue] at (-0.15,0.6) () {$A_I$};

            \filldraw[Purple] (0.2,2.25) circle (2pt) node[above] () {$\alpha'$};
            \filldraw[blue] (0,1.5) circle (2pt) node[left,blue] () {$\alpha$};
            \filldraw[blue] (0.75,0.75) circle (2pt) node[below,blue] () {$\beta$};
            \filldraw[blue] (-0.75,-0.75) circle (2pt) node[above,blue] () {$\gamma$};
            \filldraw[blue] (0,-1.5) circle (2pt) node[right,blue] () {$\delta$};
            
    \node at (3,0) (fff) {};
    \end{scope}
	\end{tikzpicture}
	\caption{The triality defect associators. Morphism junctions appear in purple and are adjacent to a local fusion junction.  From left to right: $[F^{U(1)_N}_{{D}_3,D_3,D_3}]_{U(1)_N\overline{D}_3,U(1)_N\overline{D}_3}$, $[F^{D_3}_{{D}_3,\overline{D}_3,D_3}]_{\mathbb{1},\mathbb{1}}$ and $[F^{D_3}_{\overline{D}_3,D_3, D_3}]_{\bi,U(1)_N\overline{D}_3}$.}
	\label{fig: triality associators}
\end{figure}
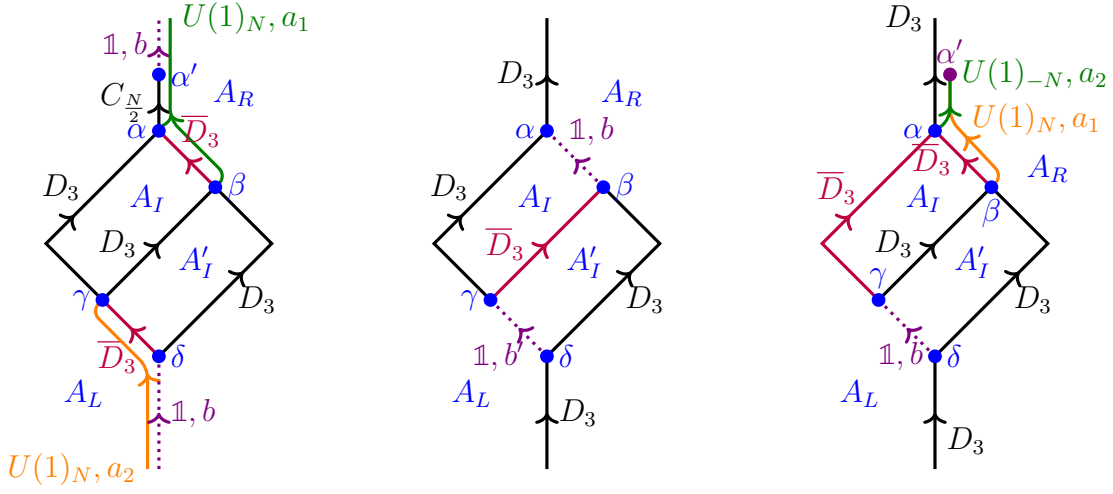

In this subsection, we focus on the first three associators among the following defect trios, with morphism junctions to the identity added whenever it is possible:
\begin{equation}
    D_3\times D_3\times D_3\,, \quad D_3\times \overline{D}_3\times D_3\,, \quad D_3\times D_3\times \overline{D}_3\,, \quad \overline{D}_3\times D_3\times D_3\,.
\end{equation}
These are drawn in Figure~\ref{fig: triality associators}.

\subsubsection{$\left[ F^{U(1)_N}_{D_3,D_3,D_3} \right]_{U(1)_N\overline{D}_3, U(1)_N\overline{D}_3}$}
This associator (first diagram of Figure~\ref{fig: triality associators}) relates the following two different fusions:
\begin{equation}
    \begin{aligned}
        (D_3\times D_3) \times D_3 &= U(1)_N\, \overline{D}_3\times D_3 = U(1)_N \, C_0 \to U(1)_N ~, \\
        D_3\times (D_3\times D_3) &= U(1)_N\, D_3\times \overline{D}_3 = U(1)_N \, C_{N\over 2} \to U(1)_N ~.
    \end{aligned}
\end{equation}
Using the junction analyses in the previous subsection, the boundary data at each junction is
\begin{equation}
    \begin{aligned}
        \alpha\beta &: \quad A_L-A_R, \quad N(A_I+A_R) + {N\over 2}(A_L-A_R) - b  = \text{pure $U(1)$ gauge}\\
        & \quad S(\alpha)+S(\beta) = {iN\over 4\pi}\int_{\Sigma_{\alpha\beta}} \left[A_R\wedge A_I - A_L \wedge A_R + (A_I - A_R)\wedge A_I' \right] ~, \\
        \gamma\delta:& \quad A_L-A_R , \quad b+NA_I' = \text{pure $U(1)$ gauge} ~, \\
        & \quad S(\gamma)+S(\delta) = -{iN\over 4\pi}\int_{\Sigma_{\gamma\delta}} \left[A_L\wedge A_I + A_L\wedge A_I' +A_I\wedge A_I' - 2A_R\wedge A_I' \right] ~.
    \end{aligned}
\end{equation}

After the parallel fuison of three $D_3$ defects in the middle,
\begin{equation}
    \begin{aligned}
        S &= {iN\over 2\pi}\int^\beta_\gamma\left( A_L \dr A_I + {1\over 2}A_L \dr A_L + A_I \dr A_I' + {1\over 2} A_I \dr A_I + A_I'\dr A_R +{1\over 2}A_I' \dr A_I'\right) \\
        &= {iN\over 4\pi} \int^\beta_\gamma (A_L+A_I+A_I') \dr (A_L + A_I + A_I')  \quad \leftarrow U(1)_N\\
        & \quad -{iN\over 2\pi}\int^\beta_\gamma \dr A_I' \, (A_L-A_R) \quad \leftarrow  \text{ Condensation defect } {C}_0 \\
        & \quad + {iN\over 2\pi}\int_{\Sigma_\beta-\Sigma_\gamma} \left[ -A_I'\wedge A_R + {1\over 2}\left( A_I \wedge A_L + A_I'\wedge A_L + A_I'\wedge A_I \right) \right] \quad \leftarrow S_{bdry;global}\\
    \end{aligned} \label{eq: wea}
\end{equation}
the interval hosts a $U(1)_N$ Chern-Simons theory, a condensation defect and some boundary terms. Summing the boundary terms from the upper junctions along with $S_{bdry;global}$ we get 
\begin{equation}\label{eq:bt1}
    S(\alpha)+S(\beta) + \left. S_{bdry;global}\right|_{\alpha\beta} = - \frac{iN}{4\pi}\int_{\Sigma} (A_I + A_I' + A_L)\wedge (A_R + A_I + A_I') ~,
\end{equation}
while the sum of boundary terms at the lower junctions vanishes,
\begin{equation}
    \begin{aligned}
        S(\gamma)+S(\delta) + \left. S_{bdry;global}\right|_{\gamma\delta} = 0 ~.
    \end{aligned}
\end{equation}
The total boundary term \eqref{eq:bt1} at the upper junction connects the $U(1)_N$ theories at the junction $\beta$. To see this, we notice that the $U(1)_N$ theories above $\beta$, below $\gamma$ and between them are
\begin{equation}
\begin{aligned}
    \text{Above $\beta$}\quad & {iN\over 4\pi}\int^\infty_\beta (A_R+A_I+A_I')\dr(A_R+A_I+A_I') ~, \\
    [\gamma,\beta] \text{ and } \text{Below $\gamma$} \quad &  {iN\over 4\pi}\int^\beta_{-\infty} (A_L+A_I+A_I')\dr(A_L+A_I+A_I') ~. \\
    % \text{Below $\gamma$} \quad & {iN\over 4\pi}\int^\gamma_{-\infty} (A_L+A_I+A_I')\dr(A_L+A_I+A_I')
\end{aligned}
\end{equation}
Following a similar discussion as in Section \ref{sec: u(1)N theories} (or applying the folding trick here), we see that the boundary term \eqref{eq:bt1} ensures the gauge invariance from both sides, and the gluing condition, $(A_L+A_I+A_I') - (A_I+A_I'+A_R) = \text{pure $U(1)$ gauge}$, indicates that this is a trivial interface of the $U(1)_N$ theory.
% To connect the $U(1)_N$ at the junction $\beta$, we added the boundary term on the 2nd line of Eq.~\eqref{eq: wea}.  

Thus, we are left with a 3d condensation defect on the interval. Reducing it along the interval leads to a 2d condensation defect of the bulk 1-form symmetry (which is simply a projector). Notice that this is consistent with the operator spectrum analysis. First, as follows from the boundary conditions, the $\mathbb{Z}_N$ higher quantum symmetry line $e^{i\int (A_L - A_R)}$ on the 3d condensation defect can stretch between the two boundaries. After reduction, it becomes a topological local operator and can be interpreted as the higher quantum 1-form symmetry on the 2d condensation defect. Second, in the 4d bulk, an open $\mathbb{Z}_N$ 1-form symmetry operator can be viewed as a non-genuine line operator bounding the symmetry operator; but on the condensation defect, $\mathbb{Z}_N$ is gauged and this line operator should become genuine. This is guaranteed by the e.o.m. of $A_L$ (or $A_R$), which allows us to rewrite such an open $\mathbb{Z}_N$ 1-form symmetry operator as the $A_I'$-Wilson line $e^{i\oint A_I'}$. After reduction, $e^{i\oint A_I'}$ remains a line operator by the boundary conditions, encoding the fact that an open $\mathbb{Z}_N$ 1-form symmetry operator becomes genuine in the 2d condensation defect.

We also want to highlight that the existence of local junctions $C_k \rightarrow \mathbb{1}$ in the initial diamond diagrams we considered does not necessarily imply there will be a reduction of the condensation defect in the end. This is because we will need to perform the global fusion to collapse the diagram, during which the condensation defect can become a decoupled $\mathbb{Z}_N$-gauge theory. Only when the condensation defect appears in the global fusion, will we have the interval reduction of the 3d condensation defect which leads to the 2d condensation defect.

Overall, we see that the 2d theory corresponding to the associator 1-morphism results solely from the 3d condensation defect in the interval, which reduces to a 2d condensation defect. As a result,   
\begin{equation}
    \left[ F^{U(1)_N}_{D_3,D_3,D_3} \right]_{U(1)_N\overline{D}_3, U(1)_N\overline{D}_3} = \text{2d $\bz_N$ condensation defect}\,.
\end{equation}

\subsubsection{$\left[ F^{D_3}_{D_3,\overline{D}_3,D_3} \right]_{\bi,\bi}$}
This associator relates the following fusions above and below as depicted in the second diagram of Figure~\ref{fig: triality associators}:
\begin{equation}
    \begin{aligned}
    D_3\times (\overline{D}_3\times D_3) &= D_3\times C_0 \to D_3\times \bi = D_3 ~, \\
    (D_3 \times \overline{D}_3)\times D_3 &= C_{N\over 2}\times D_3 \to \bi \times D_3= D_3 ~.
    \end{aligned}
\end{equation}
The boundary data at each junction can be collected to
\begin{equation}
    \begin{aligned}
        \alpha,\beta,\beta': \quad & b+NA_L, \quad b+NA_I', \quad A_I - A_R = \text{pure $U(1)$ gauge} ~, \\
        & S(\alpha) = -{i\over 2\pi} \int_{\Sigma_\alpha} b\wedge (A_I-A_R), \quad S(\beta) = -{iN\over 2\pi}\int_{\Sigma_\beta} A_R \wedge A_I' ~, \\
        \gamma,\gamma',\delta: \quad & N(A_I+A_I') -b' , \quad A_L-A_I',\quad b-NA_L-NA_R = \text{pure $U(1)$ gauge} ~, \\
        &S(\gamma) = -{iN\over 2\pi}\int_{\Sigma_\gamma}\left( A_I'\wedge A_I - {1\over 2}A_L \wedge A_I' \right), \quad S(\delta) = -{iN\over 4\pi}\int_{\Sigma_\delta} A_L \wedge A_I' ~.
    \end{aligned}
\end{equation}
Fusing the three defects in the middle by summing their actions result in the following actions:
\begin{equation}
    \begin{aligned}
        S &= {iN\over 2\pi}\int^\beta_\gamma \left(A_L \dr A_I +{1\over 2} A_L \dr A_L - A_I \dr A_I' - {1\over 2}A_I' \dr A_I' +A_I'\dr A_R + {1\over 2}A_I' \dr A_I'\right)\\
        &= \underbrace{{iN\over 2\pi}\int^\beta_\gamma \left(A_L \dr A_R + {1\over 2}A_L \dr A_L\right) }_{D_3} + \underbrace{{iN\over 2\pi}\int^\beta_\gamma (A_L-A_I')\dr (A_I-A_R)}_{\text{decoupled $\mathbb{Z}_N$ gauge theory}}\\
        &\quad + \underbrace{ {iN\over 2\pi}\int_{\Sigma_\beta - \Sigma_\gamma} A_I\wedge A_I' }_{S_{bdry;global}} ~.
    \end{aligned}
\end{equation}
Collecting the boundary terms, at the upper junctions
\begin{equation}
\begin{aligned}
    S(\alpha)+S(\beta) + \left. S_{bdry;global}\right|_{\alpha\beta} &= - {i\over 2\pi}\int_{\Sigma_{\alpha\beta}} (b+NA_I')\wedge (A_I-A_R) \in 2\pi i \mathbb{Z}
\end{aligned}
\end{equation}
where we used the boundary conditions, and at the lower junctions,
\begin{equation}
    \begin{aligned}
        S(\gamma) + S(\delta) + \left. S_{bdry;global}\right|_{\gamma\delta} &= 0 ~. \label{eq: wf}
    \end{aligned}
\end{equation}
Hence, we find a decoupled 3d $\mathbb{Z}_N$ on the interval. Since $A_L-A_I'$ has $(ND)$ and $A_I-A_R$ has $(DN)$ boundary conditions on $\alpha\beta$ and $\gamma\delta$ respectively, there is no untwisted local operators after reduction and the decoupled $\mathbb{Z}_N$ gauge theory becomes the trivial theory after reducing the interval $[\gamma,\beta]$.  
Hence,
\begin{equation}
   \left[ F^{D_3}_{D_3,\overline{D}_3,D_3} \right]_{\mathbb{1},\mathbb{1}} = \text{Trivial theory}\,.
\end{equation}

\subsubsection{$[F^{D_3}_{\overline{D}_3,D_3, D_3}]_{\bi,U(1)_N\overline{D}_3}$} 
Lastly, let's compute the associator in the last diagram of~\ref{fig: triality associators} which relates
\begin{equation}
    \begin{aligned}
    \text{Upper part}:\quad &\overline{D}_3 \times (D_3\times D_3) = \overline{D}_3\times \overline{D}_3~U(1)_N  = D_3~U(1)_{-N}\times U(1)_{N} \to D_3 ~, \\
    \text{Lower part}: \quad &(\overline{D}_3\times D_3)\times D_3 =C_0\times D_3 \to \mathbb{1}\times D_3 = D_3 ~.
    \end{aligned}
\end{equation}
The boundary data at each junction is
\begin{equation}
    \begin{aligned}
\alpha',\alpha,\beta: \quad & a_1 = A_I + A_I' + A_R ~, \quad  a_2 = a_L + A_I + A_R ~, \\
& \begin{cases}
    (1) & N(A_I'+A_I+A_R), \quad A_L-A_I' = \text{pure $U(1)$ gauge} ~, \\
\text{or } (2) & A_L-A_I'+2(A_I'+A_I+A_R), \quad {N\over 2}(A_L-A_I') = \text{pure $U(1)$ gauge} ~,
\end{cases} \\
    & S(\alpha) = - \frac{iN}{4\pi} \int_{\Sigma_\alpha} A_L \wedge A_R + (A_L - A_R) \wedge A_I ~, S(\beta) = \frac{iN}{4\pi} \int_{\Sigma_\beta} A_I \wedge A_R + (A_I - A_R) \wedge A_I' ~, \\
    & S(\alpha') = -\frac{iN}{4\pi} \int_{\Sigma_{\alpha'}} (A_I + A_I' + A_R) \wedge (A_L + A_I + A_R) ~, \\
\gamma,\gamma',\delta: \quad & A_L-A_I', \quad N(A_L+A_I+A_R) = \text{pure $U(1)$ gauge} ~, \\
& S(\gamma) = \frac{iN}{2\pi} \int A_I' \wedge A_I ~, \quad S(\delta) = -\frac{iN}{2\pi} A_L \wedge A_I' ~.
    \end{aligned} \label{eq: xa}
\end{equation}

Next, we fuse the three defects $\overline{D}_3\times D_3\times D_3$ in the middle by summing their actions, 
\begin{equation}
    \begin{aligned}
		S &= \frac{iN}{2\pi} \int_\gamma^\beta (- A_L d A_I - \frac{1}{2} A_I dA_I + A_I dA_I' + \frac{1}{2} A_IdA_I + A_I' dA_R + \frac{1}{2} A_I'dA_I') \\ 
        &= \frac{iN}{2\pi} \int^\beta_\gamma \left(A_L \dr A_R + \frac{1}{2}A_L \dr A_L \right) \quad (\leftarrow \, D_3)\\
		&\quad + \frac{iN}{2\pi}\int^\beta_\gamma \left[ \frac{1}{2}(A_I'-A_L) \dr( A_I'- A_L) + (A_I'-A_L)\wedge\dr (A_I + A_L + A_R) \right] \quad (\leftarrow (\mathcal{Z}_N)_N)\\
		&\quad + \frac{iN}{2\pi}\int_{\Sigma_\beta - \Sigma_\gamma} \left[ A_I' \wedge (A_I {+} \frac{1}{2}A_L) \right] \quad (\leftarrow\, S_{bdry;global}) ~,
	\end{aligned} \label{eq: bl}
\end{equation}
where the first line is the action for $D_3$ as expected from the fusion algebra, the second line is a decoupled $\mathbb{Z}_N$ gauge theory with the twist $N$ (denoted by $(\mathcal{Z}_N)_N$ and the last term is the boundary term.

The total boundary term at the upper junction is 
\begin{equation}
    \begin{aligned}
        S(\alpha)+S(\beta)+S(\alpha')+\left.S_{bdry;global}\right|_{\beta} &= 0 ~,
    \end{aligned}
\end{equation}
and the total boundary term at the lower junction is
\begin{equation}
    S(\gamma)+S(\delta)+\left. S_{bdry;global}\right|_{\gamma} = 0 ~.
\end{equation}
Therefore, we indeed find a decoupled $(\mathcal{Z}_N)_N$ theory on the interval.

On the upper junction, we have two boundary conditions (1) and (2) from condensing $U(1)_N\times U(1)_{-N}$.  Under the boundary condition (1), the upper and lower junctions have the same boundary conditions on gauge fields $A_I+A_L+A_R$ and $A_L-A_I'$ as written in Table~\ref{tab: bc1}.  Following the discussion in Section~\ref{sec: 4d method}, reducing the interval results in a 2d $\mathbb{Z}_N$ gauge theory.
\begin{table}[H]
    \centering
    \begin{tabular}{c|cc}
        \hline
     & $A_I+A_L+A_R$  & $A_L-A_I'$ \\
     \hline
     $\alpha\beta\, (1)$ & $\mathbb{Z}_N$ & pure $U(1)$ gauge\\ 
     $\gamma\gamma'$ & $\mathbb{Z}_N$ & pure $U(1)$ gauge\\
     \hline
    \end{tabular}
    \caption{Boundary conditions for the case (1) of Eq.~\eqref{eq: xa}}
    \label{tab: bc1}
\end{table}

Under the boundary condition (2), on the other hand, the gauge fields have different boundary conditions at the upper and lower junctions, as written in Table~\ref{tab: bc2}. In this case, we only find $\mathbb{Z}_2$ untwisted local operators arising from the Wilson lines $e^{i \frac{Nk}{2} \oint a_1 + a_2}$ ($k \in \mathbb{Z}_2$). Hence, reducing the interval $[\gamma,\beta]$ results in a $\bz_2$ gauge theory.

\begin{table}[H]
    \centering
    \begin{tabular}{c|cc}
    \hline
     & $(A_L-A_I') + 2(A_I+A_L+A_R)$    & $A_L-A_I'$  \\
     \hline
     $\alpha\beta\, (2)$ & pure gauge & $\mathbb{Z}_{N\over 2}$\\
    $\gamma\gamma'$ & $\mathbb{Z}_{N\over 2}$ & pure gauge\\
    \hline
    \end{tabular}
    \caption{Boundary conditions for the case (2) of Eq.~\eqref{eq: xa}}
    \label{tab: bc2}
\end{table}%

In summary,
\begin{equation}
    (\mathcal{Z}_N)_N \quad \overset{\text{reducing interval}}{\longrightarrow} \quad \begin{cases}
        (1) : &\mathbb{Z}_N \text{ gauge theory} ~,\\
        (2): & \text{$\bz_2$ gauge theory} ~.
    \end{cases}
\end{equation}

Combining the reduction of the $\bz_N$ theory on the interval and the boundary actions, there are two choices for the associator 1-morphism,
\begin{equation}
\left[ F^{D_3}_{\overline{D}_3,D_3,D_3} \right]_{U(1)_N\overline{D}_3,\mathbb{1}} = \text{$\mathbb{Z}_N$ gauge theory}  \quad \text{or}\quad \text{$\mathbb{Z}_2$ gauge theory} ~.
\end{equation}

% -------------- New Section ------------------

\section{Details on anyon condensation and domain wall fusion}\label{app:ac_dw}
In this appendix, we give some details on the anyon condensation computations and the fusion of the domain walls used in Section \ref{Sec_Group}.

Recall that we want to describe the topological operator spectrum for the following two diagrams
\begin{equation}\label{eq:fusion_dw}
\begin{tikzpicture}[baseline={([yshift=-.5ex]current bounding box.center)},vertex/.style={anchor=base,circle,fill=black!25,minimum size=18pt,inner sep=2pt},scale=1]
    \draw[line width = 0.4mm, ->-=0.5] (0,-2) -- (0,-1);
    \draw[line width = 0.4mm, ->-=0.5] (0,-1) -- (0,+1);
    \draw[line width = 0.4mm, ->-=0.5] (0,+1) -- (0,+2);
    \filldraw[black] (0,-1) circle (2pt);
    \filldraw[black] (0,+1) circle (2pt);
    \node[below] at (0,-2) {\footnotesize $\mathbb{D}^{x} \, \mathcal{T}_-[a^{(2)}]$};
    \node[above] at (0,+2) {\footnotesize $\mathbb{D}^{x} \, \mathcal{T}_+[a^{(2)}]$};
    \node[right] at (0,0) {\footnotesize $\mathbb{D}^{x} \, \mathcal{T}[a^{(2)}] \equiv \mathcal{D}_1 \times \mathcal{D}_2 \times \mathcal{D}_3$};
    \node[right] at (0,-1) {\footnotesize $\mathcal{I}_-$};
    \node[right] at (0,+1) {\footnotesize $\mathcal{I}_+$};
\end{tikzpicture} \implies \begin{tikzpicture}[baseline={([yshift=-.5ex]current bounding box.center)},vertex/.style={anchor=base,circle,fill=black!25,minimum size=18pt,inner sep=2pt},scale=1]
    \draw[line width = 0.4mm, ->-=0.5] (0,-2) -- (0,0);
    \draw[line width = 0.4mm, ->-=0.5] (0,0) -- (0,+2);
    \filldraw[black] (0,0) circle (2pt);
    \node[right] at (0,0) {\footnotesize $\mathcal{I}_{+-}[a^{(2)}] \equiv \mathbf{F}_{\mathcal{D}_1,\mathcal{D}_2,\mathcal{D}_3}$};
    \node[below] at (0,-2) {\footnotesize $\mathbb{D}^{x} \, \mathcal{T}_-[a^{(2)}]$};
    \node[above] at (0,+2) {\footnotesize $\mathbb{D}^{x} \, \mathcal{T}_+[a^{(2)}]$};
\end{tikzpicture} ~
\end{equation}
First, let us explain the line operator spectrum in the following configuration summarized by Table~\ref{tab:anyon_spec}.
\begin{equation}\label{eq:conf_11}
\begin{aligned}
    \begin{tikzpicture}[baseline={([yshift=-.5ex]current bounding box.center)},vertex/.style={anchor=base,circle,fill=black!25,minimum size=18pt,inner sep=2pt},scale=1]
        \draw[grey, thick] (-2.5,-1) -- (-1.5,-1);
        \draw[grey, thick] (-2.5,+1) -- (-1.5,+1);
        \draw[blue, thick] (-1.5,-1) -- (-1.5,+1);
        \draw[blue, thick] (+1.5,-1) -- (+1.5,+1);
        \draw[grey, thick] (+2.5,-1) -- (+1.5,-1);
        \draw[grey, thick] (+2.5,+1) -- (+1.5,+1);
        \filldraw[blue, opacity = 0.1] (-1.5,-1) -- (-1.5,+1) -- (1.5,+1) -- (1.5,-1) -- (-1.5,-1);
        \node[black] at (-2,0) {\footnotesize $\mathcal{T}_-$};
        \node[black] at (+2,0) {\footnotesize $\mathcal{T}_+$};
        \node[black, below] at (-1.5,-1) {\footnotesize $\mathcal{I}_-$};
        \node[black, below] at (+1.5,-1) {\footnotesize $\mathcal{I}_+$};
        \node[black] at (0,0) {\footnotesize $\mathcal{T}$};
    \end{tikzpicture} ~,
\end{aligned}
\end{equation}

From the math literature of anyon condensation \cite{Kong:2013aya,Kitaev:2011dxc}, the interface $\mathcal{I}_\pm$ is given the fusion category of $\mathbb{A}_\pm$-modules in $\mathcal{T}$\footnote{Here we view $\mathbb{A}_\pm$ as an algebra object in $\mathcal{T}$.}, and the theory $\mathcal{T}_\pm$ on the other side of the interface is described by the category of local $\mathbb{A}_\pm$-modules (which is a modular tensor category). The interfaces $\mathcal{I}_{\pm}$ carry the structure as bimodule categories of $\mathcal{T}$ and $\mathcal{T}_\pm$. Physically, condensing $\mathbb{A}_\pm$ means gauging the 1-form symmetry $\mathbb{A}_\pm$. The 1-form symmetry $\mathbb{A}_\pm$ can act on the line operators by fusion and by braiding, and after gauging we keep only the operators invariant under $\mathbb{A}_\pm$ action. On the interface $\mathcal{I}_\pm$, $\mathbb{A}_\pm$ can only act by fusion as $\mathcal{I}_\pm$ is 2-dimension. The simple invariant line operators on $\mathcal{I}_\pm$ must be invariant under fusion with any lines in $\mathbb{A}_\pm$, therefore must take the form
\begin{equation}
    \ell_b^{\pm} \equiv \sum_{a \in \mathbb{A}_\pm} b \times a
\end{equation}
for any simple line operator $b$ in $\mathcal{T}$. Clearly, $\ell_b^\pm$ and $\ell_{b'}^\pm$ parameterize the same invariant lines if $b^{-1}b' \in \mathbb{A}_\pm$. The $\ell_b^\pm$ constructed above is a $\mathbb{A}_\pm$-module, and all the $\ell_b^\pm$'s form the category $\mathcal{T}_{\mathbb{A}_\pm}$ of $\mathbb{A}_\pm$-modules describing the topological line operators on the interface $\mathcal{I}_\pm$. The fusion rules for those invariant lines are simply given by\footnote{Notice that the tensor product of $\mathbb{A}_\pm$-modules is the balanced tensor product $\otimes_{\mathbb{A}_\pm}$ over the algebra object $\mathbb{A}_\pm$ which gives rise to the correct multiplicity below.}
\begin{equation}
    \ell_b^{\pm} \times \ell_{b'}^\pm = \ell_{bb'}^\pm ~.
\end{equation}
It is also possible to fuse the bulk line in $\mathcal{T}$ with the interface lines with the fusion rules
\begin{equation}
    \ell_b^+ \otimes b' = \ell^+_{bb'} ~, \quad b' \otimes \ell_b^- = \ell_{b'b}^- ~, \quad \forall b' \in \mathcal{T} ~.
\end{equation}
On the other hand, in the theory $\mathcal{T}_\pm$ acquired by gauging $\mathbb{A}_\pm$, where the 1-form symmetry $\mathbb{A}_\pm$ also acts via braiding. Hence, only the $\ell_b$'s that braid trivially with $\mathbb{A}_\pm$ are invariant under the $\mathbb{A}_\pm$ action\footnote{For that, we only need to check any one of the line in the sum braids trivially with all the lines in $\mathbb{A}_\pm$. This is because in order to condense $\mathbb{A}_\pm$, every anyon in $\mathbb{A}_\pm$ must be boson and braids trivially with each other.}. Those $\ell_b$'s are known as local $\mathbb{A}_\pm$-modules, and together they form a modular tensor category (often denoted as $\mathcal{T}_{\mathbb{A}_\pm}^{loc}$) describing the 3d TQFT $\mathcal{T}_\pm$.

Next, we explain the domain wall fusion step between $\mathcal{I}_+$ and $\mathcal{I}_-$ in \eqref{eq:fusion_dw}. The outcome is a domain wall $\mathcal{I}_{-+}$ between $\mathcal{T}_-$ and $\mathcal{T}_+$, which we identify as the $F$-symbol. The result has been summarized in Table~\ref{tab:anyon_spec2}, and the result is derived from the math tools known as balanced Deligne tensor product which captures the domain wall fusion as shown in \cite{Kitaev:2011dxc}
\begin{equation}
    \mathcal{I}_{+-} = \mathcal{I}_+ \boxtimes_{\mathcal{T}} \mathcal{I_-} ~.
\end{equation}
In the special case where $\mathcal{T}$ contains only Abelian anyons, the computation can be done as follows\footnote{For experts, here we are simply using the result that, for a fusion category $\mathcal{C}$, the balanced tensor product $\mathcal{M}\boxtimes_\mathcal{C}\mathcal{N}$ between a right $\mathcal{C}$-module category $\mathcal{M}$ and a left $\mathcal{C}$-module category $\mathcal{N}$ can be computed by first viewing $\mathcal{M}\boxtimes \mathcal{N}$ as a module category over $\mathcal{C}\boxtimes \mathcal{C}^{rev}$ and taking the category of modules over the algebra $\mathcal{A} = \oplus_{X\in Irr(\mathcal{C})} X\boxtimes X^*$ in $\mathcal{M}\boxtimes \mathcal{N}$. For more details, see Remark 3.9 in \cite{Etingof:2009yvg}.}. First, one takes $\mathcal{I}_+\boxtimes \mathcal{I}_-$ corresponding to naively stacking two interfaces on top of each other. Using our previous notation, simple line operators on $\mathcal{I}_+\boxtimes \mathcal{I}_-$ can be labeled as
\begin{equation}
    \ell_{b,b'} \equiv \ell_b^+ \boxtimes \ell_b^- ~.
\end{equation}
To get the spectrum of line operators in $\mathcal{I}_{+-} = \mathcal{I}_+\boxtimes_{\mathcal{T}}\mathcal{I}_-$, one can simply make the identification
\begin{equation}
    \ell_{b,b'} \simeq \ell_{ba,a^{-1}b'} \quad \forall a \in \mathcal{T} ~.
\end{equation}
Notice that one can also fuse the lines in $\mathcal{T}_\pm$ to lines on the interface $\mathcal{I}_{+-}$, with the fusion rules
\begin{equation}
    \ell_{b_1}^+ \otimes \ell_{b_2,b_3} = \ell_{b_1 b_2,b_3} ~, \quad \ell_{b_2,b_3} \otimes \ell_{b_1}^- = \ell_{b_2, b_3 b_1} ~.  
\end{equation}
Finally, the interface $\mathcal{I}_{+-}$ will be a multi-fusion category, namely it contains local operators (corresponding to the multiple junctions between the identity lines), if $\mathbb{A}_+\cap\mathbb{A}_-$ is non-trivial. And the local operators are exactly labeled by $\mathbb{A}_+\cap\mathbb{A}_-$, corresponding to the lines stretching between two interfaces. When this happens, it usually means the 2d interface contains a sub-sector described by a decoupled 2d TQFT.

Finally, one also needs to remember to keep track of whether the resulting 2d interface $\mathcal{I}_{-+}$ couples to the bulk gauge field $a^{(2)}$. When it does, this will turn the decoupled 2d TQFT to a 2d condensation defect of the 1-form symmetry.

% -------------- New Section ------------------

\section{Duality defect associators} \label{app: duality}
In this section, we briefly explain the non-invertible duality symmetry in 4d Maxwell theory at $\tau=iN$ found in Ref.~\cite{Choi:2021kmx,Kaidi:2021xfk} and compute one particular associator of the duality defects using the Lagrangian method and group-theoretical approach.  Other associators can be computed using the same methods and the results are listed.  We find an agreement and comment on the efficiency and validity of the two approaches.

\subsection{Non-invertible duality symmetry and defect Lagrangians}
At couplings $\tau=iN$, there is a duality symmetry constructed as follows.  We first gauge a $\mathbb{Z}_{N}^{(1)}$ subgroup of the electric 1-form symmetry in half space (with Dirichlet boundary conditions for the corresponding discrete gauge field), resulting in a topological interface between the theory with coupling $\tau=iN$ and the theory with coupling $i/N$. However, since under $\mathbb{S}$-duality the coupling $i/N$ maps to $iN$, the two sides are in fact equivalent and we get a defect in a single theory, obtained by performing $\mathbb{Z}_{N}^{(1)}$ gauging and $\mathbb{S}$-duality in half space. As shown in \cite{Choi:2021kmx}, one can describe this defect, which is denoted by $D_2$, by the following action, 
\begin{equation}
\label{D2}
D_{2}:\quad S=\frac{N}{4\pi}\int_{x<0}dA_{L}\wedge\star dA_{L}+\frac{N}{4\pi}\int_{x>0}dA_{R}\wedge\star dA_{R}+\frac{iN}{2\pi}\int_{x=0}A_{L}\wedge dA_{R} \, 
\end{equation}
where $A_L$ and $A_R$ denote the Maxwell gauge field from the two sides of the defect, which is located at $x=0$.  %This action is justified by the defect equations of motion, $F_R = {i\over N}\star (NF_L)$.

The corresponding orientation reversal defect, denoted by $\overline{D}_{2}$, is given by
\begin{equation}
\label{D2b}
\overline{D}_{2}:\quad S=\frac{N}{4\pi}\int_{x<0}dA_{L}\wedge\star dA_{L}+\frac{N}{4\pi}\int_{x>0}dA_{R}\wedge\star dA_{R}-\frac{iN}{2\pi}\int_{x=0}A_{R}\wedge dA_{L}~.
\end{equation}
Together and with the condensation defect $C_0$, they satisfy the non-invertible fusion algebra
\begin{equation}
\begin{gathered}
    D_2\times \overline{D}_2 = C_0=  \overline{D}_2\times D_2, \quad D_2\times D_2= U \times C_0 = \overline{D}_2\times \overline{D}_2, \\
    D_2\times C_0=C_0\times D_2=D_2,\quad \overline{D}_2\times C_0=C_0\times \overline{D}_2=\overline{D}_2    
\end{gathered}
 \label{eq: duality fusion}
\end{equation}
where $U$ is the charge conjugation defect given by 
\begin{equation}
    U :\quad \frac{N}{4\pi} \int_{x<0} dA_L \wedge \star dA_L + \frac{N}{4\pi} \int_{x>0} dA_R \wedge \star dA_R + \frac{i}{2\pi}\int db\wedge (A_L + A_R) ~.
\end{equation}

Let us first analyze the structure of the $F$-symbol by the number of incoming invertible charge conjugation defects $U$. First, when there are two incoming $U$ defects, then the global fusion in the middle contains no decoupled TQFT or condensation defect supported on the interval; hence the $F$-symbol must be the trivial theory. Second, when there is one incoming $U$ defect and two non-invertible defects, then the global fusion in the middle always contains a condensation defect supported on the interval; under the interval reduction, this always leads to a 2d condensation of the $\mathbb{Z}_N$ 1-form symmetry. Thus, we can now focus on the cases where all three incoming twist defects are non-invertible symmetries. We find
\begin{equation}
    \left[ F^{\overline D_2}_{D_2,D_2,D_2} \right]_{U,U}, \quad \left[ F^{ D_2}_{\overline D_2,D_2,D_2} \right]_{\bi,U},\quad \left[ F^{ D_2}_{D_2,\overline D_2,D_2} \right]_{\bi,\bi},\quad \left[ F^{ D_2}_{D_2,D_2,\overline D_2} \right]_{U,\bi} = \text{Trivial TQFT}~, \label{eq: duality associators}
\end{equation}
and the remaining ones are related to the above ones via orientation reversal. Below, we will only demonstrate explicitly the computation for $\left[ F^{ D_2}_{D_2,\overline D_2,D_2} \right]_{\bi,\bi}$ using the Lagrangian method since the other ones can be computed similarly. Notice that unlike the triality case, here the approach holds for all $N$.

\subsection{Lagrangian method for $\left[F^{D_2}_{D_2,\overline{D}_2,D_2}\right]_{\mathbb{1},\mathbb{1}}$}
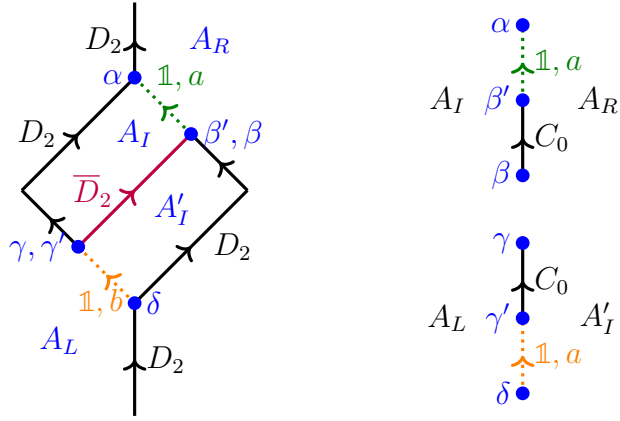
\begin{figure}[H]
	\centering
	\begin{tikzpicture}
		\begin{scope}[very thick,decoration={
				markings,
				mark=at position 0.5 with {\arrow{>}}}
			]
            \draw[postaction={decorate}] (0,1.5)--(0,2.5) node[midway,left] () {$D_2$}; % a(bc)
            \draw[postaction={decorate}] (-1.5,0)--(0,1.5) node[midway,left] () {${D}_2~$}; % a
            \draw[postaction={decorate},dotted,Green] (0.75,0.75)--(0,1.5) node[at end,right] () {$~\mathbb{1},a$}; % upper bc 
            \draw[postaction={decorate}] (1.5,0)--(0.75,0.75); % upper arm of c 
            \draw[postaction={decorate},purple] (-0.75,-0.75)--(0.75,0.75) node[midway,left] () {$\overline{D}_2~$}; % b
            \draw[postaction={decorate}] (0,-1.5)--(1.5,0) node[midway,right] () {$~{D}_2$}; % c lower arm
            
            \draw[postaction={decorate},dotted,orange] (0,-1.5)--(-0.75,-0.75) node[at start,left] () {$\mathbb{1},b$}; %ab
            \draw[postaction={decorate}] (-0.75,-0.75)--(-1.5,0);
            \draw[postaction={decorate}] (0,-3)--(0,-1.5) node[midway,right] () {$D_2$}; %(ab)c
            			
	\node[blue] at (-1,-2) () {$A_L$};
        \node[blue] at (1,2) () {$A_R$};
        \node[blue] at (0.5,-0.25) () {$A_I'$};
        \node[blue] at (0,0.75) () {$A_I$};

            \filldraw[blue] (0,1.5) circle (2pt) node[left,blue] () {$\alpha$};
            \filldraw[blue] (0.75,0.75) circle (2pt) node[right,blue] () {$\beta',\beta$};
            \filldraw[blue] (-0.75,-0.75) circle (2pt) node[left,blue] () {$\gamma,\gamma'$};
            \filldraw[blue] (0,-1.5) circle (2pt) node[right,blue] () {$\delta$};
    \node[] at (3,0) (f) {};
        \end{scope}
        \begin{scope}[shift={($(f.east)+(2,1.2)$)},very thick,decoration={
				markings,
				mark=at position 0.5 with {\arrow{>}}}
			] 
			\draw[postaction={decorate}] (0,-1)--(0,0) node[midway,right] () {$C_0$};
            \draw[postaction={decorate},dotted,Green] (0,0)--(0,1) node[midway,right] () {$\mathbb{1},a$};
            \node at (-1,0) () {$A_I$};
            \node at (1,0) () {$A_R$};
            \filldraw[blue] (0,0) circle (2pt) node[left,blue] () {$\beta'$};
            \filldraw[blue] (0,1) circle (2pt) node[left,blue] () {$\alpha$};
            \filldraw[blue] (0,-1) circle (2pt) node[left,blue] () {$\beta$};
\end{scope}
\begin{scope}[shift={($(f.east)+(2,-1.7)$)},very thick,decoration={
				markings,
				mark=at position 0.5 with {\arrow{>}}}
			] 
			\draw[postaction={decorate},dotted,orange] (0,-1)--(0,0) node[midway,right] () {$\mathbb{1},a$};
            \draw[postaction={decorate}] (0,0)--(0,1) node[midway,right] () {$C_0$};
            \node at (-1,0) () {$A_L$};
            \node at (1,0) () {$A_I'$};
            \filldraw[blue] (0,0) circle (2pt) node[left,blue] () {$\gamma'$};
            \filldraw[blue] (0,1) circle (2pt) node[left,blue] () {$\gamma$};
            \filldraw[blue] (0,-1) circle (2pt) node[left,blue] () {$\delta$};
\end{scope}
	\end{tikzpicture}
	\caption{Left: Diagram for $\left[F^{D_2}_{{D_2},\overline{D}_2,{D}_2}\right]_{\mathbb{1},\mathbb{1}}$. Right: Resolution of the morphism junctions $\beta'$ and $\gamma'$.}
	\label{fig: FD2-D2D2bD2}
\end{figure}

In this subsection, we compute the associator 1-morphism $\left[F^{D_2}_{D_2,\overline{D}_2,D_2}\right]_{\mathbb{1},\mathbb{1}}$ for any $N$ using the Lagrangian method in Section ~\ref{sec: 4d method}.
To obtain $\left[F^{D_2}_{D_2,\overline{D}_2,D_2}\right]_{\mathbb{1},\mathbb{1}}$, we collapse the diamond-shaped region in Figure~\ref{fig: FD2-D2D2bD2} to a 2d associator manifold.  We begin by analysing each fusion/splitting and morphism junctions.

\paragraph{$\beta$: $D_2\times \overline{D}_2= C_0$} The action along the fused part is
\begin{equation}
	\begin{aligned}
		S &=  {iN\over 2\pi}\int^\infty_\beta \left(-A_I'\dr A_I + A_I'\dr A_R \right)\\
        &= {iN\over 2\pi}\int^\infty_\beta \dr(-A_I')\wedge(A_I-A_R) + {iN\over 2\pi}\int_{\Sigma_\beta}  A_I'\wedge (A_R - A_I)
	\end{aligned}
\end{equation}
where $\int_\beta^\infty$ is the 3d integral above the $\beta$ junction and $\Sigma_\beta$ is the 2d manifold supporting the $\beta$ junction. Therefore, at this junction, the boundary term is
\begin{equation}
  \beta: \quad  \text{No boundary condition}, \quad S(\beta) = {iN\over 2\pi}\int_{\Sigma_\beta}  A_I'\wedge (A_R - A_I)~. 
\end{equation}

\paragraph{$\beta': C_0\to \mathbb{1}$ junction} Since we are interested in the trivial element of the Witt class, we add another junction $\beta'$ between $C_0 \to \mathbb{1}$. The corresponding action is
\begin{equation}
    S = {iN\over 2\pi}\int^{\beta'}_\beta \dr(-A_I')\wedge (A_I-A_R) + {i\over 2\pi}\int^\infty_\beta \dr a \wedge (A_I - A_R)~.
\end{equation}
Using the analysis in Section~\ref{sec: codensation morphisms}, especially Eq.~\eqref{eq: bj}, we impose the boundary condition and add the boundary term
\begin{equation}
  \beta': \quad \left.a + NA_I'\right|_{\beta'} ~, \quad \left.A_I - A_R\right|_{\beta'} =\text{pure $U(1)$ gauge} , \quad S(\beta')=0~.
\end{equation}

\paragraph{$\alpha$:  $D_2\times \mathbb{1}\to D_2$}
\begin{equation}
    \begin{aligned}
        S&= {i\over 2\pi}\int^\infty_\alpha NA_L \dr A_I + \dr a(A_I-A_R)\\
        &= \underbrace{{i\over 2\pi}\int^\infty_\alpha (a+NA_L)\dr (A_I-A_R)}_{\text{decoupled trivial TQFT}} + \underbrace{{iN\over 2\pi}\int^\infty_\alpha A_L \dr A_R}_{D_2} \underbrace{- {i\over 2\pi}\int_{\Sigma_\alpha} a\wedge (A_I-A_R)}_{S(\alpha)}~. 
    \end{aligned}
\end{equation}
Following the discussion in Section~\ref{sec: top construction of local fusion junction 4d}, the decoupled trivial TQFT fields are condensed on the boundary.  Hence,
\begin{equation}
  \alpha:\quad   \left. a+NA_L\right|_\alpha , \quad \left. A_I-A_R \right|_\alpha = \text{pure $U(1)$ gauge}, \quad S(\alpha) = - {i\over 2\pi}\int_{\Sigma_\alpha} a\wedge (A_I-A_R)~.
\end{equation}

\paragraph{$\gamma$: $C_0\to D_2\times \overline{D}_2$}
The fusion outcome is $C_0$ without other terms.
\begin{equation}
    S = {iN\over 2\pi}\int^\gamma_{-\infty}(A_L-A_I')\dr A_I~.
\end{equation}

\paragraph{$\gamma'$: $\mathbb{1}\to C_0$}
\begin{equation}
    S_d = {i\over 2\pi}\int^\gamma_{\gamma'} N \dr A_I(A_L-A_I') + {i\over 2\pi}\int^{\gamma'}_{-\infty} \dr b (A_L - A_I')~.
 \end{equation}
Again, using Eq.~\eqref{eq: bj},
\begin{equation}
  \gamma':\quad   \left. A_L-A_I', \quad b-NA_I \right|_{\gamma'} = \text{pure $U(1)$ gauge}, \quad S(\gamma')=0~.
\end{equation}

\paragraph{$\delta$: $D_2 \to \mathbb{1}\times D_2$}
\begin{equation}
    \begin{aligned}
        S &= {i\over 2\pi}\int^\delta_{-\infty} \left[ \dr b (A_L-A_I') + N A_I' \dr A_R \right]\\
        &= \underbrace{{i\over 2\pi}\int^\delta_{-\infty} (A_L-A_I') \dr (b-NA_R)}_{\text{decoupled trivial TQFT}} + \underbrace{{iN\over 2\pi}\int^\delta_{-\infty}A_L \dr A_R}_{D_2}~.
 \end{aligned}
\end{equation}
The boundary conditions are
\begin{equation}
    \delta: \quad \left. A_L-A_I' \quad b-NA_R\right|_\delta = \text{pure $U(1)$ gauge}, \quad S(\delta) = 0~.
\end{equation}

\paragraph{Merging the upper and lower junctions}
As the next step, we collapse all the upper junctions into one and combine the boundary data, and similarly for the lower junctions. We get
\begin{equation}
    \begin{aligned}
        \alpha,\beta,\beta': \quad & a+NA_L, \quad a+NA_I', \quad A_I-A_R=\text{pure $U(1)$ gauge}\\
        & S_\alpha = -{i\over 2\pi}\int_{\Sigma_\alpha}a \wedge (A_I - A_R) ,\quad S_\beta={iN\over 2\pi}\int_{\Sigma_\beta}A_I' \wedge (A_R-A_I)\\
        \gamma, \gamma',\delta: \quad & b=NA_I, \quad A_L-A_I',\quad b-NA_R= \text{pure $U(1)$ gauge}\\
        & S_\gamma=S_{\gamma'}=S_\delta =0~.
    \end{aligned}
\end{equation}

\begin{figure}[H]
	\centering
	\begin{tikzpicture}
		\begin{scope}[very thick,decoration={
				markings,
				mark=at position 0.5 with {\arrow{>}}}
			] 
			\draw[postaction={decorate}] (-2,-1)--(-2,2) node[at start,left] () {${D}_2$};
			\draw[postaction={decorate}] (0,-1)--(0,2) node[at start,right] () {$\overline{D}_2$};
			\draw[postaction={decorate}] (2,-1)--(2,2) node[at start,right] () {$D_2$};
			\node at (-2.5,0.3) () {$A_L$};
			\node at (-1,0.3) () {$A_I$};
			\node at (1,0.3) () {$A_I'$};
			\node at (2.5,0.3) () {$A_R$};
		\end{scope}
	\end{tikzpicture}
	\caption{Fusing ${D}_2,\overline{D}_2,D_2$ at $[\gamma,\beta]$.}
	\label{fig: D2-D2b-D2 fusion}
\end{figure}
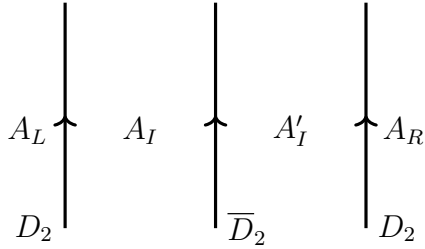

\paragraph{Global fusion} Next, we fuse the three defects $D_2,\overline{D}_2, D_2$ in the middle (Figure~\ref{fig: D2-D2b-D2 fusion}).  
The corresponding action is the sum of these defect actions, which can be rewritten as
\begin{equation}
    \begin{aligned}
        S&= {iN\over 2\pi}\int^\beta_\gamma (A_L-A_I')\dr A_I + A_I'\dr A_R\\
        &= \underbrace{{iN\over 2\pi}\int^\beta_\gamma (A_L - A_I') \dr (A_I-A_R)}_{\mathbb{Z}_N \text{ gauge theory}} + \underbrace{{iN\over 2\pi}\int^\beta_\gamma A_L \dr A_R}_{D_2}~.
    \end{aligned} \label{eq: ye}
\end{equation}
The first term corresponds to the $\mathbb{Z}_N$ gauge theory. Collecting the boundary terms at the upper junctions,
\begin{equation}
    \begin{aligned}
        S(\alpha)+S(\beta)  &= {i\over 2\pi}\int_{\Sigma_{\alpha\beta}} (a+NA_I')\wedge (A_R-A_I) \in 2\pi i \mathbb{Z} ~,
    \end{aligned}   
\end{equation}
where we used the boundary conditions $a+NA_I'$ and $A_R-A_I$ being pure $U(1)$ gauge at the upper junction. And the total boundary term at the lower junction simply vanishes. The boundary conditions of each conjugate are
\begin{equation}
    \begin{aligned}
        \alpha\beta:& \quad N(A_L-A_I') = \text{pure $U(1)$ gauge} \, &(N), \quad &&A_I-A_R = \text{pure $U(1)$ gauge} \, &(D) ~,\\
        \gamma\delta:& \quad A_L-A_I' = \text{pure $U(1)$ gauge} \, &(D), \quad &&N(A_I-A_R) = \text{pure $U(1)$ gauge} \, &(N) ~.
    \end{aligned}
\end{equation}
Following the discussion in Sections~\ref{sec: 4d method}, there is no untwisted local operator, and hence the 2d effective theory for the 3d $\mathbb{Z}_N$ gauge theory on the interval with the above boundary conditions is a trivial theory.  Thus we find
\begin{equation}
    \left[F_{D_2,\overline{D}_2,D_2}^{D_2}\right]_{\mathbb{1},\mathbb{1}} = \text{Trivial TQFT}~. \label{eq: yc}
\end{equation}

% ---------- new subsection ----- 

\subsection{Group-theoretical approach}

\subsubsection{Group theoretical construction}
For some special values of $N$, the non-invertible duality can be constructed group-theoretically~\cite{Sun:2023xxv,Kaidi:2021xfk,Choi:2022zal}.  In particular, if $N$ is odd \textit{and} if $\exists ~p$ such that $p^2=-1$ mod $N$\footnote{Notice that when $N$ is odd and such $p$ exists, we can always and will choose $p$ to be even.}, the non-invertible duality defect in a QFT $\mathcal{X}$ becomes invertible after the $ST$ gauging with the following discrete torsion.
\begin{equation}
    Z_{\widetilde{\mathcal X}}[C^{(2)}] = \sum_{c^{(2)}\in H^2(M_4,\bz_N)}Z_{\mathcal X}[c^{(2)}]~e^{{2\pi i \over N}\int_{M_4}\left( {N+1\over 2}p~c^{(2)}\cup c^{(2)} + c^{(2)}\cup C^{(2)} \right)}
\end{equation}
where $\widetilde{\mathcal{X}}$ is the QFT obtained by the above $ST$ gauging in QFT $\mathcal X$ and $C^{(2)}$ is the 2-form background field of $\bz_N$ 1-form symmetry. Utilising the duality
\begin{equation}
    Z_{\widetilde{\mathcal{X}}}[C^{(2)}] = \sum_{c^{(2)}\in H^2(M_4,\bz_N)}Z_{\mathcal X}[c^{(2)}]~e^{{2\pi i \over N}\int_{M_4}c^{(2)}\cup C^{(2)} } ~,
\end{equation}
it follows that
\begin{equation}
    Z_{\widetilde{\mathcal{X}}}[C^{(2)}]=Z_{\widetilde{\mathcal{X}}}[pC^{(2)}]~e^{{2\pi i \over N}{N+1\over 2}p \int_{M_4}C^{(2)}\cup C^{(2)}}~. \label{eq: yd}
\end{equation}
% \begin{equation}\label{eq:d_o_it}
%     Z_{\widetilde{\mathcal{X}}}[B^{(2)}] = Z_{\widetilde{X}}[p B^{(2)}]\exp\left(\frac{2\pi i}{2} \frac{N+1}{2} p \int B^{(2)}\cup B^{(2)} \right) ~.
% \end{equation}
%
This implies that the non-invertible duality defect in $\mathcal{X}$ is mapped to a $\mathbb{Z}_4^{(0)}$ invertible $0$-form symmetry in the theory $\widetilde{\mathcal{X}}$, which is an automorphism of the dual $1$-form symmetry sending $C^{(2)} \mapsto p C^{(2)}$.  The additional phase in \eqref{eq: yd} represents the mixed anomaly between the $0$-form symmetry and the $1$-form symmetry.  Let $\mathbb{D}$ be the generator of the $\mathbb{Z}_4^{(0)}$ symmetry.  As for the triality case, we dress the minimal TQFT to make $\mathbb{D}$ a gauge invariant operator, consistent of gauging the dual 1-form symmetry in $\widetilde{\mathcal{X}}$,
\begin{equation}
    D_2 = \mathbb{D}_2 \mathcal{A}^{N,p}[c^{(2)}]  \label{eq: yf}
\end{equation}
where $c$ is the gauge field for the bulk 1-form symmetry. Applying the relation \eqref{eq: yf} twice 
\begin{equation}
     Z_{\widetilde{\mathcal{X}}}[C^{(2)}] = Z_{\widetilde{\mathcal{X}}}[-C^{(2)}] ~,
\end{equation}
we see that $\mathbb{D}^2$ acts as charge conjugation of the dual $1$-form symmetry without a mixed anomaly in $\widetilde{\mathcal{X}}$.  Finally, the orientation reversal of $D_2$ is given by
\begin{equation}
    \overline{D}_2 =  \mathcal{A}^{N,-p}[c^{(2)}]{\mathbb{D}}^{-1}~.
\end{equation}
It is straightforward to check the above defects do reproduce the generic fusion rules \eqref{eq: duality fusion} following a similar approach in the triality case.

We can then proceed to work out the local fusion junctions and compute the $F$-symbols, and we find the results agree with the Lagrangian method. We will skip the details here, and only briefly list the results. For simplicity, we will choose $N = 5$ (and take $p = 2$), but it is completely straightforward to acquire the result for the generic cases. With this choice,
\begin{equation}
    D_2 = \mathbb{D} \mathcal{A}^{5,2}[c^{(2)}], \quad \overline{D}_2 = \mathcal{A}^{5,-2}[c^{(2)}]\mathbb{D}^{-1} \label{eq: ya}
\end{equation}
and the invertible 0-form symmetry $\bd$ acts on the 1-form symmetry operators as the $\bz_4$ automorphism
\begin{equation}
\bd \ca^{5,\pm 2}[c^{(2)}]\bd^{-1} = \ca^{5,\pm 2}[-2c^{(2)}] ~, \quad \bd^{-1}\ca^{5,\pm 2}[c^{(2)}]\bd = \ca^{5,\pm 2}[2c^{(2)}]~.
\end{equation}

\subsubsection{Fusion and condensation}
The 3d $\mathbb{Z}_5$ gauge theory with 0-twist coupled to the external dynamical 1-form $\mathbb{Z}_5$ gauge field $c\in H^2(M_4,\mathbb{Z}_5)$ with the following Lagrangian
\begin{equation}
    \lag_{(\mathcal{Z}_{5})_0}[c^{(2)}] = \frac{2\pi i}{5}a^{(1)} \cup \delta b^{(1)} + \frac{2\pi i}{5} a^{(1)} \wedge c^{(2)} \label{eq: ZN}
\end{equation}
represents the condensation defect $C_0$ in the path integral.  To see this, we integrate $b^{(1)}$ out to limit $a\in H^1(M_3,\mathbb{Z}_5)$ and replace $a^{(1)}$ with its Poincar\'e dual $\int_{M_3} a^{(1)} \cup \cdots = \int_{\Sigma_2}\cdots$ in the integral.  The path integral is then
\begin{equation}
 \sum_{a\in H^1(M_3,\mathbb{Z}_5)} e^{\frac{2\pi i}{5}\int_{M_3}  a^{(1)}\cup c^{(2)}} = \sum_{\Sigma_2\in H_2(M_3,\mathbb{Z}_5)}e^{{2\pi i  \over 5}\oint_{\Sigma_2} c^{(2)}} =  \sum_{i,\{m_i \in \mathbb{Z}_5\}}e^{{2\pi i \over 5}m_i\oint_{\Sigma_{2,i}} c^{(2)}}
\end{equation}
where $\{\Sigma_{2,i}\}$ are basis of the non-contractible 2-cycles on $M_3$, and we normalised $c$ in the integral so that the 1-form charge is integer quantised.  In this form, the path integral is manifestly the 1-form symmetry condensation defect.

If one condenses $a$ by restricting to a trivial cohomology element in the path integral, we see that the constrained path integral is trivial, i.e. it is the identity defect.  Using this, the morphism $C_0\to \mathbb{1}$ is implemented by condensing $a$ at the morphism junction (and along $\mathbb{1}$).

With all necessary ingredients at hand, we discuss the anyon condensation at the local fusion and morphism junctions.

\paragraph{$D_2\times \overline{D}_2=C_0\to \mathbb{1}$} Consider a fusion
\begin{equation}
    \begin{aligned}
        D_2\times \overline{D}_2 = \ca^{5,2}_{a_1}[-2c^{(2)}] \otimes \ca^{5,-2}_{a_2}[-2c^{(2)}]
    \end{aligned}
\end{equation}
where we used the $\bd$ action on the minimal TQFTs~\eqref{eq: ya}, and $a_{1,2}$ denotes the generating anyons of the minimal TQFT in each theory. To see if the theory on the RHS is the condensation defect $\mathbb{Z}_N$ gauge theory~\eqref{eq: ZN}, we first note that $a=(a_1)^2 (a_2)^2$ is coupled to the external field $c^{(2)}$ as in ~\eqref{eq: ZN} and has the topological spin $h[a]=0$.  This means that it is enough to find $b$ such that $a$ and $b$ have the correct topological spin and braiding to be the pure electric line and the pure magnetic line in $(\mathcal{Z}_N)_0$.  Let $b=(a_1)^m(a_2)^n$.  Then $h[b]=0$ and $B[a,b]=\exp\left({2\pi i\over 5}\right)$ for $b=(a_1)^{-2}(a_2)^2$.  Hence the fusion outcome is $C_0$ as expected.  Then the morphism junction $C_0\to \mathbb{1}$ is obtained by condensing the Wilson lines in $\langle (a_1)^{-2}(a_2)^{-2}\rangle =\langle a_1 a_2\rangle$, i.e.
\begin{equation}
    D_2\times \overline{D}_2 = C_0 \to \mathbb{1} \quad \text{by condensing $\langle a_1 a_2\rangle$}~. \label{eq: dd2}
\end{equation}
\paragraph{$\overline{D}_2\times {D}_2=C_0\to \mathbb{1}$} Following the same discussion, we can confirm the fusion algebra $\overline{D}_2\times D_2 =C_0$ with $a=a_1 a_2$ and $b=a_1(a_2)^{-1}$.  Hence, at the morphism junction,  
\begin{equation}
    \overline{D}_2\times {D}_2=C_0\to \mathbb{1} \quad \text{by condensing $\langle a_1 a_2\rangle$}~. \label{eq: d2d}
\end{equation}
\paragraph{$D_2\times D_2=U\times C_0 \to U$} This time, consider the fusion
\begin{equation}
    \begin{aligned}
        D_2\times \overline{D}_2 = \bd^2 \ca^{5,2}_{a_1}[2c^{(2)}] \otimes \ca^{5,2}_{a_2}[c^{(2)}]~.
    \end{aligned}
\end{equation}
We note that $\mathbb{D}^2=U$ is the charge conjugation.  Following the same argument, the minimal TQFT part corresponds to $C_0$ with $a=(a_1)^2a_2$ and $b=(a_1)^2(a_2)^{-1}$.  Again, $C_0\to \mathbb{1}$ is done by condensing $\langle (a_1)^2 a_2\rangle$. 
\begin{equation}
   D_2\times D_2=U\times C_0 \to U \quad \text{by condensing $\langle (a_1)^2 a_2\rangle$} ~.
\end{equation}
\paragraph{$\overline{D}_2\times \overline{D}_2=U\times C_0 \to U$} 
The above discussion straightforwardly applies to this fusion and morphism junctions, with $a=a_1(a_2)^2$ and $b=(a_1)^{-2}a_2$.
\begin{equation}
\overline{D}_2\times \overline{D}_2=U\times C_0 \to U \quad \text{by condensing $\langle a_1 (a_2)^2\rangle$}~.
\end{equation}

\subsubsection{Associator}
\begin{figure}[H]
	\centering
	\begin{tikzpicture}
		\begin{scope}[very thick,decoration={
				markings,
				mark=at position 0.5 with {}}
			] 
			\draw(0,1)--(0,2);
            \draw (-1,0)--(0,1);
            \draw (1,0)--(0,1);
            
            \draw (0,-1)--(-1,0);
            \draw (0,-1)--(1,0);
            \draw (-0.5,-0.5)--(0.5,0.5);
            \draw (0,-1)--(0,-2);
            
            \filldraw[blue] (0,1) circle (2pt) node[left,blue] () {1(23)}; 
        \filldraw[blue] (0.5,0.5) circle (2pt) node[right, blue] () {(23)}; 
        \filldraw[blue] (-0.5,-0.5) circle (2pt) node[left, blue] () {(12)}; 
        \filldraw[blue] (0,-1) circle (2pt) node[right, blue] () {(12)3}; 
	
    \node at (2,0) (f) {$\to$};
		\end{scope}
		\begin{scope}[shift={($(f.east) + (1.5,0)$)},very thick,decoration={
				markings,
				mark=at position 0.5 with {\arrow{>}}}
			] 
			\draw (0,-2)--(0,2);
            \filldraw[blue] (0,1) circle (2pt) node[left,blue] () {$\mathcal{I}_+$} node[right,blue] () {$\mathbb{A}_+$};
			\filldraw[blue] (0,-1) circle (2pt) node[left,blue] () {$\mathcal{I}_-$} node[right, blue] () {$\mathbb{A}_-$}; 
			\node at (1.5,0) (ff) {$\to$};
		\end{scope}
        \begin{scope}[shift={($(ff.east) + (1.5,0)$)},very thick,decoration={
				markings,
				mark=at position 0.5 with {\arrow{>}}}
			] 
			
			\draw (0,-2)--(0,2);
			
            \filldraw[blue] (0,0) circle (2pt) node[left,blue] () {$\mathcal{I}$};
		\end{scope}
	\end{tikzpicture}
	\caption{Anyon condensation in the associator 1-morphism of defects $1,2,3$}
	\label{fig: anyon condensation}
\end{figure}
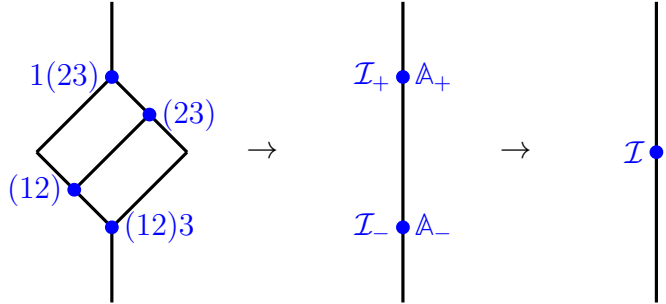

    In this subsection, we compute the associator 1-morphism for the three defects, say $1,2,3$, by first collapsing the upper junctions $(23)$ and $1(23)$ into $\mathcal{I}_+$ and similarly for the lower junctions (first arrow in Figure~\ref{fig: anyon condensation}), and finally reducing the interval to merge $\mathcal{I}_\pm \to \mathcal{I}$.  At this final step, we identify the resulting 2d theory on the associator 1-morphism using the spectrum of the local and line operators.  
To be specific, let's first denote by $\ba_{ij}$ the condensed anyons at the local fusion $i\times j$ junction.  For example, anyons in $\ba_{23}$ are condensed at the junction $(23)$ and above, and anyons in $\ba_{12}$ are condensed at the junction $(12)$ and below.  After merging the two upper junctions $1(23)$ and $(23)$ into $\mathcal{I}_+$, the anyons in $\ba_+=\ba_{1(23)}\times \ba_{23}$ are condensed and similarly at the lower junctions with $\ba_- = \ba_{12}\times \ba_{(12)3}$.

The lines that are condensed at both $\mathcal{I}_\pm$ (i.e. $\ba_+\cap \ba_-$) stretch between the two junctions.  Hence, after merging $\mathcal{I}_\pm \to \mathcal{I}$, they become a vertex operator. 
       
The line operators on $\mathcal{I}$ are the lines on three defects after quotient out $\mathbb{A}_+$ and $\mathbb{A}_-$. The remaining lines can be in the 3d minimal TQFT coupling to the bulk fields $c^{(2)}$, which are neutral under $\mathbb{A}_+ \times \mathbb{A}_-$. Quotient out those lines, we then acquire the lines in the decoupled 2d theory.

In summary, the spectrum of the decoupled 2d theory on $\mathcal{I}$ after the reduction is
       \begin{equation}
       \begin{aligned}
           \text{Vertex operators}: \quad & \ba_+\cap \ba_- \\
           \text{Line operators}: \quad & (\text{neutral lines under $\ba_+\times \ba_-$}) \backslash \bz_5\times \bz_5\times \bz_5 / (\ba_+\times \ba_-)
       \end{aligned} \label{eq: yb}
       \end{equation}
\paragraph{Example:     $\left[F^{D_2}_{D_2,\overline{D}_2,D_2}\right]_{\bi,\bi}$}
Using Eqs.~\eqref{eq: dd2} and \eqref{eq: d2d}, we identify 
    \begin{equation}
        \ba_{23}= \langle a_2a_3\rangle, \quad \ba_{1(23)}=0, \quad \ba_{12}=\langle a_1a_2\rangle, \quad \ba_{(12)3}=0
    \end{equation}
    and hence
    \begin{equation}
        \ba_+ =\langle a_2a_3\rangle, \quad \ba_{-}=\langle a_1a_2\rangle~.
    \end{equation}
Since $\ba_+\cap \ba_-=0$, there is no vertex operator on $\mathcal{I}$.  The lines that are neutral under $\ba_+\times \ba_-$ are those in $\langle a_1 a_2 a_3\rangle$ which has no overlap with $\ba_+\times \ba_-$, the lines condensed on $\mathcal{I}$.  Therefore, there is no line left on the 2d theory in the double coset~\eqref{eq: yb}.  In other words, 
\begin{equation}
    \left[F^{D_2}_{D_2,\overline{D}_2,D_2}\right]_{\bi,\bi} = \text{Trivial TQFT}~.
\end{equation}
This is consistent with the result using the Lagrangian method~\eqref{eq: yc} with $N=5$.

\paragraph{Other associators} Using the same procedures, one can compute the other associators, which are all trivial TQFT, matching the results of the Lagrangian method,
\begin{equation}
    \left[ F^{\overline D_2}_{D_2,D_2,D_2} \right]_{U,U}, \quad \left[ F^{ D_2}_{\overline D_2,D_2,D_2} \right]_{\bi,U},\quad \left[ F^{ D_2}_{D_2,\overline D_2,D_2} \right]_{\bi,\bi},\quad \left[ F^{ D_2}_{D_2,D_2,\overline D_2} \right]_{U,\bi} = \text{Trivial TQFT}~.
\end{equation}

\paragraph{Lagrangian vs group-theoretical methods} While the group-theoretical approach is simpler than the Lagrangian method, it is not available for some $N$ (of $\bz_N$ 1-form symmetry) while the Lagrangian method applies to all $N$.

\newpage
\bibliographystyle{bibstyle2017}
\bibliography{defectrefs}

\end{document}